\documentclass[12pt]{article}
\pdfoutput=1

\usepackage{cite}
\usepackage{booktabs}
\usepackage[english]{babel}
\usepackage{amsmath,amssymb,amsbsy,amstext, amsthm, simplewick}
\usepackage{hyperref}
\usepackage{graphicx}
\usepackage{amsfonts}
\usepackage{amssymb}
\usepackage[small]{caption}
\usepackage{upgreek}
\usepackage[svgnames,dvipsnames,x11names,table]{xcolor}
\usepackage{multirow} 
\usepackage{geometry}
\usepackage[hang,flushmargin]{footmisc}
\usepackage{bm}
\usepackage{braket}
\usepackage{subcaption}
\usepackage{mathtools}
\usepackage{setspace}
\usepackage{cleveref}
\usepackage{comment}
\usepackage{scalerel}
\usepackage[normalem]{ulem}
\usepackage{slashed}

\hypersetup{
    colorlinks=true,
    linkcolor={red!50!black},
    citecolor={blue!50!black},
    urlcolor={blue!80!black}
}

\usepackage{colortbl}

\makeatletter
\newlength{\apb@width}
\newcommand{\autoparbox}[2][c]{\settowidth{\apb@width}{#2}\parbox[#1]{\apb@width}{#2}}

\makeatother

\definecolor{lightgray}{gray}{0.9}

\usepackage[framemethod=default]{mdframed}
\newmdenv[skipabove=7pt,
skipbelow=7pt,
rightline=false,
leftline=false,
topline=false,
bottomline=false,
backgroundcolor=gray!10,
linecolor=gray,
innerleftmargin=5pt,
innerrightmargin=5pt,
innertopmargin=5pt,
innerbottommargin=5pt,
leftmargin=0cm,
rightmargin=0cm,
linewidth=4pt]{eBox}

\usepackage[most]{tcolorbox}
\tcbset{colback=white, colframe=black,
        highlight math style= {enhanced, 
            colframe=red,colback=red!10!white,boxsep=0pt}
        }

\crefname{table}{Table}{Tables}
\crefname{equation}{Eq.}{Eqs.}
\crefname{appendix}{App.}{Apps.}
\crefname{section}{Sec.}{Secs.}
\crefname{figure}{Fig.}{Figs.}

\numberwithin{equation}{section}

\def\beq{\begin{equation}}
\def\eeq{\end{equation}}

\def\bea{\begin{eqnarray}}
\def\eea{\end{eqnarray}}

\def\dd{{\rm d}}

\def\beq{\begin{equation}}
\def\eeq{\end{equation}}
\def\bea{\begin{eqnarray}}
\def\eea{\end{eqnarray}}

\def\dd{{\rm d}}

\def\O{{\cal O}}

\def\Mpl{M_{\rm pl}}

\def\H{{\cal H}}

\def\k{{\vec{\scaleto{k}{7pt}}}}
\def\ksub{{\vec{k}}}
\def\kp{{\!\!\vec{{\scaleto{\,\, k}{7pt}}^{\s\prime}}}}

\def\q{{\vec q}}
\def\p{{\vec p}}

\def\x{{\vec x}}

\def\y{{\vec y}}

\def\t{\texttt{t}}

\DeclareRobustCommand{\SkipTocEntry}[4]{}

\newcommand{\s}{\hspace{0.8pt}}

\definecolor{colorTC}{rgb}{.2,.7,.2}

\definecolor{blue3}{RGB}{31, 119, 180}
\definecolor{red3}{RGB}{	214, 39, 40}
\definecolor{orange3}{RGB}{255, 127, 14}
\definecolor{green3}{RGB}{44, 160, 44}

\begin{document}

\begin{titlepage}
\setcounter{page}{1} \baselineskip=15.5pt 
\thispagestyle{empty}
$\quad$
\vskip 50 pt

\begin{center}
{\fontsize{18}{18} \bf Soft de Sitter Effective Theory}
\end{center}

\vskip 20pt
\begin{center}
\noindent
{\fontsize{12}{18}\selectfont  Timothy Cohen$^{\s 1}$ and Daniel Green$^{\s 2}$ }
\end{center}

\begin{center}
\vskip 4pt
\textit{ $^1${\small Institute for Fundamental Science, University of Oregon, Eugene, OR 97403, USA}
}
\vskip 4pt
\textit{ $^2${\small Department of Physics, University of California at San Diego,  La Jolla, CA 92093, USA}}

\end{center}

\vspace{0.4cm}
 \begin{center}{\bf Abstract}
 \end{center}
\noindent  Calculating the quantum evolution of a de Sitter universe on superhorizon scales is notoriously difficult.
To address this challenge, we introduce the Soft de Sitter Effective Theory (SdSET).
This framework holds for superhorizon modes whose comoving momentum is far below the UV scale, which is set by the inverse comoving horizon. 
The SdSET is formulated using the same approach that yields the Heavy Quark Effective Theory.
The degrees of freedom that capture the long wavelength dynamics are identified with the growing and decaying solutions to the equations of motion.
The operator expansion is organized using a power counting scheme, and loops can be regulated while respecting the low energy symmetries.
For massive quantum fields in a fixed de Sitter background, power counting implies that all interactions beyond the horizon are irrelevant.  
Alternatively, if the fields are very light, the leading interactions are at most marginal, and resumming the associated logarithms using (dynamical) renormalization group techniques yields the evolution equation for canonical stochastic inflation.  
The SdSET is also applicable to models where gravity is dynamical, including inflation.
In this case, diffeomorphism invariance ensures that all interactions are irrelevant, trivially implying the all-orders conservation of adiabatic density fluctuations and gravitational waves.  
We briefly touch on the application to slow-roll eternal inflation by identifying novel relevant operators.
This work serves to demystify many aspects of perturbation theory outside the horizon, and has a variety of applications to problems of cosmological interest.
\end{titlepage}

\setcounter{page}{2}

\restoregeometry

\begin{spacing}{1.4}
\newpage
\setcounter{tocdepth}{2}
\tableofcontents
\end{spacing}

\setstretch{1.1}
\newpage

\section{Introduction}
Understanding the quantum behavior of the universe on superhorizon scales is one of the great outstanding problems in theoretical physics.  
Conceptual confusions abound, as there is no rigorously defined observable quantity that a general formulation of such a theory should compute~\cite{Bousso:1999cb,Strominger:2001pn,Witten:2001kn,Mazur:2001aa,Maldacena:2002vr,Alishahiha:2004md}.
One physical manifestation arises when attempting to quantify the predictions of eternal inflation~\cite{Freivogel:2011eg}.  
That is not to diminish the significant progress that has been made when analyzing situations that are under better control, such as extracting correlations that are generated during inflation~\cite{Salopek:1990jq,Maldacena:2002vr,Cheung:2007st,Senatore:2009cf,Senatore:2012ya,Assassi:2012et} and exploring the dynamics of quantum fields on a fixed de Sitter (dS) background~\cite{Ford:1984hs,Antoniadis:1985pj,Starobinsky:1986fx,Starobinsky:1994bd,Tsamis:1994ca, Tsamis:1996qm,Tsamis:1997za,Weinberg:2005vy,Weinberg:2006ac,Seery:2010kh,Burgess:2010dd,Rajaraman:2010xd,Marolf:2010zp, Marolf:2011sh,Marolf:2012kh,Beneke:2012kn,Akhmedov:2013vka,Anninos:2014lwa, Burgess:2015ajz,Akhmedov:2017ooy,Hu:2018nxy,Akhmedov:2019cfd,Gorbenko:2019rza,Baumgart:2019clc,Mirbabayi:2019qtx,Green:2020txs}.  
Yet, while many of these results can be explained using simple intuitive arguments, the mathematical derivations are often complex, which obscures the elegance of the final result.  
This situation demands a framework where such conclusions are obvious from the outset.

Beyond issues of theoretical curiosity, there is a pragmatic application of dS quantum field theory that is critical to the success of the inflation.
The interpretation of cosmological measurements in our Universe are consistent with the existence of so-called adiabatic modes, $\zeta$, that necessarily experienced many $e$-folds of superhorizon evolution~\cite{Hu:1996yt,Spergel:1997vq,Dodelson:2003ip}.  
The standard paradigm relies on the conservation of $\zeta$, which was first proven classically~\cite{Salopek:1990jq}, and more recently was demonstrated to all-loop-order using perturbation theory~\cite{Senatore:2012ya,Assassi:2012et}. 
Since the adiabatic modes did not evolve on superhorizon scales, the initial conditions for cosmological observables that were set by quantum fluctuations generated during inflation can be used to make predictions at much later times.
Furthermore, the conservation of $\zeta$ underpins the single-field consistency conditions~\cite{Maldacena:2002vr}, which are one of the most promising probes of inflation for future cosmological surveys~\cite{Dalal:2007cu,Alvarez:2014vva}.  
Given the central role that the adiabatic modes play in the history of our Universe, a simple all-orders proof of their conservation (that does not depend on somewhat involved diagrammatic arguments) would serve to strengthen our understanding of this fundamental fact.

The physical setting can be stated simply: we wish to understand the behavior of quantum fields in a (near) dS background.
It is therefore surprising that a more intuitive and efficient approach to calculations has thus far remained obscure.
In particular, the setup provides a natural candidate ultraviolet (UV) scale, the inverse comoving dS horizon $\Lambda_\text{UV} \sim a(t) H$, where $a(t)$ is the scale factor and $H$ is the Hubble constant.
The subject of this paper is to demonstrate that an Effective Field Theory (EFT)~\cite{Weinberg:1980wa} description emerges in the infrared (IR) when the comoving momentum $\big|\s\k\s\big| \ll \Lambda_\text{UV}$.
We will isolate the propagating degrees of freedom and determine the symmetry transformations that govern the physics in this soft limit.
These provide the necessary inputs with which to build a ``continuum EFT''~\cite{Georgi:1994qn}.

The resulting framework, the Soft de Sitter Effective Theory,\footnote{This name was proposed in~\cite{Baumgart:2019clc}.  They observed that dS perturbation theory simplifies in the soft limit, and suggested that it should be possible to discover an EFT description for the long wavelength modes.} (SdSET) provides a variety of benefits over working with the UV theory directly.
It will allow us to organize the operator structure as an expansion governed by a so-called power counting parameter, $\lambda \sim \big|\s\k\s\big|/[a(t)\s H]$.
As a modern incarnation of dimensional analysis, power counting is an extremely useful tool that enables the user to infer which operators contribute to an observable at a given order in perturbation theory, making calculations much easier to organize.
The SdSET also clarifies the behavior of quantum corrections, by naturally accommodating  ``dynamical dimensional regularization'' (dyn dim reg) to tame divergent integrals while simultaneously preserving the EFT symmetries.
Finally, by converting IR divergences of the full theory into UV divergences of the EFT, the resummation of the logarithms that appear in perturbation theory is trivialized through the use of the (Dynamical) Renormalization Group ((D)RG)~\cite{Tanaka:1975ti,Boyanovsky:1998aa,Boyanovsky:2003ui}.
We will demonstrate the power of this new EFT by performing a number of calculations in detail that will demystify known results, and will show that going to higher order in perturbation theory is significantly simplified. 
By ensuring that power counting is manifest at every step, the SdSET makes the physics transparent while simultaneously improving calculational efficiency.

By integrating out short distance physics, modern EFT approaches are useful for reducing complicated multi-scale problems to simpler descriptions.  
A technical benefit is that the connection between RG flow and dimensional analysis becomes manifest.\footnote{While this essential result was long understood~\cite{Glimm:1973kp,Appelquist:1974tg}, the arguments required complex derivations in terms of Feynman diagrams,~\emph{e.g.}~using Weinberg's theorem~\cite{Weinberg:1959nj}.  Polchinski~\cite{Polchinski:1983gv} simplified this understanding by reducing properties of renormalization to dimensional analysis from the outset.
} 
The behavior of a dS universe at long wavelengths is more than qualitatively similar to renormalization.\footnote{The connections to the RG can also be made precise in the context of dS holography (dS/CFT)~\cite{Witten:2001kn,Strominger:2001pn,Mazur:2001aa,Maldacena:2002vr,Maldacena:2011nz}, where time evolution in dS (or inflation) can be interpreted as the RG flow within a quantum field theory~\cite{Henningson:1998gx,Bianchi:2001de,Bianchi:2001kw,Strominger:2001gp,McFadden:2009fg}.}
Cosmological correlators in dS are conformally invariant; the scaling with time in the long wavelength limit plays the role of the operator scaling dimension~\cite{Creminelli:2011mw}.  
Then the question of whether quantum corrections can cause dramatic effects in the IR is equivalent to studying the loop corrections to these scaling dimensions.  
Since detailed calculations tell us that the corrections to the time evolution are at most logarithmic~\cite{Weinberg:2005vy,Weinberg:2006ac}, one would like a formalism with which to apply DRG techniques in order to resum these effects, thereby restoring the convergence of perturbation theory.
The SdSET makes the connection with the DRG plain.

Unlike conventional EFTs, where the power counting $\lambda \ll 1$ is typically determined by a ratio of physical scales, the SdSET is formulated as an expansion in terms of the physical wavenumber in Hubble units, $\lambda \sim \big|\s\k\s\big|/[a(t)\s H]$.  
Relativistic quantum fields in dS do not have well defined scalings in terms of $\lambda$.  
Instead, solving the equations of motion in an expanding background reveals a ``growing mode" and a ``decaying mode" that evolve differently as a function of time.  
These will be the degrees of freedom that we will use to formulate the SdSET, and both modes are needed if we wish to specify the dynamics in terms of a local action.  
We will be able to derive the EFT using a ``one-to-many'' mode expansion by splitting the original field into these two modes plus a hard mode that models the short distance physics.
By integrating out the hard mode, we will derive a first order EFT action, which is consistent with the number of propagating IR degrees of freedom.  
Then the fact that they have a common origin as a single UV field is encoded as a reparametrization symmetry, which relates the coefficients of operators in our effective action.  
This treatment shares several features with Heavy Quark Effective Theory~\cite{Isgur:1989vq, Eichten:1989zv, Georgi:1990um, Grinstein:1990mj}, while also introducing some novel aspects such as the need for classical boundary conditions which set the EFT correlators at the horizon.

One benefit of having manifest power counting is that the behavior of late time contributions to correlation functions is determined by the scaling dimension a given operator.
Operators that scale with negative, zero, or positive powers of $\lambda$ can be characterized as being relevant, marginal or irrelevant respectively, as is familiar from renormalizing a quantum field theory. 
Theories with massive scalars with $m^2 \sim H^2$ contain only irrelevant operators, and thus the only quantum corrections that contribute at late times are those that renormalize the parameters (and perhaps the scaling dimensions) within our EFT. 
Theories with derivatively coupled fields, in particular those that include metric fluctuations, similarly only allow for irrelevant operators.  
We will show how this simple consequence of power counting and symmetries underlies the all orders conservation of $\zeta$ and gravitational waves.  
For contrast, in the case of light scalars in dS, there can be marginal operators, which result in non-trivial DRG flow, as was also suggested previously~\cite{Podolsky:2008qq,Burgess:2010dd,Dias:2012qy}, see also~\cite{Burgess:2015ajz,Mirbabayi:2015hva,Gorbenko:2019rza,Baumgart:2019clc,Mirbabayi:2019qtx,Green:2020txs} for related recent work.  
This RG can be converted to the formalism for stochastic inflation~\cite{Starobinsky:1986fx,Starobinsky:1994bd}.  
By regulating integrals using dyn dim reg such that the symmetries of the problem are maintained at all intermediate steps, we can show that the corrections to stochastic inflation are limited to higher orders in the marginal coupling, \emph{i.e.}, they are not generated by derivative or other irrelevant corrections.

Some aspects of what we present below was previously known from investigations within the UV description.
These calculations proceed by restricting the time and loop integration range to only include the superhorizon momenta~\cite{Weinberg:2005vy,Weinberg:2006ac}.  
However, this hard cutoff approach (apparently) yields large contributions from the regulator.  
The interpretation is particularly confusing, since the cutoff is simultaneously applied to both time and momenta.  
In addition, IR divergences plague these loops.  
Computing with the SdSET resolves these issues by construction.
Since the UV dynamics have been expanded away at the level of the action, one is free to integrate over the full range of time and momentum by relying on a critical feature of (dyn) dim reg, the vanishing of scaleless integrals.
This underlies the connection between regulating the integrals and preserving the EFT symmetries.
It then becomes straightforward to interpret divergences: power-law divergences can be absorbed by matching the full theory onto the EFT, while the full theory IR divergences, \emph{e.g.} those that appear in the case of light fields in dS, become identified with UV divergences in the SdSET, so that it is straightforward to apply the DRG to resum these logs.
All of this will be accomplished without introducing any ad-hoc choices that break the low energy rules.

This paper is organized as follows.  We begin by identifying the soft modes and understanding their properties in~\cref{sec:SoftModes}.  We then identify these soft modes as the building blocks for the SdSET in \cref{sec:EFT}.  The in-in perturbation theory framework used to calculate observables, along with a discussion of how to regulate divergent integrals, is the subject of~\cref{sec:Calc}. In \cref{sec:qft}, we specialize to the problem of describing the quantum dynamics of a scalar field on a fixed dS background, and perform some example tree-level, one-loop, and two-loop calculations.  In \cref{sec:metric}, we extend the description to allow for metric fluctuations, and explore the implications for inflation.  Finally, we conclude in~\cref{sec:conclusions} and discuss some future outlook.  A brief Appendix provides some details for how to regulate the divergences that appear in an example correlator calculation.

\section{Soft Modes}
\label{sec:SoftModes}
In order to setup up an EFT, we must first specify the low energy degrees of freedom and the symmetries that emerge in the IR.
To this end, we will find it useful to decompose a single full theory field into multiple modes\footnote{This approach was first developed for Heavy Quark Effective Theory~\cite{Isgur:1989vq, Eichten:1989zv, Georgi:1990um, Grinstein:1990mj}, for reviews see~\cite{Georgi:1991mr, Neubert:1993mb, Shifman:1995dn, Wise:1997sg, Manohar:2000dt}.}
\beq
\phi(\x,t) = \phi_S(\x, t) + \Phi_H(\x,t) \ ,
\label{eq:ModeDecomp}
\eeq
where $\phi_S$ are ``soft modes'' and $\Phi_H$ are ``hard modes,'' whose separation is determined by $k_{\rm physical} \lesssim H$ and $k_{\rm physical} \gtrsim H$ respectively.  The symmetries of the soft modes will be inherited from the underlying isometries of spacetime.
That these modes factorize is related to the intuitive story we tell about modes freezing out as they cross the horizon, and our EFT description will make this manifest.  
In this section, we will use the canonical description of a UV scalar field, $\phi(\x,t)$, in a dS spacetime and will derive the consequences for the soft modes, $\phi_S(\x,t)$.

Our goal will be to derive a description of the $\phi_S$ modes using the so-called continuum EFT framework~\cite{Georgi:1994qn}, see~\cite{Manohar:1995xr, Kaplan:1995uv, Rothstein:2003mp, Kaplan:2005es, Petrov:2016azi, Manohar:2018aog, Cohen:2019wxr} for some reviews.
In particular, the fact that these soft and hard fields only have support for a limited range of momenta might naively seem to introduce a hard cutoff, in the Wilsonian EFT sense.
However, this is not the case because we will formulate the theory such that divergent integrals will be regulated using dim reg and variants thereof.  
The non-trivial fact that scaleless integrals vanish in dim reg allows us to treat both the $\Phi_H$ and $\phi_S$ fields as being defined for the full range of momenta (and times), since the regions of loop momentum that have been integrated out will make scaleless contributions to the EFT.\footnote{One insightful way to understand the interplay of full theory and EFT integrals is to appeal to the ``method of regions''~\cite{Beneke:1997zp, Smirnov:2002pj}.}
This is how continuum EFT allows one to decouple heavy physics without using an explicit cutoff, which in turn allows integrals to be regulated such that the low energy symmetries are preserved.
While this is perhaps less intuitive than a Wilsonian approach that relies on a hard cutoff, there are tremendous computational and conceptual benefits to the continuum EFT formulation, as we will see in what follows.

\subsection{Isolating the Soft Modes}
\label{sec:ScalarsIndS}
A simple context within which we can develop the SdSET is the theory of a real massive scalar field in dS spacetime.  
We write the explicit action for a free theory in dS using the FRW slicing:
\beq
\dd s^2 = -\dd t^2  + a^2(t)\s \dd\x^{\s 2} = a^2 (\tau)\big(-  \dd\tau^2 +  \dd\x^{\s 2}\,  \big) \ ,
\label{eq:lineElement}
\eeq
where $t$ is the proper time, and $\tau$ is the conformal time. The scale factor $a$ is related to the Hubble constant $H$ via Einstein's equations:\footnote{The two ways of expressing the time coordinate are related to each other using $\tau = - \exp(-H\s t)/H$.  While we will typically use $\t$, there are key points in what follows where the calculations are significantly simplified by working with $\tau$, see \emph{e.g.}~\cref{sec:TreeMatching}.} 
\begin{align}
a(\t) =  e^{\t }\qquad \Longleftrightarrow \qquad a(\tau) = -\frac{1}{H\tau} \ ,
\label{eq:ConfTime}
\end{align}
where, in order to make it manifest that all dimensionful scales are set by $H$ (we will use $H$ to characterize dS, in place of the cosmological constant), we have defined
\begin{align}
\hspace{10pt}\t \equiv H t \qquad\Longleftrightarrow \qquad \frac{\partial}{\partial \t}\s a \equiv \dot a = a \ .
\end{align}
The free-theory action including the canonical coupling to this background geometry is then
\begin{align}
S_\phi &= \int  \dd^3 x\,\dd \t\, \frac{[a(\t)\s H]^3}{H^4} \bigg[ - \frac{1}{2} \nabla_\mu \phi \nabla^\mu \phi- \frac{1}{2} m^2 \phi^2 \bigg]  \ .
\label{eq:UV_free}
\end{align}

We can develop our intuition for which modes survive in the long wavelength limit by studying the free-theory equations of motion:
\beq
\frac{1}{H^2}\Big(- \nabla_\mu  \nabla^\mu + m^2 \Big)\phi(\x,\t) =\ddot \phi(\x,\t) + 3\s \dot \phi(\x,\t) -  \frac{1}{[a\s H]^{2}}\s \partial_i \partial^i\phi(\x,\t) +  \frac{m^2}{H^2} \phi  (\x,\t) = 0 \ ,
\label{eq:EOMphiFullPos}
\eeq
where $\dot \phi$ denotes a derivative with respect to $\t$, and $i$ is a spatial index. 
Obviously, time and space must be treated independently due to the non-trivial time dependence of $a(\t)$.
The equations of motion manifest an $\x$\s -\s translation symmetry, so it is useful to Fourier transform the spatial coordinates to express \cref{eq:EOMphiFullPos} in terms of the comoving momentum $\ksub$: 
\beq
\ddot \phi\big(\k,\t\big) + 3\s \dot \phi\big(\k,\t\big) + \frac{k^2}{[a\s  H]^2} \phi\big(\k,\t\big) +  \frac{m^2}{H^2} \phi\big(\k,\t\big) = 0 \ ,
\label{eq:EOMphiFullMom}
\eeq
where $k =\big|\s\vec k\s\big|$.

Our interest is in the solutions to this equation of motion with $ k/[a\s H] \ll 1 $, which we will refer to as {\it superhorizon} modes $\phi_S\big(\k,\t\big)$; these are in contrast to the {\it subhorizon} modes $\Phi_H\big(\k,\t\big)$ with $ k/[a\s H] \gg 1$ that we want to integrate out, see \cref{eq:ModeDecomp}.
For the superhorizon modes, we can solve \cref{eq:EOMphiFullMom} to zeroth order in $\ksub$ by simply neglecting the explicit $k$-dependent term.
Using the following change of variables 
\begin{align}
\phi_S\big(\k, \t\big) = [a(\t)\s H]^{-3/2 + \nu} \varphi_S\big(\k,\t\big)\ ,
\label{eq:phiSSol}
\end{align}
it is clear that $\nu$ is determined by a $\ksub$-independent equation:
\beq
\nu^2 -\frac{9}{4}+ \frac{m^2}{H^2} = 0  \ ,
\label{eq:EOMrho}
\eeq
whose solution is 
\begin{align}
\nu =   \pm\s \sqrt{\frac{9}{4} - \frac{m^2}{H^2}} \ ,
\label{eq:rhoSol}
\end{align}
to zeroth order in $k$.  
We see that $\phi_S$ unsurprisingly has two solutions since \cref{eq:EOMrho} is quadratic, implying the existence of two independent modes.
For simplicity, will define $\nu > 0$ to be the positive solution and keep track of the negative solution with an explicit minus sign.
By construction, the solution to the equations of motion in the limit $k \to 0$ corresponds to $\varphi_S\big(\k, \t\big)  \to  \varphi_S\big(\k\s \big)$ a ($\k$\s -\s dependent) constant.

Assuming $m^2>0$, both of these superhorizon modes decay exponentially in time.  
However, for $m^2 < 9\s H^2/4$, the solution with $+\nu$ decays more slowly than the solution with $-\nu$, and so we will call them the ``growing'' and ``decaying'' modes respectively. 
The growing mode is usually the solution of observational interest, as it will dominate the late time correlations.  
Yet, we will see that both solutions are required to write down a local SdSET.

To summarize, we denote the $\ksub$\s -\s dependent part of $\phi_S$ that corresponds to the two choices in \cref{eq:rhoSol} as $\varphi_\pm\big( \k\s \big)$, where the ``+'' mode is ``growing'' while the ``$-$'' mode is ``decaying.''
Since taking $\nu < 0$ simply flips the role of the growing and decaying modes, without loss of generality, we will always take $\nu \geq 0$ such that\footnote{In many calculations that follow, we will find it more convenient write expressions using 
\begin{align}
\alpha = \frac{3}{2} -\nu \ , \qquad \text{and} \qquad \beta = \frac{3}{2} + \nu \ ,
\label{eq:DefAlphaBeta}
\end{align}
such that
\begin{align}
\hspace{14pt} \alpha + \beta = 3 \ , \qquad \text{and} \qquad \alpha - \beta = -2\s \nu \ .
\end{align}
In \cref{sec:BottomUp}, we derive the constraint $\alpha + \beta = 3$ within the EFT without appealing to the UV. 
}
\begin{align}
\phi_S\big(\s \k, \t\big) &= H \Big( [a(\t)\s H]^{-3/2 + \nu} \varphi_+(\k,\t) + [a(\t)\s H]^{-3/2 - \nu} \varphi_-\big(\s \k,\t\big) \Big) \notag\\[8pt]
&= H \Big( [a(\t)\s H]^{-\alpha} \varphi_+(\k,\t) + [a(\t)\s H]^{-\beta} \varphi_-\big(\s \k,\t\big) \Big) \ .
\end{align}
The convention to include an overall factor of $H$ implies that the power counting dimension and mass dimension of $\varphi_+$ are the same; this is consistent with the UV field having unit mass dimension.
In addition, we are restricting our scope to the parameter space with $m/H \leq 3/2$, although we do not anticipate any obstructions to generalizing to under-damped solutions with $m/H > 3/2$.\footnote{When $m/H>3/2$, $\alpha$ and $\varphi_+$ are complex.  Since $\phi$ is a real field, it follows that $\beta =\alpha^*$ and $\varphi_- = \varphi_+^*$. To our knowledge, SdSET (as constructed in this paper) is still well defined provided one is careful to impose these reality constraints.}

Now we have everything we need to write a long wavelength EFT purely in terms of the superhorizon (soft) modes by integrating out the subhorizon (hard) modes $\Phi_H(\x,\t)$.  
Furthermore, well-defined power counting will make the scaling behavior with $a(t)\s H$ manifest.
This is a feature of the EFT, as it is specifically not a property of the full theory in~\cref{eq:UV_free}; the equation of motion for $\phi\big(\s \k,\t\big)$ gives rise to two solutions with different scaling behaviors with respect to time and, as such, we cannot assign a single scaling with $a\s H$ to an operator built using $\phi\big(\s \k,\t\big)$.
In~\cref{sec:EFT}, we derive the SdSET by integrating out $\Phi_H$.

\subsection{Deriving Stochastic Initial Conditions}
\label{sec:TreeMatching}
One novel aspect of the SdSET is that it requires specifying initial conditions for the correlation functions of our EFT fields.\footnote{This is typically referred to as ``matching'' in the cosmology community, in the sense of matching modes as they cross the horizon.  To avoid confusion, we will reserve that term for its use in the EFT context in this paper, namely matching a UV theory onto an SdSET.}
For a free field, this means specifying the two-point function.
This input is responsible for the non-trivial leading-order nearly-scale-invariant power-spectrum that is a cornerstone of the inflationary paradigm.
Up to the precise numerical coefficients, we will see that these boundary conditions follow from the symmetry and scaling behavior expected from the bottom-up description in the next section.  
We will calculate these coefficients by taking the long wavelength limit of the two-point function, assuming the Bunch-Davies solution for the field.

The need to specify initial conditions results from the fact that physical wavelengths undergo cosmological redshifting.  
Our interest is in the long-wavelength modes at some specific time $t$, such that $k/[a(t) H] \ll 1$;  there always exists some earlier time $t'$ such that $k/[a(t')H] = 1$.  
Prior to $t'$, the mode is considered to be ``hard," and therefore the SdSET only characterizes mode evolution starting from initial conditions given at $t'$ or later. 
Fortunately, by construction, we have factored out the time-dependence of the long-wavelength modes such that solving the quadratic equations of motion yields constant $\varphi_\pm$.  
Therefore, at the quadratic level, the initial conditions can be fixed at any time.

To determine the correct choice of initial conditions, we write the field operator, $\phi(\x,t)$, in terms of classical solutions to the equations of motion, $\bar \phi\big(\s \k,\tau\big)$, via
\beq
\phi(\x,\tau) = \int \frac{\dd^3 k}{(2\s\pi)^3}\s e^{i\s \ksub\cdot \x} \left(\bar \phi\big(\s \k,\tau\big) a_{\ksub}^\dagger + \bar \phi^*\big(\s \k,\tau\big) a_{-\ksub} \right) \ ,
\eeq
where $a_\ksub^{\dagger}$ and $a_\ksub$ are the canonical creation and annihilation operators respectively that satisfy $\big[a_\ksub^{\dagger}\s,\s a_{\ksub'} \big] = (2\s \pi)^3\s \delta\big(\k-\kp\big)$. 
This decomposition of $\phi$ is real by construction.  
Working in the Bunch-Davies vacuum, one finds 
\beq\label{eq:mode_m}
\bar \phi\big(\s\k,\tau\big) =-i\s e^{i\s\left(\nu+\frac{1}{2}\right) \frac{\pi}{2}} \frac{\sqrt{\pi}}{2} H(-\tau)^{3 / 2} H_{\nu}^{(1)}(-k\s \tau) \ ,
\eeq
where $H_{\nu}^{(1)}$ is a Hankel function of the first kind.
Taking the limit $k\s\tau \ll 1$, we find
\beq
\bar \phi_S \big(\s\k,\tau\big)  = C_\alpha\s e^{i\s \delta_{\nu}} \frac{H}{\sqrt{2}\s k^{\frac{3}{2}-\alpha}} [a\s H]^{-\alpha} +i\s e^{-i\s \delta_\nu}  D_\beta \frac{H}{\sqrt{2}\s k^{\frac{3}{2}-\beta}} [a\s H]^{-\beta} \ ,
\label{eq:BarPhi}
\eeq
where $\delta_\nu =\frac{\pi}{4}(-3+2\s\nu)$, we used $\tau  = -1/[a\s H]$, and we defined 
\begin{subequations}
\begin{align}
C_\alpha &= 2^{1-\alpha} \s \frac{ \Gamma\big(\frac{3}{2} - \alpha\big) }{\sqrt{\pi}}
\label{eq:Calpha} \\[5pt]
D_\beta &=-2^{1-\beta }  \s \frac{\sqrt{\pi }  }{\cos (\pi\s \beta )\s\Gamma \big(\beta -\frac{1}{2}\big)} \ .
\label{eq:Cbeta}
\end{align}
\label{eq:CalphaAndCBeta}%
\end{subequations}
For reference, $C_0 =C_1 = 1$, while $D_3 = 1/3$ and $D_2 = -1$, which correspond to $m = 0$ and $m = \sqrt{2}\, H$ respectively.

Next, we want to determine the boundary conditions for the EFT fields $\varphi_{\pm}\big(\s\k,\tau\big)$.
We first notice that the field operator takes the form
\begin{align}
\phi_S(\x,t) \simeq \!\int \!\frac{\dd^3 k}{(2\s\pi)^3}\s e^{i\s \ksub\cdot \x}  \bigg[[a\s H]^{-\alpha}\s \bar{\varphi}_+ \Big(e^{i\s \delta_\nu} a^{\dagger}_\ksub + e^{-i\s \delta_\nu} a_{-\ksub} \Big) + i\s [a\s H]^{-\beta} \bar{\varphi}_- \Big(e^{-i\s \delta_\nu} a^{\dagger}_\ksub - e^{i\s \delta_\nu} a_{-\ksub}\Big)\bigg]\, ,\notag\\[2pt]
\label{eq:phiExpansion2}
\end{align}
where we have defined
\begin{subequations}
\begin{align}
\bar \varphi_+ &= C_\alpha  \frac{1}{\sqrt{2}\s k^{\frac{3}{2}-\alpha}}  \\[5pt]
\bar \varphi_- &= D_\beta \frac{1}{\sqrt{2}\s k^{\frac{3}{2}-\beta}} \ .
\end{align}
\end{subequations}
We then perform a Bogoliubov transformation (see \emph{e.g.}~\cite{Grishchuk:1990bj} for related discussion) on the creation and annihilation operators of the form
\begin{subequations}
\begin{align}
\tilde a_\ksub &= e^{i\s \delta_\nu} a^{\dagger}_\ksub + e^{-i\s \delta_\nu} a_{-\ksub} \\[7pt]
\tilde b_\ksub &=i\Big( e^{-i\s \delta_\nu} a^{\dagger}_\ksub - e^{i\s \delta_\nu} a_{-\ksub}\Big) \ .
\end{align}
\end{subequations}
These operators are real and their commutators vanish 
\begin{align}
\Big[\tilde a_\ksub^\dag \s,\s \tilde a_\ksub\Big] = \Big[\tilde b_\ksub^\dag \s,\s \tilde b_\ksub\Big]  =0\ .
\label{eq:tildecomm}
\end{align}
Then it is straightforward to derive the following vacuum expectation values
\begin{subequations}
\begin{align}
\Big\langle \tilde a_\ksub\, \tilde a_{\ksub'} \Big\rangle &= (2\pi)^3\s \delta\Big(\k+\kp\Big) \\[5pt]
\Big\langle \tilde b_\ksub\, \tilde b_{\ksub'} \Big\rangle &= (2\pi)^3\s \delta\Big(\k+\kp\Big) \ ,
\end{align}
\label{eq:CreAnnStocRanVarCondition}%
\end{subequations}
where $\langle .. \rangle \equiv \langle 0 | .. | 0 \rangle $, and $|0\rangle$ is the vacuum that is annihilated by $a_{\vec k}$.
These represent the quantum fluctuations that become the classical initial conditions for the long wavelength fields, \emph{i.e.},  $\tilde a_\ksub$ and $\tilde b_\ksub$ will be interpreted as stochastic random variables in the SdSET.  
Using the mode functions for the power spectrum in \cref{eq:CreAnnStocRanVarCondition}, the field operators take the form
\begin{subequations}
\begin{align}
\varphi_+\big(\x, t\big) &= \int \frac{\dd^3 k}{(2\s\pi)^3} e^{i\s \ksub\cdot \vec x} \s \bar{\varphi}_+\big(\s\k,t\big) \tilde{a}_{\k} \\[8pt]
\varphi_-\big(\s\k, t\big) &= \int \frac{\dd^3 k}{(2\s\pi)^3}e^{i\s \ksub\cdot \vec x}\s \bar \varphi_-\big(\s\k,t\big) \tilde{b}_{\k} \ ,
\end{align}
\end{subequations}
with (classical) power spectra 
\begin{subequations}
\begin{align}
\Big\langle \varphi_+\big(\s\k \,\big)\, \varphi_+\big(\,\kp\s \big) \Big \rangle &=\frac{C_\alpha^2}{2} \frac{1}{k^{3-2\s\alpha}}  \s (2\s\pi)^3 \s\delta\Big( \k+\kp \Big) \\[8pt]
\Big\langle \varphi_-\big(\s\k \,\big)\, \varphi_-\big(\,\kp\s \big) \Big\rangle &=\frac{D_\beta^2}{2} \frac{1}{k^{3-2\s\beta}}  \s (2\s\pi)^3 \s\delta\Big( \k+\kp \Big)\ .
\end{align}
\end{subequations}
Note that in the massless limit $\alpha \to 0$, we get the famous scale invariant power spectrum.  
In the opposite limit, $\alpha \to 3/2$, the power spectrum diverges; this divergence is proportional to $k^0$ and is a pure contact term; the sub-leading piece in this limit gives a $\log k$ power spectrum.

Additionally, we see 
\beq
\Big[\tilde a_\ksub\s ,\s \tilde b_{\ksub'} \Big] = -  2\s i\s \cos (\pi\s \alpha) \s (2\pi)^3\s \delta\Big(\k +\kp\Big) \ ,
\eeq
so that 
\begin{align}
\big[\varphi_+ (\x,\t) \s,\s \varphi_-(\x^{\s \prime},\t)\big] &=  \int \frac{\dd^3 k \,\dd^3 k'}{(2\s\pi)^6}\s e^{i\s \ksub \cdot x}\s e^{i\s \ksub' \cdot \x^{\s\prime}}\s \frac{ C_\alpha D_\beta}{2} \s\Big[ \tilde a_{\ksub}\s ,\s \tilde b_{\ksub'} \Big] \notag \\[5pt]
&= i \s C_\alpha\s D_\beta\s  \cos(\pi \alpha)\s \delta(\x-\x^{\s\prime}) \notag\\[5pt]
&=  -  \frac{i}{2\s\nu}\s  \delta(\x-\x^{\s\prime}) \ .
\label{eq:varphiCanonCom}
\end{align}
This tells us that $\varphi_-$ is the conjugate momentum to $\varphi_+$.

\subsubsection*{Non-Gaussian Correlations}
The presence of interactions in the UV theory can give rise to non-Gaussian correlations at tree level.  
This would induce higher-point statistics for $\varphi_\pm$ from boundary conditions.  
Although the overall scaling behavior is fixed, it is not restrictive enough to determine the functional form of these correlations.  
For example, an $N$-point correlation of $\varphi_+\big(\s\k,\tau\big)$ is only constrained to take the form 
\beq
\big\langle \varphi_{+}\big(\s \k_1\big) \ldots  \varphi_{+}\big(\s \k_N\big) \big\rangle =  \mathcal{K}^{-3\s(N-1) + N\s \alpha} F\big(\{\q_i \} \big) \s (2\s\pi)^3\s \delta\Big(\sum \k_i\Big) \ ,
\label{eq:NpointCorr}
\eeq 
where $\mathcal{K}$ is some reference momentum scale and $F(\{\q_i \})$ is an undetermined function of the dimensionless vectors $\q_i = \vec k_i / \mathcal{K}$.  
For calculating these correlation functions, it is useful to define
\begin{align}
\langle \s.\s.\s \rangle'\equiv \frac{1}{(2\s\pi)^{3}\s \delta\Big(\sum \k_i\Big)} \langle \s.\s.\s \rangle \ ,
\label{eq:DefPrime}
\end{align}
such that we drop the overall momentum conserving $\delta$-function, or   
\beq\label{eq:NG_scaling}
\big\langle \varphi_{+}\big(\s \k_1)\s \ldots \s \varphi_{+}\big(\s \k_N\big) \big\rangle' =  \mathcal{K}^{-3\s(N-1) + N\s \alpha}\s F\big(\{\q_i \}\big) \  .
\eeq
For quantum field theory in fixed dS, $F(\q_i)$ is additionally constrained by invariance under the group of dS isometries~\cite{Creminelli:2011mw,Maldacena:2011nz,Kehagias:2012pd,Mata:2012bx,Arkani-Hamed:2018kmz,Baumann:2019oyu,Sleight:2019mgd,Sleight:2019hfp} as well as analyticity and factorization~\cite{Arkani-Hamed:2015bza,Arkani-Hamed:2018bjr,Benincasa:2018ssx,Baumann:2020dch}.  
In general, one should also derive initial conditions for correlators involving $\varphi_-$.  However, for the focus of this paper, determining the leading contributions to $\varphi_+$ correlators, we will not need to know the initial conditions for $\varphi_-$.

\section{Soft de Sitter Effective Theory}\label{sec:EFT}
In this section, we will develop the structural aspects of our EFT for long-wavelength modes.
The EFT approach is particularly relevant when one is studying a physical setting that involves a large separation of scales, which can be leveraged to determine a parametrically small ``power counting'' parameter $\lambda \ll 1$.
For field theories of the long-wavelength modes in (quasi) dS space, we will show that the comparison between the physical momentum of a long wavelength mode,  $k_{\rm physical} = k /a$, and the scale of the horizon set by the Hubble constant $H$ (or approximately so in the case of slow-roll inflation) are exactly what is needed to setup our EFT:  $\lambda \sim k_\text{physical}/H = k/[a\s H] \ll 1$.  
It is often convenient to express this condition in terms of the comoving momentum, $k$, which is conserved by translation invariance (in the FRW slicing of dS).
Then $a\s H$ (the inverse comoving horizon) will serve as the dimensionful scale for the EFT, and we will expand in small $k$.

We will begin laying out the degrees of freedom and symmetries that characterize this EFT in \cref{sec:Rules}.  
We will then derive the quadratic action for the SdSET both from the top down by integrating out $\Phi_H$ in \cref{sec:TopDown}, and from the bottom up by simply relying on symmetries and power counting in \cref{sec:BottomUp}.
We will then turn to a discussion of how to consistently include interactions in the EFT in \cref{sec:Interactions}.
Before moving to some detailed examples, we will quickly summarize the leading power SdSET action in \cref{sec:LeadingPowerAction}.

\subsection{Defining the SdSET}
\label{sec:Rules}
In this section, we will derive the quadratic action for the SdSET, following the standard approach.
Starting from the full theory action, the first step is to specify a mode expansion of our full theory field as a separation into soft and hard modes. 
The soft modes $\varphi_\pm$ are identified as the degrees of freedom that persist to the IR.
Then we will introduce a power counting scheme and will identify a set of symmetry transformations that are inherited from the UV theory.
These tools can be used to derive the action from the bottom up, and to determine the form that interactions can take.
We will apply these tools to arrive at the quadratic action up to $\mathcal{O}(\lambda^2)$, see \cref{eq:quadActionEFT}.

\subsubsection*{Degrees of Freedom}
The solutions to the equations of motion for $\phi$ in the long-wavelength limit provide a natural candidate building block for the SdSET.  
In particular, we will identify $\phi_S$ defined in \cref{eq:ModeDecomp} with the solutions presented in \cref{eq:phiSSol}.
Explicitly, we will decompose the fundamental scalar field into three modes:\footnote{To our knowledge, splitting the soft mode into multiple fields was first introduced for NRQCD in~\cite{Luke:1999kz}.}
\begin{align}
\phi(\x,\t) &= \phi_S(\x, \t) + \Phi_H(\x,\t) \notag\\[6pt]
&= H \Big[ [a(\t)\s H]^{-3/2+\nu} \varphi_+(\x, \t) + [a(\t)\s H]^{-3/2-\nu} \varphi_-(\x, \t) \Big]+ \Phi_H(\x,\t) \ ,
\label{eq:dof_split}
\end{align}
where $0\leq \nu \leq 3/2$. 
From this top-down perspective, we split $\phi$ into $\Phi_H$ and $\varphi_{\pm}$ by working in momentum space using comoving wavenumbers: $\Phi_H\big(\p,\t\big)$ is a hard mode with support for $p/[a\s H]\gtrsim 1$ while $\varphi_{\pm}\big(\s \k,\t\big)$ have support for $k/[a\s H] \lesssim 1$.\footnote{Since $\nu > 0$, \cref{eq:dof_split} implies that $\varphi_+$ will dominate at late times.
It is therefore tempting to integrate-out $\varphi_-$ as well.  
However, correlations of $\varphi_-$ only fall off as a power law, so that integrating this field out would produce a non-local action.
}  
Then we generate an EFT by integrating out $\Phi_H$.
Critically when deriving the EFT action, we will assume that $\Phi_H$ and $\varphi_\pm$ are three-momentum eigenstates, which implies that momentum conservation should be applied at the level of the Lagrangian.
This eliminates any quadratic operators that mix the light and heavy modes, \emph{e.g.} $\mathcal{L}\, \slashed{\supset} \,\varphi_\pm\s \Phi_H$; variations with derivatives also vanish.

\subsubsection*{Power Counting}
The UV scale 
\begin{align}
\Lambda_\text{UV}(\t) = a(\t)\s H \ ,
\label{eq:LambdaUV}
\end{align}
is time dependent, which is an unusual feature of our EFT.
Fortunately, we are interested in determining equal-time correlation functions such that $\Lambda_{\rm UV}$ if fixed for a given observable.
Then we power count factors of momenta with respect to $\Lambda_\text{UV}$, by expanding in
\begin{align}
\lambda \sim \frac{k}{\Lambda_\text{UV}} = \frac{k}{a\s H} \ .
\label{eq:PowerCounting}
\end{align}
Power counting is a prescription for tracking the expansion order of observables as a Taylor series in terms of quantities that are parametrically small with respect to the scale $\Lambda_\text{UV}$.  For our purposes here, the appropriate scalings are
\begin{subequations}
\begin{align}
\t &\sim 1\\[3pt]
\x &\sim 1/\lambda \\[3pt]
\k &\sim \lambda \\[3pt]
\varphi_+(\x,\t) & \sim \lambda^{3/2-\nu} = \lambda^\alpha \\[3pt]
\varphi_-(\x,\t) & \sim \lambda^{3/2+\nu} = \lambda^\beta \ .
\end{align}
\label{eq:powercounting}%
\end{subequations}
The power counting of the fields follows from dimensional analysis when taking $\Lambda_\text{UV}$ as the fundamental scale, see~\cref{eq:dof_split}.
\subsubsection*{Symmetries}
There are two kinds of symmetries that will persist to the IR and provide non-trivial constraints on the form EFT operators can take.
The first is a remnant of the UV Lorentz symmetry, and the second is a consequence of the one-to-many mode expansion taken in~\cref{eq:dof_split}.

Spacetime symmetry is broken by our power counting, in that time and space scale differently with $\lambda$, see \cref{eq:powercounting}.
However, we still must enforce $O(3)$ invariance on any three-vector quantity, since there is no preferred spatial direction.
Another remnant of the broken spacetime symmetry is a rescaling  that trivially leaves the metric invariant:
\begin{subequations}
\begin{align}
\t &\s\s\to\s\s \t \\[3pt]
x_i &\s\s\to\s\s\frac{1}{\eta}\s x_i  \\[3pt]
a(\t) &\s\s\to\s\s \eta\s a(\t)\\[3pt]
\ksub &\s\s\to\s\s \eta\s \ksub\\[3pt] 
\varphi_+(\x,\t) &\s\s\to\s\s \eta^{3/2-\nu}\s \varphi_+(\x,\t) = \eta^\alpha\s \varphi_+(\x,\t)\\[3pt]
\varphi_-(\x,\t) &\s\s\to\s\s \eta^{3/2+\nu}\s \varphi_-(\x,\t) =  \eta^\beta\s \varphi_-(\x,\t) \ .
\end{align}
\label{eq:RescaleSym}%
\end{subequations}
This rescaling symmetry is inherited by the EFT action, and is responsible for determining where the factors of $a(\t)$ appear.
This transformation leaves UV action in~\cref{eq:UV_free} invariant when noting $\phi \to \phi$ under this rescaling, which respects the relation between $\phi$ and $\varphi_\pm$ given in~\cref{eq:dof_split}.

Additionally, when working in a \textit{fixed dS background}, the geometry possesses isometries that can be characterized in the UV theory by a constant three-vector $b_i$.  
At linear order in $b_i$, the coordinates transform as\footnote{ These isometry transformations are not homogenous in power counting.  As a result, their application will enforce relationships between operators at different orders in $\lambda$, see~\cite{Green:2020ebl} for related discussion.  
This can be interpreted as the UV symmetry being non-linearly realized within the EFT.} 
\begin{subequations}
\begin{align}
x^{i} &\s\s\to\s\s x^{i}-2\left(b_{j}\s x^{j}\right) x^{i}+b^{i}\left(\sum_{j}\big(\s x^{j}\s\big)^{2}-[a(\t)\s H]^{-2} \right) \\ 
\t &\s\s\to\s\s \t+2\s b_{j}\s x^{j} \ ,
\end{align}
\label{eq:dSIsos}%
\end{subequations}
and $b_i$ has units of $1/x_i$.
At short distances, the quadratic term is negligible, and these act like Lorentz boosts.  
In the superhorizon limit, $a\to \infty$ and one cannot neglect the $(x^j)^2$ term; the transformations in \cref{eq:dSIsos} behave like special conformal transformations.

The UV scalar fields transform trivially since only the coordinates change, $\phi(x,\t) \to \phi(x',\t')$. 
As a consequence, the SdSET fields must transform as
\begin{align}
\varphi_\pm(\x,\t) &\s\s\to\s\s \left[1-2\s\Delta_\pm\s x_i\s b^i + \big(x^{2}-[a\s H]^{-2}\big)\s b_i\s \partial^{i}-2\s x^{i}\s \vec{x} \cdot \vec{\partial} + 2\s b_i\s x^i\s \partial_{\t} \right]\s \varphi_\pm(\x,\t) \ ,
\label{eq:sc_iso}
\end{align}
where $\Delta_+ \equiv \alpha$ and $\Delta_- \equiv \beta$.  
Then taking the limit $a\to \infty$ with $\dot \varphi_\pm \to 0$, the transformation of $\varphi_\pm$ in~\cref{eq:sc_iso} is equivalent to a special conformal transformation where $\varphi_\pm$ has scaling dimension $\Delta_\pm$.  
This connection has been used to compute various correlation functions of fields in fixed dS backgrounds and small deformations thereof~\cite{Creminelli:2011mw,Maldacena:2011nz,Kehagias:2012pd,Mata:2012bx,Arkani-Hamed:2018kmz,Baumann:2019oyu,Sleight:2019mgd,Sleight:2019hfp}.  
Note also that this symmetry is generally broken during inflation; we will be careful to point out when results depend critically on constraining the properties of the EFT using this isometry.

The second kind of symmetry due to~\cref{eq:dof_split}, the fact that $\varphi_\pm$ originated from a single UV degree of freedom $\phi$, which is a ``reparametrization'' symmetry.  
In particular, we notice that this mode expansion $\phi \to \varphi_\pm$ is invariant under the transformation\footnote{Technically speaking, there is also a symmetry where we switch $\varphi_+ \leftrightarrow \varphi_-$ in \cref{eq:RPITrans}, but this provides no additional information.}
\begin{subequations}
\begin{align}
\varphi_{+} &\s\s\to\s\s \varphi_{+} +\epsilon\s [a\s H]^{\alpha-\beta}\s \varphi_{-} \\[5pt]
 \varphi_{-} &\s\s\to\s\s (1-\epsilon)\s\varphi_{-} \ .
\end{align}
\label{eq:RPITrans}%
\end{subequations}
In particular, while it respects the rescaling rule for $\varphi_\pm$ in \cref{eq:RescaleSym}, this symmetry mixes operators that are different orders in the $\lambda$ power counting expansion, see~\cref{eq:powercounting}.
The presence of such a redundancy is familiar from other examples of one-to-many mode expansions that yield EFTs, \emph{e.g.} the ReParametrization Invariance (RPI) of Heavy Quark Effective Theory~\cite{Luke:1992cs}. 
From the bottom-up, we will enforce that the SdSET respects RPI.
As we will see, RPI will be critical to our argument that the quadratic action is unique, see~\cref{sec:BottomUp}.

\subsubsection*{Stochastic Initial Conditions}

The initial conditions for the fields $\varphi_\pm$ obey classical statistical fluctuations.  In particular, in the absence of interactions (see the discussion around \cref{eq:NpointCorr} above), the power spectra are given by classical correlations
\begin{subequations}
\begin{align}
\Big\langle \varphi_{+}\big(\s\k\,\big)\s \varphi_{+} \big(\,\kp\s\big) \Big\rangle_c &= \frac{C_\alpha^2 }{2}  \frac{1}{k^{3-2\alpha}} \s(2\s\pi)^3\s \delta\big(\k+\kp\big) \label{eq:TreePowerSpecPhiPlus}  \\[8pt]
\Big\langle \varphi_{-}\big(\s\k\,\big)\s \varphi_{-} \big(\,\kp\s\big) \Big\rangle_c &= \frac{D_\alpha^2 }{2}  \frac{1}{k^{3-2\beta}} \s(2\s\pi)^3\s \delta\big(\k+\kp\big) \ .
\end{align}
\label{eq:pluspower}%
\end{subequations}
While the scaling behavior of these boundary conditions are consistent with our power counting (as required), the precise coefficients $C_\alpha$ and $D_\beta$ must be determined from the UV theory, see \cref{eq:CalphaAndCBeta}.
Again, we emphasize that UV interactions will typically generate additional non-Gaussian contributions to the initial conditions for $\varphi_+$, \emph{e.g.}~see~\cref{sec:NGIC} below for an example.

\subsection{Free SdSET From the Top Down}
\label{sec:TopDown}
Now we can apply these rules to derive the EFT from the top down.
The starting point is to plug the mode decomposition in \cref{eq:dof_split} into the full theory Lagrangian in \cref{eq:UV_free}.
Recall that the $\phi_S$ and $\Phi_H$ fields are momentum eigenstates.
We invoke this property to eliminate any terms that violate momentum conservation, \emph{e.g.} there are no terms in the quadratic action that mix $\phi_S$ with $\Phi_H$.
Hence, the only allowed terms in the quadratic action for $\varphi_{\pm}$ are
\begin{align}
\hspace{-10pt} S_{2,\pm} &= \int \dd^3 x\, \dd \t\, \frac{[a\s H]^3}{H^4}\s \frac{1}{2}\s\bigg[ [a\s H]^{-2\alpha} \big(\dot \varphi_+ -\alpha\s \varphi_+\big)^2   - m^2 [a\s H]^{-2\s\alpha}\s\varphi_+^2  \notag\\[4pt]
& \hspace{120pt}+ [a\s H]^{-2 \beta} \big(\dot \varphi_- -\beta\s \varphi_-\big)^2  - m^2\s [a\s H]^{-2\beta}\s\varphi_-^2 \notag\\[4pt]
& \hspace{120pt} + 2\s [a\s H]^{-\alpha-\beta} \Big( \big(\dot \varphi_+ -\alpha\s \varphi_+\big)\big(\dot \varphi_- -\beta\s \varphi_-\big)-  m^2\s \varphi_+\s \varphi_-  \Big) \notag\\[4pt]
& \hspace{120pt} - \partial_i \Big( [a\s H]^{-\alpha}\s \varphi_+ + [a\s H]^{-\beta} \s\varphi_- \Big)  [a\s H]^{-2} \notag\\[4pt]
&\hspace{175pt} \times \partial^i \Big( [a\s H]^{-\alpha}\s \varphi_+ + [a\s H]^{-\beta}\s \varphi_- \Big) \bigg] \ .
\label{eq:S2pmStart} 
\end{align}
We can use various relations to simplify this action.  
In the first line, we can rewrite $\varphi_+\s \dot \varphi_+ = 2\s \dd(\varphi_+)^2/\dd \t$ and use integration by parts, $H^2\s \big( \alpha^2 -3\s \alpha\big) +m^2 = 0$, and $\dot H = 0$; the second line can be simplified using the same steps and $H^2\s \big( \beta^2 -3\s \beta\big) +m^2 = 0 $.
In the third line, we use $H^2\s \alpha\s \beta -m^2=0$ and $\alpha +\beta = 3$, which follow directly from \cref{eq:rhoSol,eq:DefAlphaBeta}, to remove all the $\varphi_+\s \varphi_-$ terms.  
Finally, we use \cref{eq:DefAlphaBeta} to express all the exponents in terms of $\nu$.  
This yields the following simpler expression of the action that is equivalent to \cref{eq:S2pmStart}:  
\begin{align}
S_{2,\pm} = \int \dd^3 x\, \dd \t\,  \frac{1}{2}\s\bigg[ &[a\s H]^{2\s\nu}\s\dot \varphi_+ ^2  + [a\s H]^{-2\s\nu}\s \dot \varphi_-^2  
  + 2\s  \dot \varphi_+ \s\dot \varphi_- - 2\s\nu\s \big(\dot \varphi_+\s \varphi_- - \varphi_+\s\dot\varphi_-\big) \notag \\[2pt]
& -[a\s H]^{2\s\nu-2}\s \partial_i\s  \varphi_+\s\partial^i\s  \varphi_+-[a\s H]^{-2\s\nu-2}\s \partial_i \s \varphi_-\s\partial^i \s \varphi_--2\s [a\s H]^{-2} \s  \partial_i  \s\varphi_+\s\partial^i\s \varphi_- \bigg] \ .\notag\\[2pt]
\label{eq:S2pmSomeSimp}
\end{align}
Then we power count using \cref{eq:powercounting}; we will keep terms up to $\mathcal{O}(\lambda^2)$.
Due to the symmetry relations in \cref{eq:RescaleSym}, power counting is equivalent to simply tracking explicit factors of $[a\s H]$.
Noting that $\nu \geq 0$, we drop any terms that fall faster than $[a\s H]^{-2}\s$:
\begin{align}
S_{2,\pm} &= \int \dd^3 x\, \dd \t\,  \frac{1}{2}\s\bigg[[a\s H]^{2\s\nu}\s\dot \varphi_+ ^2  + 2\s  \dot \varphi_+\s \dot \varphi_- - 2\s\nu\s \big(\dot \varphi_+\s\varphi_- -  \varphi_+\s\dot\varphi_-\big) \notag\\[3pt]
&\hspace{90pt}-[a\s H]^{2\s\nu-2}\s  \partial_i \s \varphi_+\s\partial^i \s \varphi_+ -2\s [a\s H]^{-2} \s  \partial_i\s  \varphi_+\s\partial^i\s  \varphi_- \bigg] \ .
\end{align}
Next, we use integration by parts to rewrite some of the terms:
\begin{align}
S_{2,\pm} = \int \dd^3 x\, \dd \t\,  \frac{1}{2}\s\bigg[&-\frac{1}{2}\s\varphi_+\s [a\s H]^{2\s\nu}\s\bigg(2\s \ddot{\varphi}_+ +4\s \nu\s  \dot{\varphi}_+  - 2\s[a\s H]^{-2}\s \partial^2\s \varphi_+\bigg) \notag\\[4pt]
& - \varphi_- \bigg( 2\s\ddot{\varphi}_+ +4\s \nu\s \dot{\varphi}_+ - 2\s[a\s H]^{-2}\s\partial^2\s \varphi_+\bigg) \bigg] \ .
\end{align}
Then we perform a field redefinition on $\varphi_-$:\footnote{The benefit of this field redefinition can be seen by noticing that $[a\s H]^{2\s\nu} \s \dot\varphi_+^2$ only contributes to the $\varphi_-$ equations of motion.  Removing these terms from the action by a field redefinition makes their impact on correlators of $\varphi_+$ apparent from power counting.}
\begin{align}
\varphi_- \s\s\to\s\s \varphi_- + \frac{1}{2}\s [a\s H]^{2\s\nu}\s \varphi_+ \ ,
\end{align}
which yields the following form of the action after some integration by parts:
\begin{align}
S_{2,\pm} = \int \dd^3 x\, \dd \t\,  \bigg[ \dot{\varphi}_+\s \dot{\varphi}_- -\nu\s \big(\dot \varphi_+\s\varphi_- -  \varphi_+\s\dot\varphi_-\big) - [a\s H]^{-2}\s\partial_i\s\varphi_+\s\partial^i\s \varphi_- \bigg] \ .
\end{align}
Finally, we note that the $\varphi_+$ and $\varphi_-$ fields each represent one degree of freedom, such that their equations of motion should only contain terms with single time derivatives, see~\cite{Weinberg:2008hq} for a nice discussion \emph{e.g.}~footnote 1.
This implies that we should interpret $\dot{\varphi}_- \dot{\varphi}_+$ as a contribution to interactions, and in fact as we will show below this term is power suppressed and appears at $\mathcal{O}(\lambda^4)$.
By construction, this immediately implies that the leading power equations of motion are simply
\begin{align}
\hspace{-9.5pt}\dot{\varphi}_+ = 0 \qquad \text{and}\qquad \dot{\varphi}_- = 0 \ .
\label{eq:leadingEOM}
\end{align}
This is a manifestation of the expected superhorizon behavior, confirming that we have identified the long wavelength modes of interest.
Including the leading power suppressed term in the action, the equations of motion for $\varphi_+$ and $\varphi_-$ are then
\begin{align}
\dot{\varphi}_+ = \frac{1}{2\s \nu\s [a\s H]^2}\s \partial^2\s \varphi_+ \qquad \text{and} \qquad \dot{\varphi}_- = -\frac{1}{2\s \nu\s [a\s H]^2}\s \partial^2\s \varphi_- \ .
\label{eq:EOMvarphipm}%
\end{align}
which immediately tells us that 
\begin{align}
S_{2,\pm} \supset \int \dd^3 x\, \dd \t\, \dot{\varphi}_+\s \dot{\varphi}_-  = -\int \dd^3 x\, \dd \t\, \frac{1}{4\s \nu^2\s [a\s H]^4}\s \partial^2\s \varphi_+ \s\partial^2\s \varphi_- \sim \mathcal{O}(\lambda^4) \ .
\end{align}
Finally, we get the following form for the quadratic action including terms up to $\mathcal{O}(\lambda^2)$:
\begin{align}
\tcboxmath{
S_{2,\pm} = \int \dd^3 x\, \dd \t\,  \bigg[ -\nu\s\big(\dot \varphi_+\s\varphi_- -  \varphi_+\s\dot\varphi_-\big) - \frac{1}{ [a\s H]^{2}}\s\partial_i\s\varphi_+\s\partial^i\s \varphi_- \bigg] +\mathcal{O}(\lambda^4) \ .}
\label{eq:quadActionEFT}
\end{align}
Note that we could choose to rescale the unobservable field $\varphi_- \to \varphi_-/(2\s\nu)$ to canonicalize the kinetic term, but we will leave this normalization as is because it leads to simpler notation in what follows.\footnote{By inspection of~\cref{eq:quadActionEFT}, something non-trivial is occurring when $\nu \to 0$.
From \cref{eq:rhoSol}, we see that this corresponds to the critical mass $m = (3/2) H$ where the $\varphi_\pm$ modes become degenerate.
For larger masses, $\nu$ becomes imaginary and we recognize this as the (real) non-relativistic action for a single complex scalar, after imposing $\varphi_- =\varphi_+^*$.  
While we see no obstruction to deriving a version of the SdSET that is valid in this regime, doing so is beyond the scope of this work.}

\subsubsection*{Canonical Commutation Relations}
Given the quadratic action in~\cref{eq:quadActionEFT}, it is straightforward to derive that $\varphi_-$ is the conjugate momentum to $\varphi_+$.
Then one simply applies the standard techniques of second quantization, which yields the canonical commutation relations that we found from the top down in~\cref{eq:varphiCanonCom}.
For the derivation above, we appealed to the commutators for the creation and annihilation operators.  
Alternatively, one can work with the fields directly, starting with the commutator for the UV fields,
\beq
\big[\phi(\x,t) \s,\s \dot \phi(\y,t) \big] \sim \frac{1}{\sqrt{-g}}\s \delta(\x-\y\s) \ ,
\eeq
where the $\sqrt{-g}$ is forced on us by diffeomorphism invariance or, more simply, by requiring consistency with the scaling relations in~\cref{eq:RescaleSym}: $a\to \lambda\s a$, $\x \to \lambda^{-1}\s \x$, and $\phi \to \phi$.  
Using the mode expansion in~\cref{eq:dof_split}, this commutator becomes 
\beq
\big[\phi(\x,t) \s,\s \dot \phi(\y,t) \big] \simeq  [a\s H]^{-\alpha -\beta}\s (\beta - \alpha)\s H^3\s \big[ \varphi_+\s,\s \varphi_-\big] = 2\s \nu\s H^3\s [a\s H]^{-\alpha-\beta}\s \frac{i}{2\s\nu}\s \delta(\x-\y\s) \ .
\eeq
This provides a nice confirmation that our approach is self-consistent.

\subsection{Free SdSET From the Bottom Up}
\label{sec:BottomUp}
In \cref{sec:TopDown}, we derived the low-energy action starting from a specific UV theory (the free real scalar field in a dS background), plugging in a mode decomposition, and integrating out the hard fluctuations.  
Any self-respecting EFT should also be constructable from the bottom-up directly without appealing to the UV, by identifying the correct low energy degrees of freedom, and imposing a power counting prescription and a set of symmetries.
We will show how to derive the quadratic action~\cref{eq:quadActionEFT} from the bottom up in what follows.

\subsubsection*{$\bm{\alpha +\beta = 3}$}
We have already seen how critical the EFT constraint $\alpha +\beta = 3$ has been in deriving the SdSET quadratic action. 
While it was easy to see how this arose as a consequence of the UV description, it is natural to wonder if it is an all-orders statement within the EFT.  
In particular, $\alpha$ and $\beta$ can be changed through integrating out the short distance modes, for example though a correction to the effective mass of $\phi$ or an anomalous dimension, see \emph{e.g.}~\cite{Green:2020txs}.
To this end, our first goal towards understanding the bottom-up structure is to show that we can derive this fact within the EFT, which implies it cannot be altered by matching or running effects.

To see this constraint from the bottom up, we will assume that $\alpha + \beta \neq 3$, and will then show we can redefine $\alpha$ and $\beta$ such that $\alpha+\beta =3$.  
We start with the dominant kinetic term (which as we will see below can be derived from the bottom-up):  
\beq
{\cal S} \supset \int \dd^3 x \, \dd \t \, \rho\s [a\s H]^{3-\alpha-\beta}\s \big[\dot \varphi_+\s \varphi_- -  \varphi_+\s\dot{\varphi}_-\big] \ ,
\eeq
where $\rho$ is a constant.
We notice that the equations of motion from this term alone are
\beq
2\s\dot{\varphi}_{\pm} + (3-\alpha -\beta)\s H\s  \varphi_{\pm} = 0 \ .
\eeq
Our goal in defining $\alpha$ and $\beta$ was to factor out the time dependence such that when $k\to 0$, $\dot \varphi_\pm =0$.  
Note that we have the freedom to redefine the fields 
\beq
\varphi_{\pm} = \widetilde \varphi_{\pm}\s [a\s H]^{-(3-\alpha -\beta)/2} \ ,
\label{eq:FieldRedef}
\eeq
such that
\beq
\tilde \alpha = \alpha + \frac{(3-\alpha -\beta)}{2} \qquad \text{and} \qquad \tilde \beta =  \alpha + \frac{(3-\alpha -\beta)}{2}  \ .
\eeq
For this choice of fields, we see that $\tilde \alpha + \tilde \beta = 3$.  
Furthermore, under this change, the action becomes
\beq
{\cal S} \supset \int \dd^3 x \, \dd \t \, \rho\s [a\s H]^{3-\alpha-\beta}\s \big[\dot \varphi_+\s \varphi_- - \varphi_+ \s\dot\varphi_-\big]  =  \int \dd^3 x \, \dd \t \, \rho\s\big[\dot{\widetilde \varphi}_+\s \widetilde \varphi_- - \widetilde \varphi_+ \s \dot{ \widetilde \varphi}_-\big] \ ,
\eeq
and thus this field redefinition does not introduce any new terms in the actions for $\widetilde \varphi_\pm$.  
In addition, \cref{eq:FieldRedef} preserves the RPI symmetry in~\cref{eq:RPITrans}; specifically, $\tilde \alpha -\tilde \beta$ is unchanged by the rescaling.  
As result, we are free to enforce $\alpha +\beta  =3$ as we derive the quadratic action for the SdSET from the bottom up.

\subsubsection*{Quadratic Action}
The EFT degrees of freedom are $\varphi_\pm$.  
In addition, we have the rescaling symmetry in \cref{eq:RescaleSym} that should leave the action invariant.  
Under this symmetry, $\varphi_+(\x,t)$ and $\varphi_-(\x,t)$ are assigned scaling dimensions $\alpha$ and $\beta$ with $\alpha +\beta =3$.  
Using these rules, we can writing down the full set of consistent terms that could contribute to the effective action truncated to two fields, and up to two temporal or spatial derivatives
\begin{align}
\hspace{-8pt} S_2 = \int \dd^3 x\, \dd \t\, \frac{[a\s H]^3}{H^4}\s \Bigg[ &[a\s H]^{3-2\s \alpha} \bigg(\frac{m_+^2}{2\s H^2}\s \varphi_+^2 + \frac{\chi_+}{2}\s \dot \varphi_+^2 \bigg) +[a\s H]^{1-2\s \alpha}\s \frac{\kappa_{+}}{2}\s \partial_i \varphi_+\s \partial^i \varphi_+ \notag \\  
&+ [a\s H]^{3-2\s \beta }\s \bigg(\frac{m_-^2}{2\s H^2}\s \varphi_-^2+ \frac{\chi_-}{2}\s\dot \varphi_-^2\bigg)  + [a\s H]^{1-2\s \beta}\s \frac{\kappa_{-}}{2} \s \partial_i \varphi_-\s \partial^i \varphi_-  \notag \\
&+\frac{m_{\pm}^2}{H^2}\s \varphi_+\s \varphi_-   + \rho\s \dot \varphi_+\s \varphi_- + \chi_\pm \s\dot{\varphi}_+\s \dot{\varphi}_- + [a\s H]^{-2}\s \kappa_{\pm}\s  \partial_i \varphi_+\s \partial^i \varphi_-  \Bigg]\ ,
\label{eq:bottomupS2Start}
\end{align}
where we have used integration by parts to combine redundant terms; the free parameters are $m_+^2$, $\chi_+$, $\kappa_+$, $m_-^2$, $\chi_-$, $\kappa_-$, $m^2_\pm$, $\rho$, $\chi_\pm$, and $\kappa_\pm$ (note that we use the ``$\pm$'' subscript here to denote a single parameter that multiplies a term with $\varphi_+$ and $\varphi_-$).
This action has many more parameters than the action we derived from the top down.  
For example, we see that mass parameters $m^2_+$, $m^2_-$ and $m^2_\pm$ appear to be allowed, even though they vanished above.
Constraints like this one should be realized within the EFT without appealing to the particulars of a UV completion.

Our first step is to notice we have not fully accounted for the symmetries of our EFT.  
As mentioned in \cref{sec:Rules} above, our fields $\varphi_\pm$ respect an RPI symmetry that further constrains the form of the quadratic action.  
The explicit transformations are given in \cref{eq:RPITrans}, and they relate terms that are different orders in the $\lambda$ power counting expansion.  
To linear order in $\epsilon$, we find
\begin{subequations}
\begin{align}
[a\s H]^{3-2\s \alpha}\s\frac{m_{+}^2}{2\s H^2}\s \varphi_{+}^2 &\s\s\to\s\s [a\s H]^{3-2\s \alpha}\s\frac{m_{+}^2}{2\s H^2}\s \big(\varphi_{+}^2 + 2\s\epsilon [a\s H]^{-\beta +\alpha}\s \varphi_+ \varphi_-\big) \\[8pt]
[a\s H]^{3-2\s \beta }\s\frac{m_{-}^2}{2\s H^2}\s \varphi_{-}^2 &\s\s\to\s\s [a\s H]^{3-2\s \beta }\s\frac{m_{-}^2}{2\s H^2}\s (1-2\s\epsilon)\s \varphi_-^2 \\[8pt]
\frac{m_{\pm}^2}{H^2}\s \varphi_{+}\s \varphi_{-} &\s\s\to\s\s \frac{m_{\pm}^2}{H^2} \Big((1-\epsilon)\s \varphi_{+}\s \varphi_- + \epsilon\s [a\s H]^{-\beta +\alpha}\s \varphi_- ^2 \Big) \ .
\end{align}
\end{subequations}
Using $\alpha + \beta = 3$, we see that invariance of the action requires $m_{\pm}^2 = m_+^2 = m_-^2 \equiv m^2$.  
Identical reasoning applied to the two-spatial-derivative terms implies $\kappa_{\pm} =  \kappa_+ = \kappa_- \equiv \kappa$.

Next, we perform this same exercise for the terms with time derivatives.
Again truncating to linear order in $\epsilon$, we find
\begin{subequations}
\begin{align}
[a\s H]^{3-2\s \alpha}\s\frac{\chi_+}{2} \dot \varphi_{+}^2 &\s\s\to\s\s [a\s H]^{3-2\s \alpha}\s\frac{\chi_+}{2} \s \Big(\dot \varphi_{+}^2 + 2\s\epsilon [a\s H]^{\alpha-\beta}\s \dot \varphi_+\s \dot \varphi_- + 2\s \epsilon\s (\alpha-\beta)\s [a\s H]^{\alpha-\beta } \dot \varphi_+\s \varphi_-\Big) \notag\\[1pt]
\\[1pt]
 [a\s H]^{3-2\s \beta}\s\frac{\chi_-}{2}\s \dot\varphi_{-}^2 &\s\s\to\s\s [a\s H]^{3-2\s \beta }\s\frac{\chi_-}{2}\s (1-2\s\epsilon)\s \dot\varphi_-^2 \\[9pt]
\chi_{\pm}\s \dot\varphi_{+}\s \dot\varphi_- &\s\s\to\s\s \chi_{\pm}\s\Big( (1-\epsilon)\s \dot\varphi_{+}\s \dot\varphi_- +  \epsilon\s [a\s H]^{\alpha-\beta }\s \dot\varphi_-^2 + \epsilon\s (\alpha-\beta)\s[a\s H]^{\alpha-\beta }\s  \dot\varphi_-\s \varphi_- \Big)\\[9pt]
\rho\s \dot\varphi_+\s \varphi_- &\s\s\to\s\s \rho\s \Big((1-\epsilon) \dot\varphi_+\s \varphi_-  + \epsilon\s [a\s H]^{\alpha-\beta}\s \dot{\varphi}_-\s \varphi_- + \epsilon\s(\alpha-\beta)\s [a\s H]^{\alpha-\beta}\s \varphi_-^2\Big) \ . \label{eq:rhoTermVar}
\end{align}
\end{subequations}
Comparing the two time derivative terms, we see that we need $\chi_{\pm} = \chi_{+} =  \chi_- \equiv \chi$.  
The $\dot{\varphi}_+\s \varphi_-$ terms imply that $\rho = (\alpha-\beta)\s \chi$, and the second two terms in \cref{eq:rhoTermVar} combine into a total derivative since
\begin{align}
[a\s H]^{\alpha-\beta}\s\dot\varphi_-\s \varphi_-&= [a\s H]^{\alpha-\beta}\s\frac{1}{2}\s \partial_t\s \big(\varphi_-^2\big) \notag\\
&= - \frac{1}{2}\s (\alpha-\beta)\s [a\s H]^{\alpha-\beta}\s \varphi_-^2 + \text{total derivative} \ .
\end{align}
We conclude that imposing the symmetry in \cref{eq:RPITrans} tells us that there are only three independent coefficients, $m^2$, $\chi$, and $\kappa$.

Having fixed these relations between coefficients, we can now return to the non-zero mass term $m^2$.  
Noting that we have not placed any constraints on $\alpha$ and $\beta$ beyond $\alpha+\beta = 3$, we will now show that we can always absorb the mass parameter $m^2$ into a redefinition of $\alpha$ and $\beta$, using 
\begin{align}
\alpha \s\s\to\s\s \alpha + \vartheta \qquad \text{and} \qquad \beta \s\s\to\s\s \beta-\vartheta \ ,
\end{align}
which respects $\alpha + \beta = 3$ and shifts the EFT fields by
\begin{align}
\varphi_{\pm} \s\s\to\s\s [a\s H]^{\pm \vartheta}\s \varphi_{\pm} \ .
\end{align}
This shifts the time dependent terms in the action by 
\begin{subequations}
\begin{align}
[a\s H]^{3-2\s \alpha}\s \frac{\chi}{2}\s \dot\varphi_{+}^2  &\s\s\to\s\s [a\s H]^{3-2\s \alpha} \frac{\chi}{2}\s  \Big(\dot\varphi_{+}^2 +  \big(\vartheta^2 +2\s \nu\s \vartheta\big)\s \varphi_{+}^2\Big) + \text{total derivative}
\\[8pt]
[a\s H]^{3-2\s \beta}\s \frac{\chi}{2}\s\dot\varphi_{-}^2  &\s\s\to\s\s [a\s H]^{3-2\s \beta} \frac{\chi}{2}\s\Big(\dot\varphi_{-}^2 +  \big(\vartheta^2 + 2\s\nu\s\vartheta\big)\s \varphi_{-}^2\Big) + \text{total derivative}\\[8pt]
(\alpha-\beta)\s \chi \dot\varphi_+\s \varphi_- &\s\s\to\s\s (\alpha-\beta+2\s \vartheta)\s \chi\s \big(\dot\varphi_+\s \varphi_- +\s \vartheta\s\varphi_+\s \varphi_-\big) \\[8pt]
\chi\s \dot\varphi_+ \dot\varphi_- &\s\s\to\s\s\chi\s\Big( \dot\varphi_+\s \dot\varphi_- - 2\s \vartheta\s \dot\varphi_+\s \varphi_- -  \vartheta^2\s \varphi_+\s \varphi_-\Big) + \text{total derivative} \ .
\end{align}
\end{subequations}
We see that all of these shifts combine into 
\begin{align}
\frac{m^2}{H^2} \s\s\to\s\s \frac{m^2}{H^2} + \chi\s \big(\vartheta^2 + 2\s\nu\s \vartheta\big) \ .
\end{align}
We conclude that an appropriate choice of $\vartheta$ can be used to set $m^2 = 0$, \emph{i.e.}, the mass parameter can be absorbed into the parameters $\alpha$ and $\beta$.
This could have been anticipated from the top down, since from this point of view the mass determines these parameters, see \emph{e.g.}~\cref{eq:rhoSol}.
Putting this together, the quadratic action of our EFT is 
\begin{align}
S_2 = \int \dd^3 x\, \dd\t\s \Bigg[ & \chi\s \bigg( [a\s H]^{-2\s \alpha+3}\s  \frac{1}{2}\s \dot\varphi_+^2 + [a\s H]^{-2\s \beta +3 }\s \frac{1}{2}\s \dot\varphi_-^2 +\dot\varphi_+\s \dot\varphi_- -\nu\s \big(\dot\varphi_+\s \varphi_- - \varphi_+\s\dot\varphi_-\big)\bigg) \notag\\[3pt]
& + [a\s H]^{-2}\s \kappa\s \bigg( \partial_i\s \varphi_+\s \partial^i \varphi_-  +[a\s H]^{-2\s \alpha}\s \frac{1}{2}\s \partial_i \varphi_+\s \partial^i \varphi_+ + [a\s H]^{-2\s \beta}\s \frac{1}{2}\s \partial_i \varphi_-\s \partial^i \varphi_- \bigg)\Bigg]  \ . \notag\\[2pt]
\end{align}

So far, this derivation applies equally well if the background geometry is fixed dS space, or if gravity is dynamical as is the case for applications to inflation.  
In a fixed de~Sitter background, the action is additionally constrained by the isometries in~\cref{eq:dSIsos}.  
In particular, the SdSET is not manifestly Lorentz invariant, and so it is natural to wonder if these additional relations between time and space can impose the constraints between the operators with temporal and spatial derivatives.
It should come as no surprise that enforcing this symmetry on the EFT will fix the relative coefficients $\chi = -\kappa$, as we expect for theories with UV Lorentz invariance, see the top down results derived above.  

It is straightforward to check that the variation of the two-time derivative terms are cancelled by the variations of the gradient terms when $\kappa = -\chi $, and so we will not show it explicitly here.  
The non-trivial check is that the variation of the single-time derivative term 
\beq\label{eq:time_iso}
 \delta \big(-  \nu\s \chi \s \big(\dot\varphi_+\s \varphi_--\varphi_+\s\dot\varphi_-\big) \big) = -2\s \nu\s b_i\s \chi\s a^{-2}\s \Big(\big(\partial^i \varphi_+\big)\s \varphi_- -  \varphi_+\big(\partial^i \varphi_-\big)\Big) \ ,
\eeq
is cancelled by the only other contribution that does not arise in a Lorentz invariant example, the scaling of $\varphi_\pm$ with a non-trivial dimension $\Delta_\pm$.  
Using $\alpha = 3/2-\nu$ and $\beta = 3/2+\nu$, we find the variation of the gradient term is
\bea
\delta \big( \kappa\s [a\s H]^{-2}\s \partial_i  \varphi_+\s \partial^i \varphi_- \big) &=&\kappa\s [a\s H]^{-2}\s \Big[ \Big( \partial_i \big(-2\s \alpha\s \vec b\cdot \x\s\big) \Big) \s\varphi_+\s  \partial^i \varphi_- +\partial_i \varphi_+ \Big(\partial^i \big(-2\s \beta\s \vec b\cdot \x\big) \Big)\s  \varphi_-  \Big] \nonumber \\[9pt]
&=& 2\s\nu\s b_i\s \kappa\s a^{-2}\s  \Big( \varphi_+\big(\partial^i \varphi_-\big)- \big(\partial^i \varphi_+\big)\s \varphi_-\Big) \ , \label{eq:space_iso}
\eea 
where we have dropped a total derivative.  
We see that canceling the $b_i$ dependent terms between Eqs.~(\ref{eq:time_iso}) and~(\ref{eq:space_iso}) requires that $\kappa = -\chi$, as expected.  
Having imposed this constraint, the quadratic action is identical to \cref{eq:S2pmSomeSimp}, and hence all the arguments used to arrive at the final form of the quadratic action in \cref{eq:quadActionEFT} follow.

\subsection{Interactions and Locality}
\label{sec:Interactions}
Now that we have the degrees of freedom and their free equations of motion, the next step is to understand the structure of the allowed interactions.
First, we will explore the types of terms that can appear without derivatives, followed by a discussion of the derivative couplings.
Then we will argue that integrating out the heavy mode $\Phi_H$ at one loop yields matching contributions to the Wilson coefficients for the interactions in the SdSET.

\subsubsection*{Structure of the Potential}
Since we are working within an EFT framework, we expect that there should be an infinite tower of interactions, which can be organized using power counting.
We will begin by exploring the structure in the simple case where the interactions are a polynomial of $\varphi_+$ and $\varphi_-$ with no derivatives.
Using the power counting rules and the rescaling symmetry given in \cref{eq:powercounting,eq:RescaleSym} respectively, we find that the leading polynomial interaction is
\begin{align}
S_\text{int} \supset - \int \dd^3 x\, \dd \t \s [a\s H]^{3- n\s \alpha}\s \frac{c_{n,0}}{n!}\s \varphi_+^n \sim \lambda^{-3+n\s \alpha} \ ,
\label{eq:phiPnInt}
\end{align}
where $c_{n,m}$ is the Wilson coefficient, where the subscripts $n$ and $m$ count the number of $\varphi_+$ and $\varphi_-$ fields respectively.
This interaction is relevant when $n\s\alpha < 3$, which naively implies that such relevant terms should be included when performing any calculation in the IR.  
However, as we will now show, the presence of this operator is an artifact of our choice of fields,\footnote{Here we are assuming a fixed dS metric.  These operators can be important when the coupling to gravity is included, as we will discuss in Section~\ref{sec:EI}.} since these terms can be absorbed into a field redefinition of $\varphi_-$.

For simplicity, we will assume that the only interaction is given by \cref{eq:phiPnInt}.
Then neglecting the power suppressed terms in \cref{eq:quadActionEFT}, the action is
\begin{align}
S \supset - \int \dd^3 x \, \dd \t \bigg[\nu\s\big(\dot \varphi_+\s\varphi_- -  \varphi_+\s\dot\varphi_- \big)  +[a\s H]^{3- n\s \alpha}\s \frac{c_{n,0}}{n!} \varphi_+^n\bigg]  \ .
\end{align}
We can redefine 
\begin{align}
\varphi_- \s\s\to\s\s \varphi_- + \frac{n\s c_{n,0}}{2\s\nu\s(3-n\s\alpha)\s n!}\s [a\s H]^{3-n\s\alpha}\s \varphi_+^{n-1} \ ,
\label{eq:varphimShift}
\end{align}
in the action, which yields
\begin{align}
S &\to - \int \dd^3 x \, \dd \t\, \bigg[ -2\s \nu\s \varphi_+\s \dot{\varphi}_- - \frac{c_n}{n!}\s [a\s H]^{3-n\s\alpha}\s \varphi_+^{n} + \frac{c_n}{(3-n\s\alpha)\s n!}\s \dot{\varphi}_+\s \varphi_+^{n-1} +[a\s H]^{3- n \alpha}\s \frac{c_n}{n!}\s \varphi_+^n\bigg] \notag\\[9pt]
& =  - \int \dd^3 x \, \dd \t\, \bigg[ -2\s \nu\s\varphi_+\s \dot{\varphi}_- - \frac{n\s (n-1)\s c_n}{(3-n\s\alpha)\s n!}\s \dot{\varphi}_+\s \varphi_+^{n-1} - n\s \frac{c_n}{n!}\s [a\s H]^{3- n\s\alpha}\s \varphi_+^n + \frac{c_n}{n!}\s [a\s H]^{3-n\s\alpha}\s \varphi_+^n \bigg] \notag\\[9pt]
& =  - \int \dd^3 x \, \dd \t\, \bigg[ -2\s \nu\s \varphi_+\s \dot{\varphi}_- -\frac{\dd}{\dd \t}\bigg( \frac{c_n\s (n-1)}{(3-n\s\alpha)\s n!} [a\s H]^{3- n\s\alpha}\s \varphi_+^{n}\bigg)  \bigg] \notag\\[9pt]
& =  - \int \dd^3 x \, \dd \t\, \Big[ \nu\s\big(\dot \varphi_+\s\varphi_- -  \varphi_+\s\dot\varphi_- \big)\Big] \ ,
\end{align}
where we used integration by parts on the kinetic term before applying \cref{eq:varphimShift} to simplify the calculation.
We conclude that the leading contribution to the potential takes the form
\begin{align}
S_\text{int} \supset - \int \dd^3 x \, \dd \t\, \Big([a\s H]^{3- n\s\alpha - m\s\beta}\s \frac{c_{n, m}}{n!\s m!}\s \varphi_+^{n}\s \varphi_-^m \Big) \sim \lambda^{(n-1)\s\alpha + (m-1)\s\beta} \ ,
\label{eq:EFTPotentialGeneral}
\end{align}
where $m \geq 1$, and we have used $\alpha + \beta = 3$ when evaluating the scaling.  
Since $0< \alpha < 3/2$ and $3/2 < \beta < 3$ for a massive particle, we conclude that these operators are obviously irrelevant as long as $n >1$ or $m>1$.
Similarly, the limits on the range of $\beta$ imply that any term with $n=0$ and $m>1$ is also irrelevant.
The edge case $n= m = 1$ was already taken into account when deriving the quadratic action above in \cref{sec:TopDown}.
We therefore conclude that for a massive particle in dS, physics outside the horizon is irrelevant.

Next, we will briefly discuss the dimensions of EFT Wilson coefficients.
Imagine that we are performing a tree-level matching calculation where the UV theory includes polynomial interactions of the form
\begin{align}
S_{\rm UV, int} \supset - \int \dd^3 x \, \dd \t\, \frac{[a\s H]^3}{H^4}\s \frac{1}{N!}\s \lambda_{\phi,N}\s \phi^N \ ,
\label{eq:SUVphiN}
\end{align}
where $\lambda_{\phi,N}$ is a coupling with mass dimension $4-N$.
To match onto the EFT, we simply plug in the mode expansion \cref{eq:dof_split}, which yields terms of the form
\beq
S_{\rm int} \supset - \int   \dd^3 x \, \dd\t\s \frac{[a\s H]^3}{H^4}\s \frac{\lambda_{\phi, N}}{\s (N-1)!}\s H^N\s [a\s H]^{(1-N)\s\alpha - \beta}\s \varphi_+^{N-1}\s \varphi_- + \dots \ ,
\eeq
where we have already performed the $\varphi_-$ field redefinition \cref{eq:varphimShift} to eliminate the $\varphi_+^N$ term, and we are neglecting operators involving $\Phi_H$, since these do not contribute at tree level; the contribution to loop-level matching from integrating out $\Phi_H$ will be discussed next.
Next, we define a dimensionless Wilson coefficient $c_{N-m,m} \equiv H^{4-N} \lambda_{\phi,N}$ such that 
\beq
S_{\rm int} \supset - \int  \dd^3 x \, \dd\t\s   \frac{c_{N-1,1}}{(n-1)!}\s [a\s H]^{(2-N)\s\alpha}\s \varphi_+^{N-1}\s \varphi_-  +\dots \ .
\eeq
As promised, $\Lambda_\text{UV} = a\s H$ is the only dimensionful scale that appears explicitly in the action, see \cref{eq:LambdaUV}.

\subsubsection*{Derivative Interactions}
Moving beyond the potential, the EFT interactions can also include derivatives.  We write this schematically as
\beq
S_{\rm int} \supset - \int  \dd^3 x\, \dd\t\s  \frac{c_{n,m}^{\{i\}}}{n!\s m!} \left(\frac{\dd}{\dd \t} \right)^r \left( \frac{1}{a\s H} \partial_i  \right)^{2\s s} \Big([a\s H]^\alpha\s \varphi_+\Big)^n\s \Big([a\s H]^\beta\s \varphi_-\Big)^m  \ ,
\label{eq:SintWithTder}
\eeq
where $r$, $s$, $n$, and $m$ are integers, and $c_{n,m}^{\{i\}}$ are the Wilson coefficients.\footnote{RPI may impose relations among these Wilson coefficient; exploring these constraints lies beyond the scope of this work.}  Here we want to consider all possible ways in which these derivatives could act on the fields, which we have labeled abstractly as $\{i\}$.  However, as long as they are not total derivatives, it will not affect our power counting.  
Recall that the equations of motion including terms of $\mathcal{O}\big(\lambda^2\big)$ relate a single time derivative term to a power suppressed spatial derivative term, see~\cref{eq:EOMvarphipm}.
Hence, in order to make power counting manifest, one should simply apply the equations of motion to eliminate the time derivatives.
This implies that the most generic structure of the derivative interactions nicely organize as a power expansion in terms of spatial derivatives.
Furthermore, one can use a field redefinition of $\varphi_-$ as in \cref{eq:varphimShift} above to eliminate terms whose only field dependence is $\varphi_+^n$.
We conclude that the derivative expansion is fully captured by operators of the form 
\begin{align}
S_{\rm int} \supset - \int \dd^3 x \, \dd\t\s \frac{c_{n,m}^{\{s\}}}{n!\s m!}\left( \frac{1}{a\s H} \partial_i  \right)^{2\s s} \Big([a\s H]^\alpha\s \varphi_+\Big)^n\s \Big([a\s H]^\beta\s \varphi_-\Big)^m  \ ,
\label{eq:OpExpansion}
\end{align}
where $s > 0$, $n \geq 0$, and $m > 0$, and we have defined a new Wilson coefficient $c_{n,m}^{\{s\}}$, which is related to a combination of the $c_{n,m}^{\{i\}}$ from \cref{eq:SintWithTder}.
Following the same argument as was made for the previous case of the potential, and including the fact that the spatial derivatives provide extra power suppression, all such interactions are irrelevant when the scalar field is massive.

\subsubsection*{Integrating Out the Hard Modes}
Now that we have explored the operator structure of the SdSET, we will argue that integrating out the short distance physics can be absorbed into the Wilson coefficients of our local operator expansion given in~\cref{eq:OpExpansion}.
We will demonstrate this at one-loop order, by showing that explicitly integrating out the hard modes $\Phi_H$ can be accounted for as a correction to the EFT parameters.
Having then established that the EFT is local, we no longer need to appeal to $\Phi_H$.
Parameters can be determined by simply matching between full theory and EFT observables.

For concreteness, we will assume that the UV quadratic action is given by~\cref{eq:UV_free} with the addition of a local $\lambda_\phi\s \phi^4$ interaction for the scalar field $\phi$, \cref{eq:SUVphiN} with $N=4$.
Critical to the calculations presented here, when we use \cref{eq:dof_split} to express $\phi$ in terms of the modes $\varphi_S$ and $\Phi_H$, we find interactions of the form $\varphi_\pm^4$, $\varphi_\pm^3 \Phi_H$, $\varphi_\pm^2 \Phi_H^2$, $\varphi_\pm \Phi_H^3$, and $\Phi_H^4$.  
Imposing momentum conservation as we did above when we derived the quadratic action implies $\mathcal{L} \slashed{\supset}\s \varphi_\pm^3 \Phi_H$. 
In addition, we will neglect the $\Phi_H^4$ and $\varphi_\pm \Phi_H^3$ terms, since they do not contribute to matching at one-loop order.  
Therefore, the only UV interaction term that contributes to matching is 
\begin{align}
S_{\rm UV, int} \supset -  \int \dd^3 x\s \dd\t\, [a\s H]^{3}\bigg[& \frac{\lambda_\phi}{2 H^2}\Big([a\s H]^{-\alpha}\s \varphi_+ +[a\s H]^{-\beta}\s \varphi_-\Big)^2\s  \Phi_H^2 \bigg]\ . 
\end{align}
Now we can integrate out the heavy modes $\Phi_H(\p, \tau)$ at one-loop.

The first step is to isolate the hard mode function $\bar{\Phi}_H$. 
This is simple to do up to one-loop order using \cref{eq:ModeDecomp}:
\beq\label{eq:hard_m}
\bar{\Phi}_H(\x,\t) = \bar{\phi}(\x,\t) - \bar{\phi}_S(\x, \t) 
\eeq
where $\bar{\phi}(\x,t)$ are the mode functions of the free theory~\cref{eq:mode_m}, and $\bar{\varphi}_S$ are the soft mode functions~\cref{eq:BarPhi}. 
The $\bar\Phi_H$ and $\bar\phi_S$ mode functions are defined for all momentum, but $\bar\Phi_H$ will have no support at low momentum by construction.  
Note that although $\bar\phi_S$ does not explicitly vanish at large momentum, the possible contributions to loop integrals are zero when we use a momentum-independent scheme like dyn dim reg, that we will describe in \cref{sec:Reg}.  
This is the method of regions in action~\cite{Beneke:1997zp, Smirnov:2002pj}.  
While \cref{eq:hard_m} is somewhat abstract, it shows that $\Phi_H$ is dominated by large momentum without imposing a hard cutoff.\footnote{Since the purpose of this section is only to demonstrate that we have a local effective action, we could simply define $\Phi_H(p,t) = \phi(p,t)$ for $p> a\s H$ and zero otherwise, as we would in a Wilsonian EFT.  Although such a procedure breaks symmetries, establishing the decoupling of heavy modes demonstrates that the SdSET is local. By emphasizing the continuum perspective, there is no question that the symmetries inherited from having a dS background in the UV are respected at low energies.}

Integrating out $\Phi_H$ at one-loop is accomplished by evaluating 
\begin{align}
\delta S_{\rm int} &= \frac{\lambda_\phi^2}{4}\int  \frac{\dd^3 x\, \dd\tau\, \dd^3 y\,  \dd\tau'}{(-\tau)^{4}\s (-\tau')^{4}} \Big((-\tau)^{\alpha}\s \varphi_+(\x,\tau) +(-\tau)^{\beta}\s\varphi_-(\x,\tau)\Big)^2\,\frac{\Phi_H^2(\x,\tau)\s   \Phi_H^2(\y, \tau'\s)}{H^4} \nonumber\\[6pt]
&\hspace{104pt} \times \Big((-\tau'\s)^{-\alpha}\s\varphi_+(\y,\tau'\s) +(-\tau'\s)^{\beta}\s \varphi_-(\y,\tau'\s)\Big)^2 \ ,
\label{eq:deltaSIntStart}
\end{align}
where to derive this formula, we have simply inserted the $\varphi_\pm^2\s \Phi_H^2$ operator twice and are integrating over the spacetime coordinates for the hard modes.
Hence, we need to evaluate the correlation function 
\begin{align}
\!\Big\langle \Phi_H^2(\x,\tau)\s \Phi_H^2(\y, \tau'\s) \Big\rangle = \!\int \frac{\dd^3 p\, \dd^3 k}{(2\s\pi)^6}\,e^{i\s \ksub\cdot (\x-\y\s)} \bar \Phi_H(\s\p-\k,\tau) \bar \Phi_H(-\p,\tau)  \bar \Phi_H(-\p+\k,\tau'\s) \bar \Phi_H(\s\p,\tau'\s)  \, .\notag\\
\end{align}
Next, we Taylor expand the integrand assuming $k/p \ll 1$: 
\begin{align}
\Big\langle \Phi_H^2(\x,\tau)\s \Phi_H^2(\y, \tau'\s)\Big\rangle &\simeq \int \frac{\dd^3 p }{(2\s\pi)^3} \bar \Phi_H(\p,\tau) \bar \Phi_H(-\p,\tau) \bar \Phi_H(-\p,\tau'\s) \bar \Phi_H(\p,\tau'\s)  \int \frac{\dd^3  k}{(2\s\pi)^3}\s e^{i\s \ksub\cdot (\x-\y\s)} \notag \\[5pt]
&= \delta^3(\x-\y\s)  \int \frac{\dd^3 p }{(2\s\pi)^3} \bar \Phi_H(\p,\tau)\s \bar\Phi_H(-\p,\tau)\s \bar\Phi_H(-\p,\tau'\s) \bar\Phi_H(\p,\tau'\s)\ ,
\label{eq:PhiH4Eval}
\end{align}
where we have truncated to zeroth order in $\ksub$ in the first line.
Keeping higher orders in the $\ksub$ Taylor expansion results in an analytic function of $\ksub$, which yield a series in terms of derivatives of $\delta$-functions upon integration over $\ksub$.

In order to argue for the form of the resulting expansion, we do not need to perform the $\p$\s-integral explicitly; all we need to know is that when $\tau' \ll \tau$, the $i\epsilon$ prescription ensures that\footnote{One way to see this is by Wick rotating the positive frequency solution for $\Phi_H$ given in \cref{eq:hard_m} via $\tau \to i \tau_E$ with $\tau_E>0$, and taking the large $|\tau_E|$ limit.  See \emph{e.g.}~\cite{Green:2013rd,Green:2020txs} for further discussions. }  
\begin{align}
\bar \Phi_H(\p,\tau) \bar \Phi_H(-\p, \tau') \sim e^{-p\s|\tau -\tau'|} \ .
\label{eq:PhiHPhiHExp}
\end{align}
This is what fundamentally underlies the decoupling of heavy modes as $\tau \to \infty$.\footnote{In fact, this scaling behavior is in exact analogy with integrating out a heavy particle in a relativistic theory, whose influence follows a Yukawa potential which drops off exponentially fast.}
An immediate consequence is that we can Taylor expand $\varphi_{\pm}(\y, \tau')$ about $\tau$:
\begin{align}
\varphi_{\pm}(\y, \tau'\s) = \varphi_{\pm}(\y, \tau )+(\tau-\tau'\s)  \frac{\dd}{\dd \tau} \varphi_{\pm}(\y,\tau) +\ldots \ ,
\end{align}
such that 
\begin{align}
\delta S_{\rm int} &\supset \frac{\lambda_\phi^2}{4}\int  \frac{\dd^3 x\, \dd\tau}{(-\tau)^{4}} (-\tau)^{i\s \alpha + j\s \beta} \s\varphi_+^i(\x,\tau)\s \varphi_-^j(\x,\tau)\s  \frac{\partial^q}{\partial \tau^q}\s \varphi_+^n(\x,\tau)\s \varphi_-^m(\x,\tau)\notag\\[4pt]
&\hspace{30pt}\times\int \frac{\dd\tau'}{(-\tau')^4}\s \frac{\dd^3 p }{(2\s\pi)^3}\s (-\tau')^{n\s\alpha+m\s\beta}\s |\tau-\tau'|^q\s  \big|\bar \Phi_H(\p,\tau)\big|^2\s \big| \bar \Phi_H(-\p,\tau')\big|^2  \ ,
\end{align}
where we have integrated over $\dd^3 y$ using the $\delta$-function appearing in \cref{eq:PhiH4Eval}, and we are using $n$ and $m$ to indicate the various polynomial terms that appear.
Next, we can change the integration variables to $\xi \equiv - p\s \tau'$ and $\chi \equiv \tau'/\tau$, which yields 
\begin{align}
\delta S_{\rm int} &\supset \delta c_{i+n,\s j+m}^{\{q\}}\int  \frac{\dd^3 x\, \dd\tau}{(-\tau)^{4}}\s[a\s H]^{-(i+n)\s\alpha-(m+j)\s\beta}\varphi_+^i\s \varphi_-^j\s  \frac{\partial^q}{\partial \t^q}\s \varphi_+^n\s \varphi_-^m \ ,
\end{align}
where 
\beq
\delta c_{i+n,\s j+m}^{\{q\}} = \frac{\lambda_\phi^2}{4}\int_1^\infty \dd\chi \int_0^\infty \frac{\xi^2 \dd \xi }{2\s\pi^2}  \s\chi^{n\s\alpha+m\s\beta}\s (\chi -1)^q\s \frac{\pi^2}{16} \s\big|\s{\cal H}_\nu^{(1)}(\xi)\s\big|^4 \ ,
\eeq
and we have defined ${\cal H}_\nu = H_\nu(\xi) - H^{\rm soft}_\nu(\xi)$, the mode function with the behavior as $\xi \to 0$ subtracted.\footnote{We emphasize again that this approach is only valid at one loop, and is presented this way for convenience.  Operationally, this trick works because there is only a single integration variable $\xi$.  The technically correct approach would be to perform the calculation in the full theory and EFT, and to take their difference to obtain the matching correction.}  
By construction this integral converges as $\xi \to 0$.  
As $\xi \to \infty$, this integral also converges exponentially quickly after implementing the $i \epsilon$ prescription.  
The integral over $\chi$ is regulated using dyn dim reg and thus is also finite, see \cref{eq:ddr_time}.

Therefore, we conclude that integrating out the heavy modes yields a local expansion of the form 
\beq
\mathcal{O}_\text{SdSET} \sim \left(\frac{\dd}{\dd \t} \right)^r \left( \frac{1}{a\s H}\s \vec \partial_i  \right)^{2\s s} \varphi_{\pm} (\x, t)\ .
\eeq
Finally, following the same arguments as were given previously in the discussion of the derivative operators, we can express this expansion only in terms of spatial derivatives, see \cref{eq:OpExpansion}.
In other words, the impact of short distance physics can be fully absorbed as a matching contribution to the SdSET Wilson coefficients.

\subsection{Leading Power SdSET Action}
\label{sec:LeadingPowerAction}
We have now constructed the action for SdSET from both a top-down and bottom-up approach.  Dropping terms of $\mathcal{O}\big(\lambda^2\big)$ (or higher) in the power counting parameter $\lambda \sim k/[a\s H]$, this action is given by
\begin{align}
\tcboxmath{
S_{\pm} = \int \dd^3 x\, \dd\t\s  \Bigg[ -\nu\s\big(\dot{\varphi}_+\s \varphi_-  - \varphi_+\s\dot{\varphi}_-\big) -  \sum_{n\geq 2}^{n_{\rm max}}\s [a\s H]^{3- n\s\alpha -\beta}\s \frac{c_{n, 1}}{n!}\s \varphi_+^{n}\s \varphi_- \Bigg] \ ,}
\label{eq:ActionEFT}
\end{align}
with $\alpha + \beta =3$ and where $n_{\rm max}$ is an integer such that $2/\alpha\leq n_{\rm max} < 2/\alpha+1$.  Rotational invariance and scaling symmetries ensure that gradient corrections are suppressed by at least $\lambda^2$ and are always irrelevant.  Operators including powers higher powers of $\varphi_-$ such as $\varphi_-^m$ with $m> 1$ are suppressed by at least $\mathcal{O}\big(\lambda^3\big)$, and are thus smaller than the leading gradient corrections. 
The rest of this paper is devoted to studying the physical implication of this EFT.

\section{Calculating Observables}
\label{sec:Calc}
Now that we have the EFT action, whose operators are organized in a well defined way using power counting, all that remains is to compute observables.  
We will be interested in the (RG improved) perturbative predictions of the theory, which are equal-time correlation functions in a fixed initial state, or in-in correlators.  
We will first review how to calculate them.
We will then show how interactions in the UV theory can give rise to non-Gaussian initial conditions for the EFT modes. 
Finally, we will explain how to regulate divergent integrals that appear using dyn dim reg.

\subsection{In-in Perturbation Theory}
Our interest is in cosmological (in-in) correlators, which can be computed in perturbation theory using~\cite{Weinberg:2005vy}
\begin{subequations}
\begin{align}
\big\langle {\cal O}(\t)\big\rangle&=\left\langle\left[\overline{T} \exp \left(i \int_{-\infty}^{\t}\dd \t'\, H_{{\rm int}}(\t') \right)\right] {\cal O}^{{\rm int}}(\t)\left[T \exp \left(-i \int_{-\infty}^{\t} \dd \t'\, H_{{\rm int}}(\t') \right)\right]\right\rangle  \label{eq:inin_real}\\[9pt]
&= \sum_{N=\s 0}^{\infty} i^{N} \int_{-\infty}^{\t} \dd \t_{N} \int_{-\infty}^{\t_{N}} \dd \t_{N-1} \cdots \int_{-\infty}^{\t_{2}} \dd \t_{1} \nonumber \\[4pt] 
&\hspace{60pt} \times\Big\langle\big[H_{{\rm int}}\big(\t_{1}\big)\s,\s\big[H_{{\rm int}}\big(\t_{2}\big)\s,\s \cdots\big[H_{{\rm int}}\big(\t_{N}\big)\s,\s {\cal O}^{{\rm int}}\big(\t\big)\big] \cdots\big]\big]\Big\rangle  \ , \label{eq:commutator}
\end{align}
\end{subequations}
where, in practice, ${\cal H}_{\rm int} = - {\cal L}_{\rm int}$, $H_{\rm int}=\int \dd^3 x\, \mathcal{H}_{\rm int}$, and ${\cal O}^\text{int}\big(\t\big)$ is a product of operators that depend on different comoving momenta all evaluated at a fixed time $\t$.  
The crucial feature of \cref{eq:commutator} is that each factor of the interaction Hamiltonian is associated with a commutator.  
One intuitive way to understand the origin of~\cref{eq:inin_real}, is to note that the interaction picture fields are given by 
\beq
\varphi_\pm(\x,\t) =  \left[\overline{T} \exp\left(i \int_{-\infty}^{\t}\dd \t'\, H_{{\rm int}}(\t') \right) \right] \, \varphi_\pm^{\rm int}(\x,\t)\, \left[T\exp\left(i \int_{-\infty}^{\t}\dd \t'\, H_{{\rm int}}(\t') \right)\right]\ .
\eeq 
Then the time ordering in~\cref{eq:inin_real} manifests as the ordering of the time dependence in the argument of the $H_\text{int}(\t)$ factors in \cref{eq:commutator}.

To understand the implications for the SdSET, recall that our action is first order in time, see \cref{eq:quadActionEFT}.
Hence, the conjugate momenta for $\varphi_+$ is
\beq
\frac{\delta {\cal L}}{\delta \dot \varphi_+} = - 2\s\nu\s \varphi_- \ .
\eeq
To canonically quantize the theory, we impose
\beq\label{eq:plusminuscomm}
\big[\varphi_+(\x,t)\s,\s \varphi_-(\y,t)\big] =-\frac{ i }{2\s\nu}\s \delta(\x-\y\s) \ ,
\eeq
and $\big[\varphi_+\s,\s \dot \varphi_+\big] = 0$, which agrees with the derivation using the top-down commutators given in~Eqs.~(\ref{eq:varphiCanonCom}) and (\ref{eq:tildecomm}) respectively.

Finally, we can check the self-consistency of the setup with a simple example calculation.
For concreteness, we can turn on a non-zero quartic interaction for the SdSET fields such as 
\begin{align}
\mathcal{H}_{\rm int} = \frac{c_{3,1}}{3!}\s[a\s H]^{-2\s\alpha}\s \varphi_+^3\s \varphi_-   \ ,
\end{align}
where we are following the notation in~\cref{eq:EFTPotentialGeneral}.
Then the in-in formula~\cref{eq:commutator} implies  
\beq
\big\langle\varphi_+(\x,\t) \s \ldots \big\rangle = \left\langle i \int^{\t}_{-\infty} \dd^3 x'\,\dd\t'\,  \frac{c_{3,1}}{3!}\s \varphi_+^3\big(\x^{\s\prime},\t'\big) \Big[ \varphi_-\big(\x^{\s\prime},\t'\big) \s,\s \varphi_+\big(\x,\t\big)\Big] \s \ldots \right\rangle \ .
\eeq
Then taking a time derivative yields
\beq
\dot \varphi_+(\x,\t) = - \frac{1}{2\s\nu}\s \frac{c_{3,1}}{3!}\s \varphi_+^3(\x, \t) \ ,
\eeq
as an operator statement; this matches the classical equations of motion as it must.

\subsection{Non-Gaussian Stochastic Initial Conditions}
\label{sec:NGIC}

As mentioned in~\cref{sec:Rules} above, non-trivial interactions in the UV theory can yield additional non-Gaussian contributions to the classical initial conditions for the SdSET fields.
For concreteness, we will show how this works in the context of a simple example.  
We assume that the UV theory contains a conformally coupled scalar field $\phi$, \emph{i.e.}, its mass is $m^2 = 2\s H^2$ such that $\nu = 1/2 \leftrightarrow \alpha = 1$.
We will give this field a $\lambda_\phi\s \phi^4$ interaction, see~\cref{eq:SUVphiN}, which will generate a non-Gaussian UV contribution to the initial conditions.
We need the mode function for $\phi$, which we get by evaluating~\cref{eq:mode_m} for $\nu = 1/2$; this yields
\beq
\bar\phi\big(\s\ksub,\tau\big)= \frac{-i\s (-H\s \tau)}{\sqrt{2\s k}}\s e^{i\s k\s\tau}  \ ,
\eeq
which holds from the UV to the IR, $k\s\tau \in (-\infty, 0]$.

At linear order in $\lambda_\phi$, applying~\cref{eq:commutator} in the limit $k\s\tau_0 \to 0$, we find 
\begin{align}
\Big\langle \phi\big(\k_1\big)\s \phi\big(\k_2\big)\s \phi\big(\k_3\big)\s\phi\big(\k_4\big) \Big\rangle' &=  \lambda_\phi\s \frac{(-H \s\tau_0)^{4}}{16\s k_1\s k_2\s k_3\s k_4 }\s 2 \s {\rm Im} \int_{-\infty}^{\tau_0} \dd\tau \, a^d(\tau) \s(-H\s \tau)^{2\s d-4} e^{i\s k_\text{tot}\s(\tau-\tau_0)} \notag \\[8pt]
&\to  \lambda_\phi  \s  \frac{(-H\s \tau_0)^{4}}{8\s k_1\s k_2\s k_3\s k_4\s k_\text{tot}} \ ,
\end{align}
where $k_\text{tot} = \sum k_i$.
This determines the classical initial conditions for the SdSET four-point function.
Recalling that we took $\alpha = 1$ here, converting to the $\varphi_+$ description requires absorbing a factor of $a(\tau_0)^{-1}= -H\s \tau_0$ for each $\varphi_+$, implying that
\beq
\Big\langle \varphi_{+}\big(\k_1\big)\s \varphi_{+}\big(\k_2\big)\s \varphi_{+}\big(\k_3\big)\s \varphi_{+}\big(\k_4\big) \Big\rangle_c'  =   \frac{ \lambda_\phi}{8\s k_1\s k_2\s k_3\s k_4\s \big(k_1+k_2+k_3+k_4\big) } \ ,
\label{eq:varphi4NGBC}
\eeq
where we have used the $\langle \dots \rangle_c$ notation to emphasize that these are additional classical inputs to the SdSET.
Note the pole in the total energy, $k_\text{tot} = \sum k_i$, is a generic consequence of contact interactions in the Bunch-Davies vacuum, and the residue is this pole encodes the flat-space $S$-matrix~\cite{Maldacena:2011nz,Raju:2012zr,Arkani-Hamed:2015bza,Arkani-Hamed:2018kmz,Green:2020whw}.  
When performing calculations,~\cref{eq:varphi4NGBC} provides a contribution that must be added to the results induced by the Gaussian correlators in~\cref{eq:pluspower}.

\subsection{Dynamical Dimensional Regularization}\label{sec:Reg}
One regularly encounters UV divergent integrals when computing in-in correlators in dS.  
One of the benefits of the framework here is that it is compatible with a dim-reg-like approach to taming UV divergences, which we will call ``dynamical dimensional regularization'' (dyn dim reg).
The insight is to analytically continue in the scaling dimensions $\alpha$, which control the dynamics.
As with dim reg, scaleless integrals vanish. 
Since we have already absorbed all the dependence on the UV physics into the SdSET Wilson coefficients, such integrals will only receive support from the IR dynamics.
One unusual feature of the SdSET is that we have independent scaleless integrals over time and over momenta.  
Due to the way time dependence is accounted for in the EFT, these integrals will typically factorize.  
As a result, it is necessary (and advantageous) to regulate them independently. 
In contrast with conventional dim reg applied to relativistic field theory, we will see that analytically continuing in the number of dimensions will not always regulate divergences in momentum integrals; the exponent for both the propagators and the measure of integration can depend on dimension.

To get a sense of how this works in practice, we can consider a typical momentum integral.
Take for example, the calculation of the power spectrum (Fourier transform of the two-point correlation function) for the composite operator $\O_{2}(\x,t) \equiv \varphi_+^2(\x, t)$ in the free theory.  
Using \cref{eq:pluspower}, this correlation function is given by
\begin{align}
\Big\langle \O_{2}(\s\k\,)  \s \O_{2}(\,\kp\s) \Big\rangle_\text{tree} &=\frac{C_\alpha^4}{2}\s  \int \frac{\dd^3 p}{(2\s\pi)^3} \frac{1}{p^{3-2\s\alpha}}\s \frac{1}{|\k -\p\s|^{3-2\s\alpha}} \s  (2\s\pi)^3\s \delta\big(\k+\kp\big) \ .
\label{eq:o2power}
\end{align}
Combining denominators by introducing Feynman parameters, we find
\begin{align}
\Big\langle \O_{2}\big(\s\k\,\big)  \s \O_{2}\big(\,\kp\s\big) \Big\rangle_\text{tree}'&= \frac{ C_\alpha^4}{2} \s   \frac{\Gamma[3 -2\s \alpha]}{\Gamma[\frac{3}{2} - \alpha]^2}\s \int_0^1 \dd x \int \frac{\dd^3 p'}{(2\s\pi)^3}\s \frac{\big(x(1-x)\big)^{1/2-\alpha}}{\big(p'^{\s 2} + x\s(1-x)\s k^2\big)^{3-2\s\alpha}} \notag\\[8pt]
&= \frac{C_\alpha^4 }{16\s \pi^{3/2}} \frac{\Gamma[\alpha]^2\s \Gamma\big[\frac{3}{2}-2\s \alpha\big]}{\Gamma[2\s\alpha]\s \Gamma\big[\frac{3}{2} - \alpha\big]^2}  \frac{1}{k^{3- 4\s \alpha}} \ , 
\label{eq:o2tree}
\end{align}
where we have evaluated this integral using the standard dim reg tricks, and we are using $\langle .. \rangle'$ as defined in \cref{eq:DefPrime}.
By dimensional analysis, we see the original integral has a logarithmic UV divergence when $2\s\alpha = 3/2$, which is made manifest by the factor $\Gamma\big[\frac{3}{2}-2\s \alpha\big]$, as is familiar from working with dim reg.  
However, unlike dim reg, the propagator scales with $d$ such that working in general dimension does not always regulate integrals, for an explicit demonstration of this behavior, see \cref{app:tri}.  
As a result, analytically continuing in $\alpha$ is both necessary and sufficient for regulating our EFT.

Next, we turn to the more novel problem of regulating the time integrals, which are forced upon us when computing cosmological (in-in) correlators.  
In our EFT, these integrals will typically yield power-law divergences when $a\s H \to 0$ ($t \to -\infty$).
This limit corresponds to the short wavelength region that we integrated out to generate the SdSET, and therefore we must define these integrals by (scaleless) analytic continuation to remove these divergences. 
A typical time integral will take the form $\int \dd\t\, [a\s H]^{\gamma}$, which we will evaluate as 
\beq\label{eq:ddr_time}
\int_{-\infty}^{H t} \dd\t \,  [a(\t)\s H]^{\gamma} = \frac{1}{\gamma}\s [a( t)\s H]^{\gamma} \ ,
\eeq
where we are using the 
implicit notation $a(\t) = e^{\t}$ and $a(t) = e^{H\s t}$.   
This expression converges when $\gamma >0$, and so we must define $\gamma < 0$ by analytic continuation.
This should be understood in the same way as the vanishing of power-law divergences in dim reg, the early time contributions are already accounted for in the initial conditions for $\varphi_+$ and $\varphi_-$. 
One major benefit of defining time integration in this way is that our results will follow the naive power counting of the integrand.

It is interesting to contrast how we calculate in SdSET against the approach taken for the UV theory.  
In the UV, integrals naively diverge at very early times, $t \to -\infty$, but are regulated by analytically continuing time into the complex plane via the $i\s\epsilon$ prescription.  
This ensures that correlators are being computed in the interacting vacuum.
However, in some cases it is advantageous to perform calculations using the commutator form of the in-in correlators, \cref{eq:commutator}, which is suited to exposing the late time behavior of the integrals as $a\s H \to \infty$.
For example, this form is usually invoked to argue that the integrals are at most log-divergent as $a\s H \to \infty$~\cite{Weinberg:2005vy,Weinberg:2006ac}.
Unfortunately, it is not known how to apply the $i\s\epsilon$ prescription while maintaining the commutator form for these calculations.
Thus, the standard approach to UV calculations fails to make the convergence at late and early times manifest simultaneously.   
This issue is trivialized by calculating in the SdSET using dyn dim reg: the convergence at late times is manifest since the hard modes have been integrated out leaving behind a local description, while the vanishing of scaleless integrals ensures convergence at early times. 
Additionally, the fact that no time derivatives appear in the EFT operator expansion, see \cref{eq:OpExpansion}, further helps to disambiguate the types of divergences that appear.

Now we have everything we need to explore the physical consequences of the SdSET.
First, \cref{sec:qft} will provide some implications for massive and massless scalar fields in static dS.
Then we turn to \cref{sec:metric}, where we allow gravity to be dynamical and make connections to inflation.

\section{Rigid Spacetime}
\label{sec:qft}
In this section, we will discuss the implications for quantum field theory in a fixed dS background.
A number of illustrative loop calculations will be presented.
We will encounter situations where potentially large logarithms appear, which will we show how to resum using the techniques of the Dynamical Renormalization Group (DRG). 
The main results derived here are as follows.  
First, we will show that loop corrections for heavy (fundamental) fields vanish in the $t\to \infty$ limit when regulating integrals using dyn dim reg such that symmetries are preserved, as discussed in \cref{sec:Reg} above. 
Next, we will show that non-trivial loop corrections to correlation functions of composite operators built out of heavy fields are captured by the SdSET; the resulting logarithms can be resummed using the DRG.
Finally, we find that loop corrections for light fields are even more interesting, due to the fact that an infinite number of operators become marginal and degenerate as the mass is taken to zero.
The logs that appear can be addressed by a non-trivial DRG equation, and to all orders in gradients, the resulting flow is described by stochastic inflation.

\subsection{Massive Theories}\label{sec:mass}
The goal here is to show that the long wavelength dynamics is trivial for scalar theories with $m^2 \simeq H^2$, in that any non-trivial UV dynamics can be encoded through matching. 
Then the fact that the massive modes decay outside the horizon with a power $ \alpha > 0$ implies that superhorizon contributions to the correlation functions of $\varphi_+$ all tend to zero in the late time limit. 

\subsubsection*{Power Spectrum to Two Loops}
As a warm-up, consider the leading one-loop correction to the $\varphi_+$ power spectrum, when the UV theory has a $\lambda_\phi\s \phi^4$ interaction, \emph{i.e.}, setting $N=4$ in \cref{eq:SUVphiN}, so that the leading SdSET interaction is 
\beq\label{eq:Hint4}
\H_{\rm int} =  \frac{c_{3,1}}{3!}\s [a\s H]^{-2\s \alpha }\s\varphi_+^3\s \varphi_- \ .
\eeq
The classical power spectrum is given by \cref{eq:pluspower}, and using Eqs.~(\ref{eq:commutator}) and~(\ref{eq:plusminuscomm}), we find that the one-loop correction is 
\beq
\Big\langle \varphi_{+} \big(\s\k, \t\big)\s \varphi_{+} \big(\,\kp, \t\big) \Big\rangle' = \frac{c_{3,1}}{8\s \nu}\s \frac{C_\alpha^2}{k^{3-2\alpha}} \int \dd \t\, [a\s H]^{- 2\s\alpha} \int \frac{\dd^3 p}{(2\s\pi)^3} \frac{C_\alpha^2}{p^{3-2\s\alpha}} =0 \ .
\eeq
This scaleless integral vanishes when using dim reg.  
Furthermore, since $\alpha > 0$, the integral converges in the IR.  
This implies that this power-law UV divergence can be absorbed by matching the UV theory onto the SdSET.

While it is a trivial consequence of computing corrections within the EFT, this observation has significant implications for the structure of perturbative calculations.  
For example, calculations performed in the full theory that regulate integrals differently can give rise to apparent $\log k / [a\s H]$ corrections to this correlator, see \emph{e.g.}~\cite{Anninos:2014lwa}.  
However, as we argued above, integrating out the hard modes $\Phi_H$ yields a purely local theory, which implies that non-analytic terms like $\log k/ [a\s H]$ can only arise from loops involving EFT modes.  
Since we can show that this integral vanishes when using a regulator that preserves the symmetries, the only possibility is that the contribution can be absorbed into a redefinition of $\alpha$ when matching onto the SdSET.

Given that the one-loop calculation was trivial, we will explore how the SdSET behaves at two-loop order.  
It is straightforward to identify the non-zero contribution by acting with the perturbation on the individual fields, such that 
\beq
\Big\langle \varphi_{+} \big(\s\k\,, t\big)\s \varphi_{+} \big(\,\kp\s, t\big) \Big\rangle'=\bigg\langle\Big(U^{\dagger}\left(t\right) \s\varphi_{+}\big(\s\k\,, t\big)\s U(t)\Big)\Big(U^{\dagger}(t)\s \varphi_{+}\big(\,\kp\s,t\big)\s  U(t) \Big)\bigg\rangle' \ ,
\eeq
where
\beq
U(t) =  T\s \exp \left(-i \int_{-\infty}^{H\s t}  \dd \t'\,H_{{\rm int}}\big(\t'\big)\right) \ .
\eeq
We refer to a given contribution as $(i,j)$ when we expand $U^{\dagger}(t)\s \varphi_{+}\big(\s\k\,,t\big)\s U(t)$ on the left to $i^\text{th}$ order in $H_{\rm int}$ and $U^{\dagger}(t)\s \varphi_{+}\big(\,\kp\s,t\big)\s U(t)$ right to $j^\text{th}$ order.\footnote{This is similar to a common convention taken for perturbation theory of large scale structure~\cite{Bernardeau:2001qr}.}

First, we consider the $(0,2)$ and $(2,0)$ diagrams.  
By construction, there must be a commutator involving $H_{\rm int}$, and hence $[\varphi_+, \varphi_-]$, that acts on an internal line.  
The resulting contribution is given by
\begin{align}
\Big\langle \varphi_{+} \big(\s\k\,, t\big)\s \varphi_{+} \big(\,\kp\s, t\big) \Big\rangle'_{(0,2)} &= \frac{c_{3,1}^2}{2} \frac{C_\alpha^6}{k^{3-2\s\alpha}}  \int \dd\t_1\, \dd\t_2\,  [a(\t_1)\s H]^{- 2\s\alpha}[a(\t_2)\s H]^{- 2\s\alpha}   \notag \\[4pt]
&\hspace{77pt} \times \frac{1}{(2\s \nu)^2}\int \frac{\dd^3 p_1\, \dd^3 p_2}{(2\s\pi)^6} \frac{1}{p_1^{3-2\s\alpha}}  \frac{1}{p_2^{3-2\s\alpha}}= 0 \ .
\end{align}
Similarly, the $(2,0)$ contribution vanishes.

The only non-trivial contribution at this order arises from the $(1,1)$ diagrams, where both commutators act on the external lines, namely on $\varphi_+(t)$, where $t$ is the time when the correlator is measured.
As a result, the two-loop correction is 
\begin{align}
\Big\langle \varphi_{+} \big(\s\k\,, t\big)\s \varphi_{+} \big(\,\kp\s, t\big) \Big\rangle'_{(1,1)} &= \frac{c_{3,1}^2}{2} \frac{C_\alpha^6 }{\nu^2 } \int \dd\t_1\, \dd\t_2 \, [a(\t_1)\s H]^{- 2\s\alpha}[a(\t_2)\s H]^{- 2\s\alpha}   \nonumber \\[7pt]
& \hspace{65pt} \times \int \frac{\dd^3 p_1\, \dd^3 p_2}{(2\s\pi)^6} \frac{1}{p_1^{3-2\s\alpha}}  \frac{1}{p_2^{3-2\s\alpha}} 
\frac{1}{|\k-\p_1-\p_2|^{3-2\s\alpha}} \ .
\end{align}
This integral is straightforward to evaluate in position space.
We notice that the perturbative evolution of the fields gives
\beq
\bigg[\varphi_+(\x,t)\s,\s \int_{-\infty}^{H\s t} \dd\t_1 \int \dd^3 x_1\, [a(\t_1)\s H]^{-2\s\alpha}\s \varphi_+^3\s\varphi_-(\x_1,\t_1)\s\bigg] = - \frac{ \varphi_+^3(\x\s)}{ 2\s\nu}  \int_{-\infty}^{H\s t} \dd\t_1 [a\s H]^{-2\s\alpha}  \ ,
\eeq
where we used the canonical commutator and time independence of the interaction picture fields to simplify this integral.  
The two-loop contribution to the two-point function is 
\beq
\Big\langle \varphi_{+} \big(\s\x\,, t\big)\s \varphi_{+} \big(\,\x\s'\s, t\big) \Big\rangle_{(1,1)} =  \frac{c_{3,1}^2}{4\s\nu^2\s (3!)^2} \left(  \int^{H\s t}_{-\infty} \dd\t\, [a\s H]^{-2\s\alpha} \right)^2 \Big\langle \varphi_+^3\big(\x\s\big)\s \varphi_+^3\big(\x^{\s\prime}\big) \Big\rangle \ ,
\eeq
where we are using $\varphi_+(\x,t) \simeq \varphi_+(\x\s)$ to leading power.
The classical two-point function of $\varphi_+$ in position space is given by the Fourier transform of~\cref{eq:TreePowerSpecPhiPlus}: 
\begin{align}
\Big\langle \varphi_{+} \big(\s\x\,, t\big)\s \varphi_{+} \big(\,\x\s'\s, t\big) \Big\rangle_c &= - C_\alpha^2 \int \frac{\dd^3 k}{(2\s\pi)^3} \s e^{i\s \ksub \cdot \x_{d}} \frac{1}{2\s k^{3-2\s\alpha}} \notag \\[5pt]
&=-   \frac{C_\alpha^2}{4\s \pi^2\s x_{d}^{2\s\alpha}} \cos(\pi\s \alpha) \Gamma[2\s\alpha-1] \ ,
\end{align}
with $\x_d \equiv \x -\x\s'$, such that 
\beq
\Big\langle \varphi_+\big(\x,t \big)\s \varphi_+\big(\x\s',t \big) \Big\rangle_{(1,1)}  =  \frac{c_{3,1}^2}{ 3!  } \s\frac{1}{x_{d}^{6\s\alpha} \s[a\s H]^{4\s\alpha}}\s\frac{C_\alpha^6}{64\s \pi^6 (4\s \nu\s \alpha)^2 } \big( -\cos(\pi\s \alpha) \Gamma[2\s\alpha-1]\big)^3 \ .
\eeq
Fourier transforming back yields the two-loop correction to the power spectrum
\begin{align}
\Big\langle \varphi_{+} \big(\s\k\,, t\big)\s \varphi_{+} \big(\,\kp\s, t\big) \Big\rangle_{(1,1)}  &= \frac{c_{3,1}^2}{3!} \frac{1}{4\s \nu^2\s \alpha^2\s [a\s H]^{4\s\alpha}}   \frac{C_\alpha^6\s R_\alpha}{16\s \pi^5 k^{3-6\s\alpha}} \nonumber  \\[6pt]
&=\frac{c_{3,1}^2}{3!} \frac{C_\alpha^4 \s R_\alpha  }{32\s \pi^5\s \nu^2\s \alpha^2}    \left( \frac{k}{a\s H} \right)^{4\s\alpha} \Big\langle \varphi_{+} \big(\s\k\,, t\big)\s \varphi_{+} \big(\,\kp\s, t\big) \Big\rangle_c   \ ,
\end{align}
where we have used \cref{eq:TreePowerSpecPhiPlus} to relate this one-loop correction to the classical two-point function, and 
\beq
R_\alpha = \big( -\cos(\pi\s \alpha)\s \Gamma[2\s\alpha-1] \big)^3 \Gamma[2-6\s\alpha] \sin(3\s\alpha\s \pi) \ .
\eeq
These corrections are suppressed by $(k/[a\s H])^{4\s\alpha}$ as expected from power counting at ${\cal O}(c_{3,1}^2)$. Alternatively, we can see where this scaling comes from by noting that the two-loop diagram is built by contracting four additional factors of $\varphi_+$ after implementing the commutators.
This shows that the SdSET simplifies the dynamics significantly.
 
\subsubsection*{Composite Operators}
Although the dynamics of $\varphi_+$ at one-loop were trivial, the application to composite operators at this order is not.
For concreteness, we will focus on the example operator 
\begin{align}
\O_{2}(\x,t) = \varphi_+(\x,t)^2 \ .    
\end{align}
As discussed in detail in~\cite{Green:2020txs} for $\alpha = 1$ (a conformally coupled scalar), this operator will acquire a one-loop anomalous scaling.  
However, as we just showed, one-loop corrections to the two-point function of $\varphi_+$ vanishes.
It is natural to wonder how the SdSET can reproduce the results from~\cite{Green:2020txs}.  
One might guess that the difference arises from the time evolution, via the interaction in \cref{eq:Hint4}.
However,
\begin{align}
\Big\langle \O_{2}\big(\s\k\,\big)  \s \O_{2}\big(\,\kp\s\big) \Big\rangle'_{\rm 1-loop} &=    \frac{c_{3,1}}{2\s\nu}\s \frac{|C_\alpha|^6}{8} \int \dd\t\, [a\s H]^{- 2\s\alpha} \int \frac{\dd^3 p_1 }{(2\s\pi)^3} \frac{1}{p_1 ^{3-2\s\alpha}} \int \frac{\dd^3 p_2}{(2\s\pi)^3} \frac{1}{p_2 ^{3-2\s\alpha} \s|\k-\p_2|^{3-2\s\alpha} } \notag \\[6pt]
&=   \frac{c_{3,1}}{2\s\nu}\s\frac{|C_\alpha|^2}{4}\s \Big\langle \O_{2}(\s\k\,)  \s \O_{2}(\,\kp\s) \Big\rangle_\text{tree}\s  \int \dd\t\, [a\s H]^{- 2\s\alpha} \int \frac{\dd^3 p_1 }{(2\s\pi)^3} \frac{1}{p_1 ^{3-2\s\alpha}} \notag\\[6pt]
&= 0 \ ,
\label{eq:VanishingOneLoopCompOp}
\end{align}
where the tree-level correlator is given in \cref{eq:o2power} above, and the scaleless $\p_1$ integral evaluates to zero.
Clearly this calculation does not explain the origin of the anomalous scaling.

In evaluating~\cref{eq:VanishingOneLoopCompOp}, we used the Gaussian correlators of $\varphi_+$ given in~\cref{eq:pluspower} to evaluate the loop correction that is generated by the coupling within the SdSET. 
This neglects the non-Gaussian initial conditions for the $\varphi_+$ correlators that were discussed in \cref{sec:NGIC} above.  
Assuming the UV theory has a $\lambda_\phi\s \phi^4$ interaction, \cref{eq:varphi4NGBC} provides the origin of the anomalous scaling.  
Focusing on a conformally coupled scalar for simplicity, the non-Gaussian correction to the $\O_2$ power spectrum is given by
\beq \label{eq:o2_3d}
\Big\langle \O_{2}\big(\s\k\,\big)  \s \O_{2}\big(\,\kp\s\big) \Big\rangle'_{\rm NG} = \int \frac{\dd^{3} p_1 \s\dd^{3} p_2}{(2\s\pi)^{6}}\s \Big\langle \varphi_{+}\big(\p_1\big)\s \varphi_{+}\big(\p_1-\k\big)\s \varphi_{+}\big(\p_2\big)\s \varphi_{+}\big(-\k-\p_2\big) \Big\rangle' \ .
\eeq
In the limit where $p_1 \gg p_2 \simeq k$, the integral becomes\footnote{This calculation can be understood as a manifestation of the method of regions~\cite{Beneke:1997zp, Smirnov:2002pj} applied to these correlators. 
This can be seen by expanding the integrand of the full theory calculation~\cite{Green:2020txs} in the long wavelength limit, which yields~\cref{eq:o2_regions}. } 
\beq\label{eq:o2_regions}
\Big\langle \O_{2}\big(\s\k\,\big)  \s \O_{2}\big(\,\kp\s\big) \Big\rangle'_{\rm NG} \simeq \frac{c_{3,1}}{8}\int \frac{\dd^{3}p_1}{(2\s\pi)^{3}}\s \frac{1}{p_1^{3}}  \int \frac{\dd^{3}p_2}{(2\s\pi)^{3}}\s \frac{1}{p_2\s \big|\k+\p_2\big|} \ , 
\eeq
where we have included a factor of two to account for the equivalent contribution from $p_2 \gg p_1$.  
We see the appearance of a logarithmic divergence, as expected.  
Naively, we would expect that working in $d$-dimensions would regulate this integral.  
Surprisingly, this divergence is present in every dimension, see~\cref{app:tri}.

When we calculated the tree-level Gaussian contribution to the power spectrum of $\O_{2}\big(\s\k\,\big)$ in \cref{eq:o2power}, it was the analytic continuation in $\alpha$ that regulated the integral.  
In the case at hand, this trick is insufficient to regulate the divergence.  
However, treating the factors of $\alpha_i$ that are associated with each $\varphi_+\big(\s\k_i\big)$ independently does regulate the integral.  
Then we simply take $\alpha_i \to 1$ (for the case of the conformally coupled scalar) after integrating.
Again, focusing on the limit $p_1 \gg p_2 \simeq k$, we will take the fields associated with the soft momenta to have arbitrary $\alpha$, while holding $\alpha=1$ for the $\varphi_+$ field that carries $p_1$ momenta.  
The power spectrum is then
\beq
\Big\langle \O_{2}\big(\s\k\,\big)  \s \O_{2}\big(\,\kp\s\big) \Big\rangle'_{\rm NG} \simeq \frac{c_{3,1}}{8} \s C_\alpha^4\s \int \frac{\dd^{3}p_1}{(2\s\pi)^{3}}\s \frac{1}{p_1^{1+2\s\alpha}}  \int \frac{\dd^{3}p_2}{(2\s\pi)^{3}}\s \frac{1}{p_2^{3-2\s\alpha}\s\big|\k+\p_2\big|^{3-2\s\alpha}} \ .
\label{eq:O2O2NGInt}
\eeq
The $p_1$ integral is logarithmically divergent as $\alpha \to 1$.  
Here we see a typical feature of continuum EFTs: a full theory IR divergence becomes an EFT UV divergence, which induces an RG flow that resums what are fundamentally IR logarithms.
Moreover, we note that the split between time and space means that IR (UV) refers to either long (short) wavelengths or times.\footnote{One of the central complications when calculating loop corrections to cosmological correlators is the non-trivial interplay between short distances and long times.  One simplification of the SdSET is that this mixing is fully captured by the time-dependence of the UV scale.} 
A novel feature here is that we have exchanged a large IR contribution in time in the full theory~\cite{Green:2020txs} for a UV divergence in EFT momentum integral.
The full integral is regulated when $p_1\simeq k >0$ and yields an IR finite result; however, we expanded away the terms that dominate in the IR by taking the simplifying limit $p_1 \gg p_2, k$.  
Nevertheless, since we are interested in extracting the UV divergence, all we need to do is regulate the IR in a convenient way.
We do this by substituting 
\begin{align}
\frac{1}{k^{2(1-\alpha)}}\s\int \frac{\dd^{3}p_1}{(2\s\pi)^{3}}\s \frac{1}{p_1^{1+2\s\alpha}} &\to \frac{1}{k^{2(1-\alpha)}}\s\int \frac{\dd^{3}p_1}{(2\s\pi)^{3}} \s\frac{1}{\big(p_1^2+K_\text{IR}^2\big)^{1/2+\alpha}} \notag \\[8pt]
& = \frac{1}{2\s\pi^2} \frac{K_\text{IR}^{2\s(1-\alpha)}}{k^{2(1-\alpha)}}\s \frac{\sqrt{\pi}\, \Gamma[\alpha -1]}{4\s \Gamma\big[\frac{1}{2}+\alpha\big]} \notag\\[8pt]
&= \frac{1}{2\s\pi^2} \left( \frac{1}{2\s(\alpha-1)} - \gamma_E + \log \frac{k}{K_\text{IR}} + \ldots \right) \ ,
\label{eq:anaomalous_K} 
\end{align}
where $K_\text{IR}$ regulates the IR divergence, thereby making the integral scaleful, and $\gamma_E$ is the Euler-Mascheroni constant.

Substituting the result back into our full expression and using~\cref{eq:o2power}, we find
\begin{align}
\Big\langle \O_{2}\big(\s\k\,\big)  \s \O_{2}\big(\,\kp\s\big) \Big\rangle'_{\rm NG} \simeq \frac{c_{3,1}}{8\s \pi^2}  \left( \frac{1}{2\s(\alpha-1)} - \gamma_E + \log \frac{k}{a\s H} + \log \frac{k}{K_\text{IR}}  \right) \times \Big\langle \O_{2}\big(\s\k\,\big)  \s \O_{2}\big(\,\kp\s\big) \Big\rangle'_{\rm G} \ , \notag\\
\label{eq:O2O2NGFinal}
\end{align}
where the $\log k/[a\s H]$ came from expanding the remaining factor of $k^{2\s(\alpha-1)}/[a\s H]^{2\s(\alpha-1)}$ before taking $\alpha \to 1$.  
It is natural to interpret the coefficient of the $1/(\alpha-1)$ pole as $2\s\gamma_{2}$ where 
\beq
\gamma_{2} = \frac{c_{3,1}}{16\s \pi^2} +{\cal O}\Big(c_{m,n}^2\Big) \ ,
\label{eq:gamma2}
\eeq
is the anomalous dimension of the $\O_{2}$ operator.  
To make this connection precise, we will apply the DRG formalism, which will allow us to resum these logarithms.

\subsubsection*{Dynamical Renormalization Group}
Having regulating the IR of the scaleless integrals in~\cref{eq:O2O2NGInt}, we isolated the UV divergent contribution to the $\big\langle \O_{2}\big(\s\k\,\big)  \s \O_{2}\big(\,\kp\s\big) \big\rangle_\text{NG}$ power spectrum, which contains a term proportional to $\log k/[a\s H]$, see \cref{eq:O2O2NGFinal}.
This logarithm can become large, and so this result suggests utilizing an RG approach to improve perturbation theory.  
However, the conventional approach to renormalization does not address logarithms of this form $\log k /[a\s H]$, since it depends explicitly on time.\footnote{The UV theory also produces logarithms of the form $\log \mu /H$ where $\mu$ is the renormalization scale~\cite{Senatore:2009cf}.  These logs come from the RG flow at scales above $H$ and thus are captured by the EFT as a contribution to the matching coefficients.  Furthermore, such terms can be consistently removed by taking the matching scale to be $\mu =H$~\cite{Green:2020txs}.} 
Instead, if we want to resum such large logs in analogy with the RG, we must use the DRG~\cite{Tanaka:1975ti,Boyanovsky:1998aa,Boyanovsky:2003ui}, see~\cite{Boyanovsky:2004gq,McDonald:2006hf,Podolsky:2008qq,Burgess:2009bs,Dias:2012qy,Green:2020txs} for previous cosmological applications.

We introduce a reference time and distance, $a_\star$ and $k_\star$, which we will use to define a subtraction point. 
We then include a counterterm $Z$ to renormalize the operator $\widetilde \O_{2} = Z\s \O_{2}$ with
\beq
Z = 1- \frac{1}{2\s(\alpha-1)} +\gamma_E - \frac{c_{3,1}}{16\s \pi^2}  \log \frac{k_\star}{a_\star\s H}  \ .
\eeq
The power spectrum of the renormalized operator is thus
\begin{align}
\Big\langle \widetilde\O_{2}\big(\s\k\,\big)  \s \widetilde\O_{2}\big(\,\kp\s\big) \Big\rangle_\text{NG}'  \simeq \left(1+\frac{c_{3,1}}{8 \pi^2} \left(  \log \frac{k /[a\s H]}{k_\star /[a_\star\s H]} + \log \frac{k}{K_\text{IR}} \right) \right) \times \Big\langle \O_{2}\big(\s\k\,\big)  \s \O_{2}\big(\,\kp\s\big) \Big\rangle'_{\rm G} \ , \notag\\
\end{align}
Observables must be independent of the unphysical $k_\star$ and $a_\star$, which implies the following differential equation must hold for the two point function:
\begin{align}
  \frac{\partial}{\partial \log\frac{k_\star}{a_\star H}}\Big\langle \widetilde\O_{2}\big(\s\k\,\big)  \s \widetilde\O_{2}\big(\,\kp\s\big) \Big\rangle_\text{NG}' 
    = -2\s \gamma_2\s \Big\langle \widetilde\O_{2}\big(\s\k\,\big)  \s \widetilde\O_{2}\big(\,\kp\s\big) \Big\rangle_\text{NG}' \ ,
\end{align}
where we have defined 
\beq
\gamma_2  = \frac{\partial Z}{\partial \log\frac{k_\star}{a_\star H} }  =\frac{c_{3,1}}{16\s \pi^2} + {\cal O}\Big(c_{m,n}^2\Big) \ .
\eeq
Here we make a crucial assumption that all the time-dependent logs can be absorbed into counterterms such that they vanish when $a_\star = a$ and $k_\star = k$.  
By construction, $k_\star/(a_\star H)$ only appears in the ratio $[k /[a\s H]]/[k_\star/(a_\star H)]$, so that we can rewrite this equation as 
\beq\label{eq:gammaRG}
\frac{\partial}{\partial \log\frac{k}{a\s H}} \Big\langle \widetilde\O_{2}\big(\s\k\,\big)  \s \widetilde\O_{2}\big(\,\kp\s\big) \Big\rangle_\text{NG}' 
    =  2\s\gamma_2 \s \Big\langle \widetilde\O_{2}\big(\s\k\,\big)  \s \widetilde\O_{2}\big(\,\kp\s\big) \Big\rangle_\text{NG}'  \ .
\eeq
Notice that since we factored out the time-dependence $[a\s H]^{-4}$ by using $\varphi_+$ fields we do not need to include classical evolution.  
Solving this equation, we conclude that the power-spectrum is given by 
\beq
\Big\langle \widetilde\O_{2}\big(\s\k\,\big)  \s \widetilde\O_{2}\big(\,\kp\s\big) \Big\rangle_\text{NG}'  =-\frac{k}{8\s \pi^2} \left(\frac{k}{a\s H} \right)^{2\s \gamma_2} \ .
\eeq
This is identical to the calculation derived in the UV description~\cite{Green:2020txs}, once an overall factor of $H^4\s [a\s H]^{-4}$ is removed in order to map from operators built using $\phi$ to $\varphi_+$.
This makes a compelling case that the SdSET captures the long wavelength physics for massive scalar fields in dS.

\subsection{Light Scalars and Stochastic Inflation}
Having worked through a number of calculations for a massive scalar, we now turn to the more interesting case of nearly massless scalars.
The key difference is already manifest at the level of power counting.  
As $\alpha \to 0$, we use \cref{eq:powercounting} to deduce that our leading interactions
\beq
{\cal H}_{\rm int} =  \sum_n \frac{c_{n,1}}{n!}\s [a\s H]^{(1-n)\s\alpha}\s \varphi_+^{n}\s\varphi_- \ ,
\label{eq:HintMassless}
\eeq
are all becoming marginal, \emph{i.e.}, they all scale as $\lambda^0$ when $m = 0$.
Since the scaling dimensions of these operators are all approaching the same value, substantial operator mixing can occur.  
More significantly, the composite operators $\varphi_+^n$ all power count as $\lambda^0$ as the mass is taken to zero, and they similarly will mix.

At tree-level, we can use the interactions in \cref{eq:HintMassless} to generate non-Gaussian correlations.  
For example, the coupling $c_{3,1}$ generates a trispectrum for $\varphi_+$, 
\begin{align}
\Big\langle \varphi_{+} \big(\s\k_1\big)\s \varphi_{+}\big(\s\k_2\big)\s \varphi_{+} \big(\s\k_3\big)\s \varphi_{+} \big(\s\k_4\big) \Big\rangle_\text{tree}' &= \frac{c_{3,1}}{6\s\big(k_1\s k_2\s k_3\big)^{3-2\s\alpha}}\s\frac{C_\alpha^6}{8\s\nu } \int \dd\t\, [a\s H]^{-2\s \alpha} + {\rm permutations} \notag\\[8pt] 
&=-c_{3,1} \s \left(\frac{k_\text{tot}}{a\s H}\right)^{2\s \alpha} \frac{ k_\text{tot}^{-2\s\alpha} \sum_{i} k_i^{3-2\s\alpha} }{6\s\big(k_1\s k_2\s k_3\s k_4\big)^{3-2\s\alpha}}\s \frac{C_\alpha^6}{16\s  \nu\s \alpha} \label{eq:tree}  \ . 
\end{align}
where $k_\text{tot} = \sum_i k_i$.  
As expected from \cref{eq:NG_scaling}, this correlator has an overall scaling of $\mathcal{K}^{-9+4\s\alpha}\s \big(k_\text{tot} /[a\s H]\big)^{2\s\alpha}$, where $\mathcal{K}$ is a factor that has the same units as momentum.  
We can take the limit $\alpha \to 0$, being careful to account for the $\alpha$-dependence of the dimensions of $\varphi_+$, to find a contribution that is proportional to $\log k_\text{tot} /[a\s H]$.\footnote{For example, we could multiply \cref{eq:tree} by $\mathcal{K}^{9-4\s\alpha}$ (some reference momentum scale) such that both sides of this equation are dimensionless before taking the $\alpha \to 0$ limit.}
As with massive theories, such logs can become large, indicating the need to improve perturbation theory using DRG techniques.

In addition to logs associated with time integrals, we also encounter divergent momentum integrals.  
This is apparent even at the level of the one-loop correction to the $\varphi_+$ power spectrum:
\beq
\Big\langle \varphi_+\big(\s\k\,\big) \varphi_+\big(\,\kp\s\big) \Big\rangle_\text{1-loop}'  = \frac{c_{3,1}}{2\s\nu} \frac{C_\alpha^2}{2\s k^{3-2\s\alpha}}  \frac{[a\s H]^{-2\s \alpha} }{2\s\alpha}  \int \frac{\dd^3 p}{(2\s\pi)^3} \frac{C_\alpha^2}{2\s p^{3-2\s\alpha}} \ .
\eeq
As $\alpha \to 0$, the integral over $\vec p$ is logarithmically divergent both in the UV and the IR.  
Since the time integral already generated a $1/\alpha$ pole, the limit $\alpha \to 0$ yields a $\log^2$, \emph{i.e.}, a sub-leading log, that should be resummed by the dynamical RG using the leading order anomalous dimension.

This leading-log behavior corresponds to operator mixing turning on as the scaling dimensions coincide.  
To determine the mixing, we consider a generic correlation function involving a local operator $\O_{n} = \varphi_+^n(\x,t)$.
Using only the Gaussian correlations in~\cref{eq:pluspower}, we see that contracting any two legs yields
\beq
\big\langle \O_n \ldots \big\rangle \supset  \big\langle \O_{n-2} \ldots \big\rangle \times  \binom{n}{2}\s \frac{C_\alpha^2}{2} \int \frac{\dd^3 p}{(2\s\pi)^3} \frac{H^{2-2\s\alpha}}{p^{3-2\s\alpha}} \ .
\eeq
As a scaleless integral, it is formally zero when regulated using dim reg.  
However, unlike in the massive case when $\alpha$ is of ${\cal O}(1)$, this integral is IR divergent since $\alpha \to 0$.  
Therefore, we need to isolate the UV divergence in order to calculate the anomalous dimension.
Regulating the IR as we did above in \cref{eq:anaomalous_K}, we find 
\beq
\big\langle \O_n(\x) \ldots \big\rangle \supset  \big\langle \O_{n-2}(\x) \ldots \big\rangle \times \binom{n}{2}\s \frac{C_\alpha^2}{4\s\pi^2} \left( - \frac{1}{2\s\alpha} - \gamma_E - \log  \frac{a\s H}{K_\text{IR}}  \right)  \ .
\label{eq:OpMix}
\eeq
We see that removing the $1/(-2\s\alpha)$ divergence requires renormalizing the operator $\O_{n-2}$, and so the DRG will mix operators.  
This was anticipated in the $\alpha \to 0$ limit, using the fact that all such operators power count as $\lambda^0$.

Unlike our expression for the anomalous dimension in \cref{eq:gammaRG}, the result in \cref{eq:OpMix} does not have an explicit logarithmic dependence on $\ksub$ or $\x$.  
This can be traced to having done this calculation in position space: the naive scaling dimension of $\O_n$ remains $\alpha$-dependent, but will give rise to the expected factors of $\log k/[a\s H]$ upon Fourier transforming.  
For our purposes, the appearance of non-trivial $a\s H$ dependence is sufficient.  
Renormalizing away the $-1/(2\s\alpha)$ factor, we can take $\alpha\to 0$ ($\nu \to 3/2$) and use $\dd\t = \dd \log a\s H$ to  find the DRG equations including operator mixing: 
\beq\label{eq:DRG_stoc}
\frac{\partial}{\partial \t}  \big\langle \O_n(\x) \ldots \big\rangle =- \frac{n}{3} \sum_{m>1}\frac{c_{m,1}}{m!}  \big\langle \O_{n-1}(\x)\s \O_m(\x) \ldots \big\rangle + \frac{n(n-1)}{8\s \pi^2} \s \big\langle \O_{n-2}(\x) \ldots \big\rangle \ ,
\eeq
where we have used the equations of motion, 
\beq
2\s \nu\s  \dot{\varphi}_+   = -\sum_{m>1}\s [a\s H]^{(2-m)\s\alpha} \s\frac{c_{m,1}}{m!}\s \varphi_+^{m} \ ,
\eeq
to evaluate the time derivative acting on $\mathcal{O}_n(\x)$.

The result in~\cref{eq:DRG_stoc} is equivalent to the equations for stochastic inflation~\cite{Starobinsky:1986fx}, that can be used to determine the classical probability distribution for $\phi(\x)$ in a dS universe.\footnote{Identifying $\langle \phi^n\rangle = \int \dd\phi\, p(\phi,t)\s \phi^n$ the DRG equation can be written as a Fokker-Planck equation for $p(\phi,t)$. See, \emph{e.g.}~\cite{Baumgart:2019clc} for a derivation.} 
While analyzing this as a diffusion-like equation is a useful way to understand the solutions to \cref{eq:DRG_stoc}~\cite{Starobinsky:1994bd,Gorbenko:2019rza}, we emphasize the derivation from a RG perspective shows that the origin of stochastic inflation is identical to the source of the anomalous scalings that manifest for massive theories, see Section~\ref{sec:mass}.  
As such, stochastic inflation should not be thought of as a special technique needed to understand light fields in dS space --- it is instead simply what falls out of the SdSET as the result of a textbook EFT analysis.

Now that we see how the canonical result arises, it is natural to wonder what the SdSET predicts for higher order corrections to~\cref{eq:DRG_stoc}. 
A primary advantage of working in the EFT description is to make power counting manifest.  
Operators of different dimension cannot mix under perturbative (D)RG flows.  
This implies that gradient terms do not contribute to the DRG evolution of $\varphi_+^n$.  
For example, take an interaction of the form 
\beq
{S}_{\rm int} \supset \int \dd^3 x\,\dd \t\, [a\s H]^{-2 -\alpha}\s\frac{c_{2,1}^{(2)}}{2}\s  \varphi_- \s\partial_i \varphi_+\s \partial^i \varphi_+ \ .
\eeq 
Then the equations of motion is 
\beq
-2\s \nu\s \dot{\varphi}_+ \supset [a\s H]^{-2 -\alpha}\s \frac{c_{2,1}^{(2)}}{2}\s \partial_i \varphi_+\s \partial^i \varphi_+ \ ,
\eeq
which would naively appear to correct the DRG given in \cref{eq:DRG_stoc}: 
\beq
\frac{\partial}{\partial \t}   \big\langle \O_n(\x) \ldots \big\rangle \stackrel{?}{\supset} - \frac{[a\s H]^{-2 - \alpha}}{3}\s \frac{c_{2,1}^{(2)}}{2}\s \big\langle \O_n(\x)\s \big(\partial_i \varphi_+\s \partial^i \varphi_+\big)(\x) \ldots \big\rangle \ .
\eeq
However, we can remove this term by redefining the composite operator 
\beq
  \O_n(\x) \s\s\to\s\s \O_n(\x)  - \frac{[a\s H]^{-2 - \alpha}}{3\s (2+\alpha) }\frac{c_{2,1}^{(2)}}{2} \big(\O_n\s \partial_i \varphi_+\s \partial^i \varphi_+\big)(\x\s) \ .
\label{eq:OnRedef}
\eeq
With this redefinition of $\O_n(\x)$, the gradient term no longer appears in the DRG equations, leaving only~\cref{eq:DRG_stoc}. %
Furthermore, the term added to $\O_n$ in~\cref{eq:OnRedef} is power suppressed by an extra factor of $\lambda^{2+2\s\alpha}$ and thus is inconsequential for the late time correlation functions of interest here.  
This freedom to remove terms that power count differently is a different way of seeing the scheme dependence that allowed us to avoid these terms in the first place though a judicious choice of regulator.  
Again, we emphasize that all we have done here is to rely on the familiar fact that non-trivial mixing can only occur between operators whose dimensions are degenerate.

Finally, we briefly comment on the possibility that other sources of non-trivial corrections to \cref{eq:DRG_stoc} exist.
First, we note that the UV theory can generate a variety of log contributions, some of which are physical while others are artifacts of how the integrals are regulated.
For example, when working in the UV, one could find corrections to $\alpha$ that are proportional to $\log k/[a\s H]$, which \emph{e.g.} can be extracted at tree-level from \cref{eq:BarPhi} by sending $\alpha \to \alpha + \delta \alpha$ and expanding for small $\delta \alpha$.
These are logs that are absorbed by matching the UV theory onto the EFT parameters, \emph{e.g.} $\alpha$, $\beta$, and/or $c_{n,m}$.
Such contributions are quite common in the cosmology literature, and often arise because the chosen regulators do not fully tame the possible divergences.  
For example, due to the form of the bulk-to-bulk propagator,\footnote{We thank Akhil Premkumar for discussions relevant to this point.} perturbative calculations of the wavefunction of the universe give rise to logs~\cite{Anninos:2014lwa} that are not generated within the EFT, and thus must be absorbed by finite shifts to $\alpha$ and $\beta$.
Other logs are predicted by the EFT, and can be summed by solving DRG equations.
We saw an example of this above when we showed that, in the limit $\alpha \to 0$, the time evolution of $\varphi_+$ is responsible for the leading-log contributions to the DRG, \emph{i.e.}, the first term on the right hand side of \cref{eq:DRG_stoc}. 
A comprehensive exploration of the possible anomalous contributions to the composite operators, like those computed in \cref{sec:mass}, that could yield additional corrections to \cref{eq:DRG_stoc} is left for future work.

\section{Dynamical Gravity and Inflation}
\label{sec:metric}
Thus far we have focused on scalar fluctuations in a fixed dS background.  
In this section, we will show how the SdSET can also be applied to settings where gravity is dynamical.
In particular, the fluctuations of the metric are of central interest to cosmologists, both observationally and theoretically.
The adiabatic fluctuations produced during inflation --- described in terms of $\zeta$, see~\cref{eq:ADM} --- seed the fluctuations observed in the CMB and the large scale structure of the Universe.  
Moreover, the conservation of these adiabatic modes is one of the central pillars with which we connect inflation to observations.
While this statement has been proven to all order in perturbation theory~\cite{Salopek:1990jq,Senatore:2012ya,Assassi:2012et}, these proofs are somewhat involved.
Tensor fluctuations are similarly expected to be conserved to all-loop orders.  
This statement has relevance not only to the predictions of inflation, but also to grander questions such as the stability of dS space itself.  
It is expected (but not explicitly demonstrated) that the proofs of the conservation of $\zeta$ can be extended to the tensors. 
Given the importance of these results to our understanding of cosmological spacetimes, our simpler demonstration serves to demystify the underlying physics, and allows us to prove the conservation of tensor modes. 
Finally, there is a dramatic possibility that the universe has undergone, or is undergoing, slow-roll eternal inflation.
However, this question remains poorly understood, due to a number of technical and conceptual challenges.

In this section, we will address all three questions in the context of the SdSET.  
We will see that the perturbative scalar and tensor fluctuations are obviously conserved as a simple consequence of power counting.  
In particular, unlike massless scalar fields in static dS, these fluctuations are constrained by additional symmetries and thus do not admit any marginal operators.  
Finally, we will show that the challenge of understanding slow-roll eternal inflation corresponds to studying the SdSET in the presence of relevant deformations.

\subsection{Metric Fluctuations During Inflation}\label{sec:pert_metric}

The first problem we will consider are the small scalar metric fluctuations during inflation.  
In the appropriate gauge~\cite{Maldacena:2002vr}, the metric can be written as 
\beq\label{eq:ADM}
\dd s^{2}=-N^{2}\s \dd t^{2}+a^2(t)\s e^{2\s\zeta(\x,t)} \left(e^{2\s\gamma(\x,t)}\right)_{ij} \left(\dd x^{i}+N^{i}\s \dd t\right)\left(\dd x^{j}+N^{j}\s \dd t\right) \ ,
\eeq
where $\gamma_i^i = \partial_i\s \gamma^{ij} = 0$ is the tensor mode and $\zeta(\x,t)$ is the adiabatic scalar fluctuation.  
Here the constrained fields $N$ and $N^i$ are the lapse and shift Lagrange multipliers of the Arnowitt-Deser-Misner formalism~\cite{Arnowitt:1959ah}.  
Now the Hubble constant depends on time, so that $\partial_t\s a(t)/ a(t)= H(t)$ and $(\partial_t H) / H^2 \ll 1 $ characterizes the geometry during inflation.
To describe metric fluctuations in pure dS, we can take  $\partial_t  H= 0$ and $\zeta = 0$.  
Inflation is characterized in part by $\partial_t H <0$, and importantly requires an additional field with a time-dependent vacuum expectation value to be active.  
This time dependence breaks the isometries of dS down to spatial translations and rotations.  
The simplest such example is a background scalar field that evolves linearly in time, $\phi(t) \simeq \partial_t \phi  \times t$ and $\partial_t^2 \phi \simeq 0$.

Our goal in this section is to show that correlation functions of $\zeta\big(\s\k\s\big)$ and $\gamma_{ij}\big(\s\k\s\big)$ are time-independent in the $k/[a\s H] \to 0$ limit.  
At the level of our quadratic action in \cref{eq:quadActionEFT}, we need to take $\alpha = 0$, and then show that there are no corrections due to interactions, \emph{e.g.}~those that contribute to the DRG.  
For intuition, we will show that in any local region, $\zeta(\x)$ and $\gamma_{ij}(\x)$ are effectively constant.
Then diffeomorphism invariance implies that constant $\zeta$ and $\gamma_{ij}$ should be locally indistinguishable from $\zeta=\gamma_{ij} = 0$.  
As such, $\partial_t \zeta$ or $\partial_t \gamma_{ij}$ could only be sourced by gradients, which vanish as $k/[a\s H] \to  0$.  
In what follows, we will show that the SdSET makes this intuition manifest.

The quadratic Lagrangian for $\zeta$ takes the form
\beq
{\cal L}_{2,\zeta} =- \frac{\Mpl^2\s \partial_t H}{H^2\s c_s^2 } \left(\partial_t \zeta^2 - a^{-2} \s c_s^2\s \partial_i \zeta\s \partial^i \zeta \right) \ ,
\eeq
where $c_s \leq 1$ is known as the  ``speed of sound" and the spatial (Latin) indices are contracted with $\delta_{ij}$.
Gravity alone will generate $\zeta$ self interactions, \emph{e.g.}~from the factor of $\sqrt{-g}$ that appears in the action. 
For contrast, the tensor fluctuations are less directly impacted by the inflationary dynamics, in that their quadratic action is not proportional to $\partial_t H$:
\beq
{\cal L}_{2,\gamma} = \frac{M_{\mathrm{pl}}^{2}}{8} \Big[\big(\partial_t{\gamma}_{i j}\big)^{2}-a^{-2} \big(\partial_k\s \gamma_{i j}\s \partial^k\s \gamma^{i j} \big)\Big]  \ .
\eeq
In other words, to lowest order, the tensor modes during inflation behave as if they are in static dS, up to slow-roll corrections proportional to $\epsilon = - \partial_t H /H^2 \ll 1$.

The action is constrained by non-linearly realized symmetries that act on $\zeta$ and $\gamma$.
These are due to large gauge transformations that leave the gauge fixed.  
In particular, when $\gamma_{ij} = 0$ there is a non-linearly realized $SO(4,1)$ symmetries~\cite{Hinterbichler:2012nm} that acts on the long wavelength $\zeta$ as %
\begin{subequations}
\begin{align} 
D_{\rm NL}&: \delta \zeta  =-1-\x \cdot \vec \partial_\x\s \zeta \\[6pt] 
K_{\rm NL}^{i}&: \delta \zeta =-2\s x^{i}-2\s x^{i}\left(\x \cdot \vec \partial_{\x}\s \zeta\right)+x^{2}\s \partial^{i}\s \zeta \ .
\end{align}
\end{subequations}
The ``dilatation," $D_{\rm NL}$, is present even when $\gamma_{ij} \neq  0$ and is easily demonstrated by checking that it is true for the metric in \cref{eq:ADM}.  
In particular, if we rescale the coordinates $ \x \to e^{-\eta}\s\x$ and simultaneously shift  $\zeta\to\zeta\s(e^{-\eta}\s \x) - \eta$, the form of the metric is unchanged.

Similarly, the tensors exhibit an infinite number of symmetry related to infinitesimal diffeomorphisms $\x \to \x + \vec \xi(\x)$ that acts on the tensor as
\beq
\delta \gamma_{i j}=\partial_{i}\s \xi_{j}+\partial_{j}\s \xi_{i} \ .
\eeq
For a polynomial of the form~\cite{Hinterbichler:2013dpa}
\beq
\xi^{(M)}_{i}=M_{i \ell_{1}}\s x^{\ell_{1}}+\frac{1}{2}\s M_{i \ell_{1} \ell_{2}}\s x^{\ell_{1}}\s x^{\ell_{2}}+\frac{1}{3 !}\s M_{i \ell_{1} \ell_{2} \ell_{3}}\s x^{\ell_{1}}\s x^{\ell_{2}}\s x^{\ell_{3}}+\ldots \ ,
\eeq
there are non-trivial choices for the tensors $M_{i \ell_1 ... \ell_n}$ that shift $\gamma$ non-linearly while keeping the gauge fixed.  
These non-linearly realized symmetries imply that a constant $\zeta$ or $\gamma$, or even some gradients thereof, have no locally measurable effects.  
In the Newtonian limit, these symmetries are just a reflection of the equivalence principle.  

What is relevant here is that these transformations should be interpreted as global symmetries that restrict the form that correlations functions of $\zeta$ and $\gamma$ can take.
These symmetries have two important consequences for our EFT: 
\begin{itemize}
\item The scaling dimension of $\zeta_+$ and $\gamma_+$, as defined in exact analogy with $\varphi_+$ in~\cref{eq:dof_split} with $\alpha = 0$, such that  $\zeta_+ \to \zeta_+ - \eta$ and $\gamma_+ \to \gamma_+ + \partial_{i}\s \xi^{(M)}_{j}+\partial_{j}\s \xi^{(M)}_{i}$ are good symmetries of the SdSET action.  
\item All interactions contain at least two derivatives and are therefore irrelevant by our power counting.  
\end{itemize}
Unlike the case of quantum field theory in fixed dS space, inflation does not respect the rescaling symmetry in \cref{eq:RescaleSym} due to the explicit time-dependence induced by the slow-rolling background~\cite{Cheung:2007st}. 
As a result, the connection between power counting and the appearance of factors of $a(t)$ can be modified by the explicit $t$ dependence.

Given this setup, all orders conservation of $\zeta_+$ and $\gamma_+$ follow trivially.\footnote{Here we are making minor assumptions that $\beta > 0$.  Because  the inflationary background can induce explicit time-dependence, we can have $\beta \neq 3$ even though $\alpha = 0$.   For example, the kinetic term normalization can be time dependent such that 
\beq
S_2 \supset \int \dd^3 x\, \dd \t\,  \rho(\t)\s   \big[\dot \varphi_+\s \varphi_- - \dot \varphi_-\s \varphi_+\big]   \ ,
\eeq
where we have assumed $\alpha+\beta =3$. In most models of inflation, this time-dependence is weak and for all practical purposes we have $\beta \simeq 3$. However, if $\rho(\t) \simeq \bar \rho \s a^{\gamma}(t)$, then we should re-scale the fields, just like in \cref{eq:FieldRedef}, so that $\varphi_\pm = \tilde \varphi_\pm [a \s H]^{-\gamma/2}$ and   
\beq
S_2   \supset \int \dd^3 x\, \dd \t\, \bar \rho\s \Big[\dot{\widetilde \varphi}_+\s \widetilde \varphi_- - \dot{ \widetilde \varphi}_-\s \widetilde \varphi_+\Big]  \ .
\eeq
The resulting dimensions, $\tilde \alpha = \alpha +\gamma/2$ and $\tilde \beta = \beta+\gamma/2$, obey $\tilde \alpha +\tilde \beta = 3 + \gamma$. Ultra-slow roll inflation~\cite{Kinney:2005vj,Namjoo:2012aa,Martin:2012pe} is a simple example that illustrates this point, since $\beta = -3$ in these scenarios. }
In particular, since $\alpha=0$ for both $\gamma_+$ and $\zeta_+$, there is no classical time dependence even after integrating out the UV modes.  
Then we need to argue that the leading contributions to $\zeta_+$ and/or $\gamma_+$ will not cause non-trivial time evolution due to interactions within the SdSET.  
Here we can rely on power counting. 
Such time dependence can only come from a marginal operator, which are limited to take the form
 \beq
 {\cal L}_{\rm int} \stackrel{?}{\supset} \zeta_+^{n-1} \zeta_- \qquad \text{and} \qquad  {\cal L}_{\rm int}\stackrel{?}{\supset} (\gamma_{+})^{n-1}_{ij}\s \gamma_-^{ij}  \ .
 \eeq
These terms violate the non-linearly realized symmetries and are thus excluded.  
As shown above, additional interactions are always suppressed by powers of $k/[a\s H]$, and thus do not contribute to the leading $\zeta_+$ correlations at late times.  
We conclude that $\zeta$ and $\gamma$ are conserved.

This argument for the conservation of $\gamma$ and $\zeta$ is, for all practical purposes, a consequence of Weinberg's initial work on the growth of quantum contributions to cosmological correlators~\cite{Weinberg:2005vy,Weinberg:2006ac}.
However, his argument was limited to the late time part of the time integration, and did not address how to interpret the regions of integration where $k|\tau| \gtrsim 1$, which is relevant since one of the integration limits is $t\to \infty$.
The problem has been solved here though the use of EFT; the short modes have been integrated out, thereby generating a description for the long wavelength dynamics alone.  
The problematic regions of integration are put on the same footing as power-law divergences in a conventional EFT: they simply renormalize the parameters of the EFT, and can be avoided altogether through the use of a scale-independent regulator like dyn dim reg.  
The all-orders proof presented in~\cite{Assassi:2012et} similarly showed that such contributions are negligible through an explicit diagrammatic argument that is reproduced by the SdSET power counting.

Finally, we note that unlike quantum field theory in fixed dS, the EFT of the physical degrees of freedom for the metric fluctuations may contain terms with inverse factors of the Laplacian, \emph{i.e.}, the theory not strictly local.  
These terms arise from integrating out $N$ and $N^i$.  
However, for our purposes, this non-locality does not impact the argument.  
The essential feature of relevance is simply that there are more derivatives in the numerator than the denominator, and so these non-local terms are power-counting suppressed.  
This feature can be checked by explicit calculation, and also follows from applying the non-linearly realized symmetries that are the result of diffeomorphism invariance.

\subsection{Slow-roll Eternal Inflation and Gravitational Backreaction}\label{sec:EI}

Thus far, we have shown that our EFT for the long wavelength modes can describe perturbative quantum field theory both for a fixed dS background and for small fluctuations of a dynamical metric. 
In particular, by applying SdSET power counting to these cases, we deduced that all EFT interactions are marginal or irrelevant.  
For the marginal interactions, we showed that some situations result in a non-trivial DRG flow, which can be calculated using perturbation theory.  
However, it was critical to these arguments that the operator 
\beq\label{eq:relevant}
S_{\rm int} \supset - \int \dd^3 x\, \dd \t\,\sqrt{-g} \, [ a \s H]^{- n\s \alpha} \frac{c_{n,0}}{n!}\s \varphi_+^n \ ,
\eeq
could be removed by a field redefinition, see \cref{sec:Interactions}.  
Said another way, these operators appear to be relevant when $\alpha <  3/ n$, but when working on a fixed dS background they only contribute to the evolution of $\varphi_-$, and thus can be removed by a field redefinition, up to a critical total derivative.  However, when the metric is dynamical, this term is no longer a total derivative and the operator in \cref{eq:relevant} gives a non-trivial coupling to the metric.
This situation is in close analogy to the cosmological constant, which does not contribute to the dynamics of the fields in a fixed background but yields profound consequences when the metric is dynamical.

The impact of \cref{eq:relevant} can be easily seen by starting from the UV description in terms of $\phi$, where the Friedman equation takes the form
\begin{align}
3\s \Mpl^2\s H^2 = T^{00} &= \frac{1}{2}\left[ \big(\partial_t \phi\big)^2 + a^{-2}\s \partial_i \phi\s \partial^i \phi \right] + V(\phi) \simeq\bar H^4\s \left(  \sum_{n=0}^\infty [a\s H]^{-n\s \alpha}\s \frac{c_{n,0}}{n!}\s \varphi_+^n \right) \ ,
 \label{eq:FriedmanEq}
\end{align}
where $\bar H^4$ is a non-dynamical value fixed at some reference time that is convenient to introduce when performing calculations, and we have neglected the power suppressed derivative terms in the second line.  
Unlike the EFT in fixed dS, the sum runs over all $n$.
When $\alpha > 0$, the leading operator is the most relevant one, namely $n =0$ or the cosmological constant. 
As $\alpha \to 0$, all $n$ operators power count as $1/\lambda^3$, \emph{i.e.}, they are relevant interactions, implying that they all contribute to the gravitational background; the same conclusion holds if $\alpha$ is negative.  
Our claim is that this tower of relevant operators are responsible for transitioning the dynamics into a novel phase known as ``slow-roll eternal inflation.''

Slow-roll eternal inflation can be obtained as a limit of conventional inflation,\footnote{A number of earlier works have demonstrated that it is possible to maintain perturbative control of such models~\cite{Creminelli:2008es,Dubovsky:2008rf,Lewandowski:2013aka}.} where one imagines the field $\phi$ has evolved to some point along the potential such that $V'(\phi) \neq 0$.
It is useful to distinguish the classical evolution $\phi_\text{C}(t)$ from the quantum fluctuations $\phi_\text{Q}$.  
In a Hubble time, the classical field moves a distance $\Delta \phi_{\rm C} = \partial_t \phi / H$, while quantum effects cause the field to fluctuate an amount of order Hubble, $\Delta  \phi_{\rm Q} \simeq H$.  
Then the intuition is simply that once $|\Delta \phi_{\rm C}| \ll |\Delta \phi_{\rm Q}|$, the dynamics that cause the end of inflation is dominated by quantum fluctuations.  
More precisely, it has been shown~\cite{Creminelli:2008es} that when  $\partial_t \phi^2 < \frac{3}{2\s\pi^2}\s H^4$, the reheating volume diverges, which is characteristic of eternal inflation.

While a complete analysis of SdSET in the eternally inflating regime is beyond the scope of this work, it is straightforward to see why it is a unique challenge that takes us beyond perturbative metric fluctuations in dS, \emph{i.e.}, classic slow-roll inflation.  
For the purpose of illustration, let us consider eternal hilltop inflation~\cite{Barenboim:2016mmw}, such that the potential has a maximum and long plateau. 
Defining the maximum of the potential to be $\phi=0$, we have $V'(0) =  0$ and $V''(0) <0$, \emph{i.e.}, $m^2 < 0$ which implies that $\alpha < 0$.  
An immediate consequence of $\alpha < 0$ is that the interactions in \cref{eq:relevant} are now more relevant than the cosmological constant, which power counts as $\lambda^{-3}$; thus the $\varphi_+^n$ interactions will dominated the late time behavior.  
Of course, this is simply the statement that the dS solution around $\phi = 0$ is unstable, and is not necessarily indicative of eternal inflation.  
Instead, external inflation arises as we take $\alpha \to 0$ and $c_{n,0} \ll 1$ such that the time evolution is dominated by the quantum fluctuations in the stochastic inflationary framework.

However, one of the main challenges in understanding stochastic inflation is the lack of a well-defined observable in the absence of  a fixed boundary~\cite{Bousso:1999cb,Strominger:2001pn,Witten:2001kn,Mazur:2001aa,Maldacena:2002vr,Alishahiha:2004md}.  We can see this problem already when using the gauge where the fluctuations are characterized by the metric fluctuation $\zeta$ in~\cref{eq:ADM}. In \cref{sec:pert_metric}, we showed that this mode is governed by irrelevant interactions outside the horizon and thus would appear to be controlled in perturbation theory, in contrast to our description in terms of stochastic inflation. This apparent contradiction is resolved as follows.  First, the metric is characterized by $a(t)\s e^{\zeta}$ and is not linear in $\zeta$. Second, fluctuations of $\zeta$ are large in the eternally inflating regime, namely
\beq
\Big\langle \zeta\big(\s\k\,\big) \zeta\big(\,\kp\s\big) \Big\rangle' = \frac{H^4}{2\s \big(\partial_t \phi\big)^2} > \frac{\pi^2}{3} \ ,
\label{eq:zetaCorrEternal}
\eeq
where this is a restatement of the condition for eternal inflation $\partial_t \phi^2 < \frac{3}{2\s\pi^2} H^4$~\cite{Creminelli:2008es}.  As a result, the fluctuations of the metric in terms $e^\zeta$ are exponentially large.  Finally, if we use a physical quantity, like the end of inflation, to define time slices on which to calculate correlation functions, the physical volumes are not only large but diverge~\cite{Creminelli:2008es,Dubovsky:2008rf,Lewandowski:2013aka}.  This is related to the larger challenge of making predictions in an eternally inflating universe~\cite{Freivogel:2011eg} (also known as the measure problem).

While the SdSET does not directly address the challenge of understanding eternal inflation, it may be useful for identifying an improved treatment by making the relevant physics manifest.  First, isolating the appropriate long-wavelength degrees of freedom may lead to a more clear understanding of observables and gauge invariance.  Second, the manifest locality of the relevant terms in the action may provide an alternate approach to describing or calculating the evolution.  We will explore this in later work.    

\section{Conclusions and Outlook}\label{sec:conclusions}

Understanding and interpreting the origin of both UV and IR divergences has been a persistent problem with the perturbation theory of cosmological correlators.  
A central difficulty is that, in contrast to quantum field theory in flat space, the scaling behavior of these loop corrections is not manifest in the microscopic description.  
The construction of the Soft de Sitter Effective Theory presented in this paper resolves this challenge for the long wavelength modes.  
As with flat space EFT, the behavior of loops is typically apparent from power counting, in that large IR effects can only appear as corrections to marginal and relevant EFT operators.  
A unique feature of SdSET is that the Lagrangian (given in \cref{eq:ActionEFT}) contains no relevant interactions in most circumstances, while marginal interactions can appear in theories with very light scalar fields.  
This feature makes manifest a number of well known results in the literature, including the (at most) logarithmic nature of the late time effects~\cite{Weinberg:2005vy,Weinberg:2006ac} and the all-orders conservation of $\zeta$~\cite{Salopek:1990jq,Senatore:2012ya,Assassi:2012et}.
Furthermore, we have trivially extended these proofs to related cases; in particular the all-orders conservation of tensor fluctuations is proven here for the first time.

A primary goal of this work was to elucidate the SdSET operator structures and their power counting.  
While we provided some simple applications above, we expect that, by isolating the important terms and more straightforwardly regulating time and momentum integrals, this EFT will simplify more complex problems.  
The framework of stochastic inflation is one area where we have already demonstrated improved insight.  
Looking forward, the SdSET approach should make identifying universal corrections more straightforward, although a complete demonstration will require a detailed matching calculation.
A case for future study where the potential improvements are less obvious is stochastic particle production due to disorder, as calculated in~\cite{Garcia:2019icv,Garcia:2020mwi}.  
While numerically it was demonstrated that the superhorizon modes undergo a log-normal random walk, showing this analytically is hampered by the same power counting and integral regulation problems described above.  
As the SdSET resolves both these problems, this framework should make an analytic demonstration more straightforward.

A more ambitious goal is to use the knowledge gained from working with the SdSET to shed light on the nature of eternal inflation.  
While eternal inflation is within the domain of our EFT, it involves a relevant coupling to the metric and thus requires a non-perturbative treatment.  
This presents an obstacle to fully describing eternal inflation, which is additionally complicated by the lack of a fixed boundary on which to define the observables. 
Nevertheless, one may hope that having identified the appropriate degrees of freedom and low energy symmetries provides a critical step towards further understanding the physics.  
At the very least, the SdSET should simplify both of these obstacles due to now having exposed that there are a limited number of relevant degrees of freedom in the asymptotic future.

Finally, the structure of our EFT offers some similarity to dS holography~\cite{Witten:2001kn,Strominger:2001pn,Mazur:2001aa,Maldacena:2002vr,Maldacena:2011nz} that could be explored in more detail.  
The degrees of freedom of the EFT, $\varphi_+$ and $\varphi_-$ are commonly used holographically to describe the source and field operators of a CFT respectively.  
Similarly, since the $\varphi_\pm$ are operators with well defined scalings, they naturally appear in methods that exploit the isometries of dS, which act as conformal transformations on the late time fields~\cite{Creminelli:2011mw,Maldacena:2011nz,Kehagias:2012pd,Mata:2012bx,Arkani-Hamed:2018kmz,Baumann:2019oyu,Sleight:2019mgd,Sleight:2019hfp}.  
Applying these methods at loop level is limited by the same technical obstructions described above for perturbation theory in dS.  
It would be interesting the develop techniques that combine the strengths of the SdSET with these holographic approaches.
With the SdSET in hand, we expect that there are many interesting aspects of physics beyond the horizon that will now be simpler to explore.

\paragraph{Acknowledgements}
We are grateful to Mustafa Amin, Daniel Baumann, Matthew Baumgart, Raphael Flauger, Marcos Garcia, Victor Gorbenko, Austin Joyce, Hayden  Lee, Aneesh Manohar, Enrico Pajer,  Guilherme Pimentel, Rafael Porto, and Akhil Premkumar for helpful discussions. We also thank Daniel Baumann, Aneesh Manohar, and Akhil Premkumar for carefully reading the manuscript, and Matthew Baumgart and Raman Sundrum for allowing us to use their name for this EFT. T.\,C.~is supported by the US~Department of Energy, under grant no.~DE-SC0011640.  D.\,G.~is supported by the US~Department of Energy under grant no.~DE-SC0019035.

\appendix

\section*{Appendix}
\addcontentsline{toc}{section}{Appendix}
\section{Regulating the Trispectrum}\label{app:tri}
In \cref{sec:mass}, we showed that log divergences appear when calculating correlation functions of the composite operator $\O_2 =\varphi_+^2$, which can be resummed using the DRG.  
In this Appendix, we will provide the details of how the relevant integrals can be regulated without using a hard cutoff. 
We will first explain why dim reg is not a useful regulator, and then demonstrate how dyn dim reg can tame the divergences.

In the interest of pedagogy, we will focus on a conformally coupled scalar in $d$-dimensions, whose mode function is
\beq
\bar{\phi}\big(\s\k,\tau\big)= \frac{(-H\s \tau)^{(d-2)/2}}{\sqrt{2\s k}} e^{i\s k\s \tau}  \ .
\eeq
Using the standard in-in formula \cref{eq:inin_real},  and taking the limit $k\s\tau_0 \to 0$, we find
\begin{align}
\Big\langle \phi \big(\s\k_1\big)\s \phi\big(\s\k_2\big)\s \phi \big(\s\k_3\big)\s \phi \big(\s\k_4\big) \Big\rangle' &=  \lambda_\phi\s \frac{(-H\s \tau_0)^{2\s d-4}}{8\s k_1\s k_2\s k_3\s k_4 } \,{\rm Im}\! \int_{-\infty}^{\tau_0} \dd\tau\,  a^d(\tau) (-H\s \tau)^{2\s d-4} e^{i\s k_\text{tot}(\tau-\tau_0)} \nonumber \\[8pt]
&\to   \lambda_\phi\s \frac{(-H\s \tau_0)^{2\s d-4}\s \Gamma[d-3]}{8\s k_1\s k_2\s k_3\s k_4\s k_\text{tot}^{d-3} } \ ,
\end{align}
where $k_\text{tot} = k_1+k_2+k_3+k_4$.  
Interpreting this as a classical boundary condition for $\varphi_+$ removes the powers of $-H\tau_0 =a^{-1}$, yielding
\beq
\Big\langle \varphi_+ \big(\s\k_1\big)\s \varphi_+ \big(\s\k_2\big)\s \varphi_+ \big(\s\k_3\big)\s \varphi_+ \big(\s\k_4\big) \Big\rangle_c' =  \lambda_\phi\s \frac{ \Gamma[d-3]}{8\s k_1\s k_2\s k_3\s k_4\s k_\text{tot}^{d-3} } \ .
\eeq
From here, we can see the origin of the anomalous scaling.  The non-Gaussian correction to the $\O_2$ one-loop power spectrum is given by 
\beq
\Big\langle \O_{2}\big(\s\k\,\big)  \s \O_{2}\big(\,\kp\s\big) \Big\rangle'_{\rm NG} = \int \frac{\dd^{d-1} p_1\, \dd^{d-1} p_2}{(2\s\pi)^{2\s d-2}} \Big\langle \varphi_+ \big(\s\p_1\big)\s \varphi_+ \big(\s\p_1 - \k\big)\s \varphi_+ \big(\s \p_2 \big)\s \varphi_+ \big(-\k-\p_2\big) \Big\rangle' \,.
\eeq
We will focus on the contribution to the integral in the limit $p_1 \gg p_2 \simeq k$:
\beq
\Big\langle \O_{2}\big(\s\k\,\big)  \s \O_{2}\big(\,\kp\s\big) \Big\rangle'_{\rm NG} \simeq \frac{\lambda_\phi}{8}\s \Gamma[d-3] \int \frac{\dd^{d-1}p_1}{(2\s\pi)^{d-1}} \frac{1}{p_1^{d-1}}  \int \frac{\dd^{d-1} p_2}{(2\s\pi)^{d-1}} \frac{1}{p_2\s \big|\k+\p_2\big|}  \ ,
\eeq
where we have included an additional factor of two that results from interchanging $p_1\leftrightarrow p_2$.  
The $\p_1$ integral reveals a logarithmic divergence; however, the divergence is not regulated by dim reg.  
The problem is that mode functions\footnote{We only showed this for the case of a conformally coupled scalar.  With more work, one can show that this problem persists for general $\alpha$ and $d$.} change with $d$ such that the integral is logarithmically divergent in every dimension.

To regulate this divergence, we will first treat each $\phi\big(\s\k_i\big)$ as though it has a mass, or equivalently $\alpha_i \neq 0$, so that $\phi\big(\s\k_i\big) \to \phi_{\alpha_i}\big(\s\k_i\big)$.  
Using \cref{eq:mode_m} for the mode function $\bar \phi$, the trispectrum takes the form 
\begin{align}
\Big\langle \phi \big(\s\k_1\big)\s \phi\big(\s\k_2\big)\s \phi \big(\s\k_3\big)\s \phi \big(\s\k_4\big) \Big\rangle' &=  \lambda_\phi \s 2 \s {\rm Im}  \int_{-\infty}^{\tau_0} \frac{\dd\tau}{(-H\s \tau)^d} \s\bar \phi_{\alpha_1}\big(\s\k_1\big)\s \bar \phi_{\alpha_2}\big(\s\k_2\big)\s \bar \phi_{\alpha_3}\big(\s\k_3\big)\s\bar \phi_{\alpha_4}\big(\s\k_4\big)  \ .\nonumber \\
\end{align}
Calculating this integral for general $\alpha_i$ and $\ksub_i$ is difficult.  
However, for the purpose of evaluating the $\O_{2}$ power spectrum, we only need the result in the limit where $k_1, k_2 \simeq p \gg k_3, k_4\simeq k$.  
Furthermore, the integral becomes exponentially suppressed when $|\tau|\s p \gg 1$ so we can assume $\tau\s p \simeq 1$ and therefore $k\s \tau \ll 1$.  
As a result, we can use the superhorizon limit for the mode functions of $\ksub_3$ and $\ksub_4$. 
This limit of the trispectrum is given by 
\begin{align}\label{eq:reg_tri}
\Big\langle \phi \big(\s\p\s\big)\s \phi\big(-\p\s\big)\s \phi \big(\s\k_3\big)\s \phi \big(\s\k_4\big) \Big\rangle'&\simeq  \lambda_\phi \s  \frac{C_{\alpha_3}^2\s C_{\alpha_4}^2\s H^{6} \big(-k_3 \s\tau_0\big)^{\alpha_3}\s \big(-k_4 \s\tau_0\big)^{\alpha_4}}{4\s \big(k_3\s k_4\big)^{3}}\s \frac{\big(-p\s \tau_0\big)^{\alpha_1}\big(-p\s \tau_0\big)^{\alpha_2}}{2\s p^3} \notag\\[4pt]
&\hspace{-48pt}\times  2\s C_{\alpha_1}\s C_{\alpha_2} \s  {\rm Im }   \int_{-\infty}^{\tau_0} \frac{\dd\tau}{(-H\s \tau)^4}\s \bar\phi_{\alpha_1}\big(\p,\tau\big)\s \bar \phi_{\alpha_2}\big(-\p,\tau\big) \s \big(-k_3 \tau\big)^{\alpha_3} \big(- k_4 \tau\big)^{\alpha_4} \ . \notag\\ 
\end{align}

Our first goal is to determine the momentum scaling of this integral to show that the UV divergence is regulated.  
Here we can use two tricks:
\begin{enumerate}
\item The integral will be peaked at $|\tau|\s \sim p^{-1}$ so that we can replace $- p\s \tau$ are effectively constants and factors of $-\tau \to p^{-1}$.
\item The mode functions are $\bar\phi_{\alpha_i}\big(\s p, \tau\big) \sim p^{-3/2} (-p\s \tau)^{3/2}\s H^{(1)}_{\nu_i}(-p\s \tau)$.  But using $-\tau\s p \simeq 1$, we have $\bar\phi_{\alpha_i}\big(\s p, \tau\big) \sim p^{-3/2}$.
\end{enumerate}
Using these arguments to isolate the dominant contribution to the time integral, we find
\begin{align}
\Big\langle \phi \big(\s\p\s\big)\s \phi\big(-\p\s\big)\s \phi \big(\s\k_3\big)\s \phi \big(\s\k_4\big) \Big\rangle' &\sim \frac{(-k_3 )^{2\alpha_3} (-k_4 )^{2\s \alpha_4}}{(k_3\s k_4)^{3}} \frac{ p^{\alpha_1} p^{\alpha_2} p^{-\alpha_3} p^{-\alpha_4}}{p^{3}} \ .
\end{align}
We see that if $\alpha_i =  \alpha$, this expression scales as $p^{-3}$ for all $\alpha$. The key to regulating this integral is therefore to have $\alpha_{1,2} \neq \alpha_{3,4}$ such that the integral can converge in the UV.  We can then regulate the IR divergence as usual, see~\cref{eq:anaomalous_K}.

From the above observation, we see that taking $\alpha_1 = \alpha_2 = 1$ and $\alpha_3 = \alpha_4 = \alpha \neq 1$ will regulate the integral.  Furthermore, this choice simplifies \cref{eq:reg_tri} by using the conformal mass mode functions for the modes carrying momentum $p$ while only keeping general $\alpha$ for the superhorizon modes.  This choice is ideal so that we do not have to integrate over any Hankel functions. Plugging into \cref{eq:reg_tri}, we have
\begin{align}
\Big\langle \phi \big(\s\p\s\big)\s \phi\big(-\p\s\big)\s \phi \big(\s\k_3\big)\s \phi \big(\s\k_4\big) \Big\rangle' &\simeq  \lambda_\phi \s  \frac{C_{\alpha}^4\s H^{4}\s (-k_3\s \tau_0)^{\alpha}\s (-k_4\s \tau_0)^{\alpha}}{4\s \big(k_3\s k_4\big)^{3}} \frac{(-H\s \tau_0)^{2}}{2\s p} \notag\\[3pt]
&\hspace{27pt}\times2\s {\rm Im } \int_{-\infty}^{\tau_0} \frac{\dd\tau}{(-H\s \tau)^4} \frac{(-H\s \tau)^2}{2\s p}   \big(-k_3\s \tau\big)^{\alpha} \big(- k_4\s \tau\big)^{\alpha} e^{2\s i\s p\s  \tau}  \nonumber \\[8pt]
&\simeq \lambda_\phi \s  \frac{C_{\alpha}^4\s H^{4} }{8\s \big(k_3\s k_4\big)^{3-2\s\alpha}} \frac{(- \tau_0)^{2+2\s\alpha}}{p^{1+2\s\alpha} }\s \Gamma[2\s\alpha -1] \ .
\label{eq:reg_tri_conformal}
\end{align}
We can now determine the $\varphi_+$ correlation function using $\tau_0 =-[a\s H]^{-1}$ and $\phi(\tau_0\to 0) \simeq H\s [a\s H]^{-\alpha}\s \varphi_+$. 
It is sufficient to drop the $\alpha$ dependent constants as they will not contribute to the divergence or logarithmic terms in the $\alpha \to 1$ limit.  
The final $\varphi_+$ correlator is then
\beq
\Big\langle \varphi_+ \big(\s\p\s\big)\s \varphi_+\big(-\p\s\big)\s \varphi_+ \big(\s\k_3\big)\s \varphi_+ \big(\s\k_4\big) \Big\rangle' =  \s  \frac{\lambda_\phi }{8\s \big(k_3\s k_4\big)^{3-2\s\alpha}} \frac{1}{p^{1+2\s\alpha} }  \ .
\eeq
The $\alpha$-dependent power of $p$ in fixed $d=4$ (spacetime) dimensions is what will regulate our correlators in the main text.

\phantomsection
\addcontentsline{toc}{section}{References}
\small
\bibliographystyle{utphys}
\bibliography{dSRefs}

\providecommand{\href}[2]{#2}\begingroup\raggedright\begin{thebibliography}{100}

\bibitem{Bousso:1999cb}
R.~Bousso, ``{Holography in general space-times},''
  \href{http://dx.doi.org/10.1088/1126-6708/1999/06/028}{{\em JHEP} {\bfseries
  06} (1999) 028},
\href{http://arxiv.org/abs/hep-th/9906022}{{\ttfamily arXiv:hep-th/9906022
  [hep-th]}}.

\bibitem{Strominger:2001pn}
A.~Strominger, ``{The dS / CFT correspondence},''
  \href{http://dx.doi.org/10.1088/1126-6708/2001/10/034}{{\em JHEP} {\bfseries
  10} (2001) 034},
\href{http://arxiv.org/abs/hep-th/0106113}{{\ttfamily arXiv:hep-th/0106113
  [hep-th]}}.

\bibitem{Witten:2001kn}
E.~Witten, ``{Quantum gravity in de Sitter space},'' in {\em {Strings 2001:
  International Conference Mumbai, India, January 5-10, 2001}}.
\newblock 2001.
\newblock
\href{http://arxiv.org/abs/hep-th/0106109}{{\ttfamily arXiv:hep-th/0106109
  [hep-th]}}.
\newblock

\bibitem{Mazur:2001aa}
P.~O. Mazur and E.~Mottola, ``{Weyl cohomology and the effective action for
  conformal anomalies},''
  \href{http://dx.doi.org/10.1103/PhysRevD.64.104022}{{\em Phys. Rev.}
  {\bfseries D64} (2001) 104022},
\href{http://arxiv.org/abs/hep-th/0106151}{{\ttfamily arXiv:hep-th/0106151
  [hep-th]}}.

\bibitem{Maldacena:2002vr}
J.~M. Maldacena, ``{Non-Gaussian features of primordial fluctuations in single
  field inflationary models},''
  \href{http://dx.doi.org/10.1088/1126-6708/2003/05/013}{{\em JHEP} {\bfseries
  05} (2003) 013},
\href{http://arxiv.org/abs/astro-ph/0210603}{{\ttfamily arXiv:astro-ph/0210603
  [astro-ph]}}.

\bibitem{Alishahiha:2004md}
M.~Alishahiha, A.~Karch, E.~Silverstein, and D.~Tong, ``{The dS/dS
  correspondence},'' \href{http://dx.doi.org/10.1063/1.1848341}{{\em AIP Conf.
  Proc.} {\bfseries 743} no.~1, (2004) 393--409},
\href{http://arxiv.org/abs/hep-th/0407125}{{\ttfamily arXiv:hep-th/0407125
  [hep-th]}}.

\bibitem{Freivogel:2011eg}
B.~Freivogel, ``{Making predictions in the multiverse},''
  \href{http://dx.doi.org/10.1088/0264-9381/28/20/204007}{{\em Class. Quant.
  Grav.} {\bfseries 28} (2011) 204007},
  \href{http://arxiv.org/abs/1105.0244}{{\ttfamily arXiv:1105.0244 [hep-th]}}.

\bibitem{Salopek:1990jq}
D.~S. Salopek and J.~R. Bond, ``{Nonlinear evolution of long wavelength metric
  fluctuations in inflationary models},''
\href{http://dx.doi.org/10.1103/PhysRevD.42.3936}{{\em Phys. Rev.} {\bfseries
  D42} (1990) 3936--3962}.

\bibitem{Cheung:2007st}
C.~Cheung, P.~Creminelli, A.~Fitzpatrick, J.~Kaplan, and L.~Senatore, ``{The
  Effective Field Theory of Inflation},''
  \href{http://dx.doi.org/10.1088/1126-6708/2008/03/014}{{\em JHEP} {\bfseries
  03} (2008) 014}, \href{http://arxiv.org/abs/0709.0293}{{\ttfamily
  arXiv:0709.0293 [hep-th]}}.

\bibitem{Senatore:2009cf}
L.~Senatore and M.~Zaldarriaga, ``{On Loops in Inflation},''
  \href{http://dx.doi.org/10.1007/JHEP12(2010)008}{{\em JHEP} {\bfseries 12}
  (2010) 008},
\href{http://arxiv.org/abs/0912.2734}{{\ttfamily arXiv:0912.2734 [hep-th]}}.

\bibitem{Senatore:2012ya}
L.~Senatore and M.~Zaldarriaga, ``{The constancy of $\zeta$ in single-clock
  Inflation at all loops},''
  \href{http://dx.doi.org/10.1007/JHEP09(2013)148}{{\em JHEP} {\bfseries 09}
  (2013) 148},
\href{http://arxiv.org/abs/1210.6048}{{\ttfamily arXiv:1210.6048 [hep-th]}}.

\bibitem{Assassi:2012et}
V.~Assassi, D.~Baumann, and D.~Green, ``{Symmetries and Loops in Inflation},''
  \href{http://dx.doi.org/10.1007/JHEP02(2013)151}{{\em JHEP} {\bfseries 02}
  (2013) 151},
\href{http://arxiv.org/abs/1210.7792}{{\ttfamily arXiv:1210.7792 [hep-th]}}.

\bibitem{Ford:1984hs}
L.~H. Ford, ``{Quantum Instability of De Sitter Space-time},''
\href{http://dx.doi.org/10.1103/PhysRevD.31.710}{{\em Phys. Rev.} {\bfseries
  D31} (1985) 710}.

\bibitem{Antoniadis:1985pj}
I.~Antoniadis, J.~Iliopoulos, and T.~N. Tomaras, ``{Quantum Instability of De
  Sitter Space},''
\href{http://dx.doi.org/10.1103/PhysRevLett.56.1319}{{\em Phys. Rev. Lett.}
  {\bfseries 56} (1986) 1319}.

\bibitem{Starobinsky:1986fx}
A.~A. Starobinsky, ``{Stochastic de Sitter (Inflationary) Stage in the Early
  Universe},''
\href{http://dx.doi.org/10.1007/3-540-16452-9_6}{{\em Lect. Notes Phys.}
  {\bfseries 246} (1986) 107--126}.

\bibitem{Starobinsky:1994bd}
A.~A. Starobinsky and J.~Yokoyama, ``{Equilibrium state of a selfinteracting
  scalar field in the De Sitter background},''
  \href{http://dx.doi.org/10.1103/PhysRevD.50.6357}{{\em Phys. Rev. D}
  {\bfseries 50} (1994) 6357--6368},
  \href{http://arxiv.org/abs/astro-ph/9407016}{{\ttfamily
  arXiv:astro-ph/9407016}}.

\bibitem{Tsamis:1994ca}
N.~C. Tsamis and R.~P. Woodard, ``{Strong infrared effects in quantum
  gravity},''
\href{http://dx.doi.org/10.1006/aphy.1995.1015}{{\em Annals Phys.} {\bfseries
  238} (1995) 1--82}.

\bibitem{Tsamis:1996qm}
N.~C. Tsamis and R.~P. Woodard, ``{The Quantum gravitational back reaction on
  inflation},'' \href{http://dx.doi.org/10.1006/aphy.1997.5613}{{\em Annals
  Phys.} {\bfseries 253} (1997) 1--54},
\href{http://arxiv.org/abs/hep-ph/9602316}{{\ttfamily arXiv:hep-ph/9602316
  [hep-ph]}}.

\bibitem{Tsamis:1997za}
N.~C. Tsamis and R.~P. Woodard, ``{Matter contributions to the expansion rate
  of the universe},''
  \href{http://dx.doi.org/10.1016/S0370-2693(98)00159-2}{{\em Phys. Lett.}
  {\bfseries B426} (1998) 21--28},
\href{http://arxiv.org/abs/hep-ph/9710466}{{\ttfamily arXiv:hep-ph/9710466
  [hep-ph]}}.

\bibitem{Weinberg:2005vy}
S.~Weinberg, ``{Quantum contributions to cosmological correlations},''
  \href{http://dx.doi.org/10.1103/PhysRevD.72.043514}{{\em Phys. Rev.}
  {\bfseries D72} (2005) 043514},
\href{http://arxiv.org/abs/hep-th/0506236}{{\ttfamily arXiv:hep-th/0506236
  [hep-th]}}.

\bibitem{Weinberg:2006ac}
S.~Weinberg, ``{Quantum contributions to cosmological correlations. II. Can
  these corrections become large?},''
  \href{http://dx.doi.org/10.1103/PhysRevD.74.023508}{{\em Phys. Rev.}
  {\bfseries D74} (2006) 023508},
\href{http://arxiv.org/abs/hep-th/0605244}{{\ttfamily arXiv:hep-th/0605244
  [hep-th]}}.

\bibitem{Seery:2010kh}
D.~Seery, ``{Infrared effects in inflationary correlation functions},''
  \href{http://dx.doi.org/10.1088/0264-9381/27/12/124005}{{\em Class. Quant.
  Grav.} {\bfseries 27} (2010) 124005},
  \href{http://arxiv.org/abs/1005.1649}{{\ttfamily arXiv:1005.1649
  [astro-ph.CO]}}.

\bibitem{Burgess:2010dd}
C.~P. Burgess, R.~Holman, L.~Leblond, and S.~Shandera, ``{Breakdown of
  Semiclassical Methods in de Sitter Space},''
  \href{http://dx.doi.org/10.1088/1475-7516/2010/10/017}{{\em JCAP} {\bfseries
  1010} (2010) 017},
\href{http://arxiv.org/abs/1005.3551}{{\ttfamily arXiv:1005.3551 [hep-th]}}.

\bibitem{Rajaraman:2010xd}
A.~Rajaraman, ``{On the proper treatment of massless fields in Euclidean de
  Sitter space},'' \href{http://dx.doi.org/10.1103/PhysRevD.82.123522}{{\em
  Phys. Rev. D} {\bfseries 82} (2010) 123522},
  \href{http://arxiv.org/abs/1008.1271}{{\ttfamily arXiv:1008.1271 [hep-th]}}.

\bibitem{Marolf:2010zp}
D.~Marolf and I.~A. Morrison, ``{The IR stability of de Sitter: Loop
  corrections to scalar propagators},''
  \href{http://dx.doi.org/10.1103/PhysRevD.82.105032}{{\em Phys. Rev.}
  {\bfseries D82} (2010) 105032},
\href{http://arxiv.org/abs/1006.0035}{{\ttfamily arXiv:1006.0035 [gr-qc]}}.

\bibitem{Marolf:2011sh}
D.~Marolf and I.~A. Morrison, ``{The IR stability of de Sitter QFT: Physical
  initial conditions},''
  \href{http://dx.doi.org/10.1007/s10714-011-1233-3}{{\em Gen. Rel. Grav.}
  {\bfseries 43} (2011) 3497--3530},
\href{http://arxiv.org/abs/1104.4343}{{\ttfamily arXiv:1104.4343 [gr-qc]}}.

\bibitem{Marolf:2012kh}
D.~Marolf, I.~A. Morrison, and M.~Srednicki, ``{Perturbative S-matrix for
  massive scalar fields in global de Sitter space},''
  \href{http://dx.doi.org/10.1088/0264-9381/30/15/155023}{{\em Class. Quant.
  Grav.} {\bfseries 30} (2013) 155023},
\href{http://arxiv.org/abs/1209.6039}{{\ttfamily arXiv:1209.6039 [hep-th]}}.

\bibitem{Beneke:2012kn}
M.~Beneke and P.~Moch, ``{On ``dynamical mass'' generation in Euclidean de
  Sitter space},'' \href{http://dx.doi.org/10.1103/PhysRevD.87.064018}{{\em
  Phys. Rev. D} {\bfseries 87} (2013) 064018},
  \href{http://arxiv.org/abs/1212.3058}{{\ttfamily arXiv:1212.3058 [hep-th]}}.

\bibitem{Akhmedov:2013vka}
E.~T. Akhmedov, ``{Lecture notes on interacting quantum fields in de Sitter
  space},'' \href{http://dx.doi.org/10.1142/S0218271814300018}{{\em Int. J.
  Mod. Phys.} {\bfseries D23} (2014) 1430001},
\href{http://arxiv.org/abs/1309.2557}{{\ttfamily arXiv:1309.2557 [hep-th]}}.

\bibitem{Anninos:2014lwa}
D.~Anninos, T.~Anous, D.~Z. Freedman, and G.~Konstantinidis, ``{Late-time
  Structure of the Bunch-Davies De Sitter Wavefunction},''
  \href{http://dx.doi.org/10.1088/1475-7516/2015/11/048}{{\em JCAP} {\bfseries
  1511} no.~11, (2015) 048},
\href{http://arxiv.org/abs/1406.5490}{{\ttfamily arXiv:1406.5490 [hep-th]}}.

\bibitem{Burgess:2015ajz}
C.~P. Burgess, R.~Holman, and G.~Tasinato, ``{Open EFTs, IR effects \&
  late-time resummations: systematic corrections in stochastic inflation},''
  \href{http://dx.doi.org/10.1007/JHEP01(2016)153}{{\em JHEP} {\bfseries 01}
  (2016) 153},
\href{http://arxiv.org/abs/1512.00169}{{\ttfamily arXiv:1512.00169 [gr-qc]}}.

\bibitem{Akhmedov:2017ooy}
E.~T. Akhmedov, U.~Moschella, K.~E. Pavlenko, and F.~K. Popov, ``{Infrared
  dynamics of massive scalars from the complementary series in de Sitter
  space},'' \href{http://dx.doi.org/10.1103/PhysRevD.96.025002}{{\em Phys.
  Rev.} {\bfseries D96} no.~2, (2017) 025002},
\href{http://arxiv.org/abs/1701.07226}{{\ttfamily arXiv:1701.07226 [hep-th]}}.

\bibitem{Hu:2018nxy}
B.-L. Hu, ``{Infrared Behavior of Quantum Fields in Inflationary Cosmology --
  Issues and Approaches: an overview},''
  \href{http://arxiv.org/abs/1812.11851}{{\ttfamily arXiv:1812.11851 [gr-qc]}}.

\bibitem{Akhmedov:2019cfd}
E.~T. Akhmedov, U.~Moschella, and F.~K. Popov, ``{Characters of different
  secular effects in various patches of de Sitter space},''
  \href{http://dx.doi.org/10.1103/PhysRevD.99.086009}{{\em Phys. Rev.}
  {\bfseries D99} no.~8, (2019) 086009},
\href{http://arxiv.org/abs/1901.07293}{{\ttfamily arXiv:1901.07293 [hep-th]}}.

\bibitem{Gorbenko:2019rza}
V.~Gorbenko and L.~Senatore, ``{$\lambda \phi^4$ in dS},''
\href{http://arxiv.org/abs/1911.00022}{{\ttfamily arXiv:1911.00022 [hep-th]}}.

\bibitem{Baumgart:2019clc}
M.~Baumgart and R.~Sundrum, ``{De Sitter Diagrammar and the Resummation of
  Time},''
\href{http://arxiv.org/abs/1912.09502}{{\ttfamily arXiv:1912.09502 [hep-th]}}.

\bibitem{Mirbabayi:2019qtx}
M.~Mirbabayi, ``{Infrared dynamics of a light scalar field in de Sitter},''
  \href{http://arxiv.org/abs/1911.00564}{{\ttfamily arXiv:1911.00564
  [hep-th]}}.

\bibitem{Green:2020txs}
D.~Green and A.~Premkumar, ``{Dynamical RG and Critical Phenomena in de Sitter
  Space},'' \href{http://dx.doi.org/10.1007/JHEP04(2020)064}{{\em JHEP}
  {\bfseries 04} (2020) 064}, \href{http://arxiv.org/abs/2001.05974}{{\ttfamily
  arXiv:2001.05974 [hep-th]}}.

\bibitem{Hu:1996yt}
W.~Hu, D.~N. Spergel, and M.~J. White, ``{Distinguishing causal seeds from
  inflation},'' \href{http://dx.doi.org/10.1103/PhysRevD.55.3288}{{\em Phys.
  Rev. D} {\bfseries 55} (1997) 3288--3302},
  \href{http://arxiv.org/abs/astro-ph/9605193}{{\ttfamily
  arXiv:astro-ph/9605193}}.

\bibitem{Spergel:1997vq}
D.~N. Spergel and M.~Zaldarriaga, ``{CMB polarization as a direct test of
  inflation},'' \href{http://dx.doi.org/10.1103/PhysRevLett.79.2180}{{\em Phys.
  Rev. Lett.} {\bfseries 79} (1997) 2180--2183},
  \href{http://arxiv.org/abs/astro-ph/9705182}{{\ttfamily
  arXiv:astro-ph/9705182}}.

\bibitem{Dodelson:2003ip}
S.~Dodelson, ``{Coherent phase argument for inflation},''
  \href{http://dx.doi.org/10.1063/1.1627736}{{\em AIP Conf. Proc.} {\bfseries
  689} no.~1, (2003) 184--196},
  \href{http://arxiv.org/abs/hep-ph/0309057}{{\ttfamily arXiv:hep-ph/0309057}}.

\bibitem{Dalal:2007cu}
N.~Dalal, O.~Dore, D.~Huterer, and A.~Shirokov, ``{The imprints of primordial
  non-gaussianities on large-scale structure: scale dependent bias and
  abundance of virialized objects},''
  \href{http://dx.doi.org/10.1103/PhysRevD.77.123514}{{\em Phys. Rev. D}
  {\bfseries 77} (2008) 123514},
  \href{http://arxiv.org/abs/0710.4560}{{\ttfamily arXiv:0710.4560
  [astro-ph]}}.

\bibitem{Alvarez:2014vva}
M.~Alvarez {\em et~al.}, ``{Testing Inflation with Large Scale Structure:
  Connecting Hopes with Reality},''
  \href{http://arxiv.org/abs/1412.4671}{{\ttfamily arXiv:1412.4671
  [astro-ph.CO]}}.

\bibitem{Weinberg:1980wa}
S.~Weinberg, ``{Effective Gauge Theories},''
  \href{http://dx.doi.org/10.1016/0370-2693(80)90660-7}{{\em Phys. Lett. B}
  {\bfseries 91} (1980) 51--55}.

\bibitem{Georgi:1994qn}
H.~Georgi, ``{Effective field theory},''
  \href{http://dx.doi.org/10.1146/annurev.ns.43.120193.001233}{{\em Ann. Rev.
  Nucl. Part. Sci.} {\bfseries 43} (1993) 209--252}.

\bibitem{Tanaka:1975ti}
F.~Tanaka, ``{Coherent Representation of Dynamical Renormalization Group in
  Bose Systems},''
\href{http://dx.doi.org/10.1143/PTP.54.289}{{\em Prog. Theor. Phys.} {\bfseries
  54} (1975) 289--290}.

\bibitem{Boyanovsky:1998aa}
D.~Boyanovsky, H.~J. de~Vega, R.~Holman, and M.~Simionato, ``{Dynamical
  renormalization group resummation of finite temperature infrared
  divergences},'' \href{http://dx.doi.org/10.1103/PhysRevD.60.065003}{{\em
  Phys. Rev.} {\bfseries D60} (1999) 065003},
\href{http://arxiv.org/abs/hep-ph/9809346}{{\ttfamily arXiv:hep-ph/9809346
  [hep-ph]}}.

\bibitem{Boyanovsky:2003ui}
D.~Boyanovsky and H.~J. de~Vega, ``{Dynamical renormalization group approach to
  relaxation in quantum field theory},''
  \href{http://dx.doi.org/10.1016/S0003-4916(03)00115-5}{{\em Annals Phys.}
  {\bfseries 307} (2003) 335--371},
\href{http://arxiv.org/abs/hep-ph/0302055}{{\ttfamily arXiv:hep-ph/0302055
  [hep-ph]}}.

\bibitem{Glimm:1973kp}
J.~Glimm and A.~M. Jaffe, ``{Positivity of the $\phi^4$ in Three-dimensions
  Hamiltonian},'' \href{http://dx.doi.org/10.1002/prop.19730210702}{{\em
  Fortsch.\ Phys.} {\bfseries 21} (1973) 327--376}.

\bibitem{Appelquist:1974tg}
T.~Appelquist and J.~Carazzone, ``{Infrared Singularities and Massive
  Fields},'' \href{http://dx.doi.org/10.1103/PhysRevD.11.2856}{{\em Phys.\
  Rev.\ D} {\bfseries 11} (1975) 2856}.

\bibitem{Weinberg:1959nj}
S.~Weinberg, ``{High-energy behavior in quantum field theory},''
  \href{http://dx.doi.org/10.1103/PhysRev.118.838}{{\em Phys.\ Rev.} {\bfseries
  118} (1960) 838--849}.

\bibitem{Polchinski:1983gv}
J.~Polchinski, ``{Renormalization and Effective Lagrangians},''
\href{http://dx.doi.org/10.1016/0550-3213(84)90287-6}{{\em Nucl. Phys.}
  {\bfseries B231} (1984) 269--295}.

\bibitem{Maldacena:2011nz}
J.~M. Maldacena and G.~L. Pimentel, ``{On graviton non-Gaussianities during
  inflation},'' \href{http://dx.doi.org/10.1007/JHEP09(2011)045}{{\em JHEP}
  {\bfseries 09} (2011) 045},
\href{http://arxiv.org/abs/1104.2846}{{\ttfamily arXiv:1104.2846 [hep-th]}}.

\bibitem{Henningson:1998gx}
M.~Henningson and K.~Skenderis, ``{The Holographic Weyl Anomaly},''
  \href{http://dx.doi.org/10.1088/1126-6708/1998/07/023}{{\em JHEP} {\bfseries
  07} (1998) 023},
\href{http://arxiv.org/abs/hep-th/9806087}{{\ttfamily arXiv:hep-th/9806087
  [hep-th]}}.

\bibitem{Bianchi:2001de}
M.~Bianchi, D.~Freedman, and K.~Skenderis, ``{How to go with an RG flow},''
  \href{http://dx.doi.org/10.1088/1126-6708/2001/08/041}{{\em JHEP} {\bfseries
  08} (2001) 041},
\href{http://arxiv.org/abs/hep-th/0105276}{{\ttfamily arXiv:hep-th/0105276
  [hep-th]}}.

\bibitem{Bianchi:2001kw}
M.~Bianchi, D.~Freedman, and K.~Skenderis, ``{Holographic Renormalization},''
  \href{http://dx.doi.org/10.1016/S0550-3213(02)00179-7}{{\em Nucl. Phys.}
  {\bfseries B631} (2002) 159--194},
\href{http://arxiv.org/abs/hep-th/0112119}{{\ttfamily arXiv:hep-th/0112119
  [hep-th]}}.

\bibitem{Strominger:2001gp}
A.~Strominger, ``{Inflation and the dS/CFT Correspondence},''
  \href{http://dx.doi.org/10.1088/1126-6708/2001/11/049}{{\em JHEP} {\bfseries
  11} (2001) 049},
\href{http://arxiv.org/abs/hep-th/0110087}{{\ttfamily arXiv:hep-th/0110087
  [hep-th]}}.

\bibitem{McFadden:2009fg}
P.~McFadden and K.~Skenderis, ``{Holography for Cosmology},''
  \href{http://dx.doi.org/10.1103/PhysRevD.81.021301}{{\em Phys. Rev.}
  {\bfseries D81} (2010) 021301},
\href{http://arxiv.org/abs/0907.5542}{{\ttfamily arXiv:0907.5542 [hep-th]}}.

\bibitem{Creminelli:2011mw}
P.~Creminelli, ``{Conformal invariance of scalar perturbations in inflation},''
  \href{http://dx.doi.org/10.1103/PhysRevD.85.041302}{{\em Phys.\ Rev.\ D}
  {\bfseries 85} (2012) 041302},
  \href{http://arxiv.org/abs/1108.0874}{{\ttfamily arXiv:1108.0874 [hep-th]}}.

\bibitem{Isgur:1989vq}
N.~Isgur and M.~B. Wise, ``{Weak Decays of Heavy Mesons in the Static Quark
  Approximation},'' \href{http://dx.doi.org/10.1016/0370-2693(89)90566-2}{{\em
  Phys. Lett. B} {\bfseries 232} (1989) 113--117}.

\bibitem{Eichten:1989zv}
E.~Eichten and B.~R. Hill, ``{An Effective Field Theory for the Calculation of
  Matrix Elements Involving Heavy Quarks},''
  \href{http://dx.doi.org/10.1016/0370-2693(90)92049-O}{{\em Phys. Lett. B}
  {\bfseries 234} (1990) 511--516}.

\bibitem{Georgi:1990um}
H.~Georgi, ``{An Effective Field Theory for Heavy Quarks at Low-energies},''
  \href{http://dx.doi.org/10.1016/0370-2693(90)91128-X}{{\em Phys. Lett. B}
  {\bfseries 240} (1990) 447--450}.

\bibitem{Grinstein:1990mj}
B.~Grinstein, ``{The Static Quark Effective Theory},''
  \href{http://dx.doi.org/10.1016/0550-3213(90)90349-I}{{\em Nucl. Phys. B}
  {\bfseries 339} (1990) 253--268}.

\bibitem{Podolsky:2008qq}
D.~I. Podolsky, ``{Dynamical renormalization group methods in theory of eternal
  inflation},'' \href{http://dx.doi.org/10.1134/S0202289309010174}{{\em Grav.
  Cosmol.} {\bfseries 15} (2009) 69--74},
\href{http://arxiv.org/abs/0809.2453}{{\ttfamily arXiv:0809.2453 [gr-qc]}}.

\bibitem{Dias:2012qy}
M.~Dias, R.~H. Ribeiro, and D.~Seery, ``{The $\delta$N formula is the dynamical
  renormalization group},''
  \href{http://dx.doi.org/10.1088/1475-7516/2013/10/062}{{\em JCAP} {\bfseries
  10} (2013) 062}, \href{http://arxiv.org/abs/1210.7800}{{\ttfamily
  arXiv:1210.7800 [astro-ph.CO]}}.

\bibitem{Mirbabayi:2015hva}
M.~Mirbabayi and M.~Simonovi\'c, ``{Effective Theory of Squeezed Correlation
  Functions},'' \href{http://dx.doi.org/10.1088/1475-7516/2016/03/056}{{\em
  JCAP} {\bfseries 03} (2016) 056},
  \href{http://arxiv.org/abs/1507.04755}{{\ttfamily arXiv:1507.04755
  [hep-th]}}.

\bibitem{Georgi:1991mr}
H.~Georgi, ``{Heavy quark effective field theory},'' in {\em {TASI 91}},
  pp.~0589--630.
\newblock 8, 1991.

\bibitem{Neubert:1993mb}
M.~Neubert, ``{Heavy quark symmetry},''
  \href{http://dx.doi.org/10.1016/0370-1573(94)90091-4}{{\em Phys. Rept.}
  {\bfseries 245} (1994) 259--396},
  \href{http://arxiv.org/abs/hep-ph/9306320}{{\ttfamily arXiv:hep-ph/9306320}}.

\bibitem{Shifman:1995dn}
M.~A. Shifman, ``{Lectures on heavy quarks in quantum chromodynamics},'' in
  {\em {TASI 95}}, pp.~409--514.
\newblock 10, 1995.
\newblock \href{http://arxiv.org/abs/hep-ph/9510377}{{\ttfamily
  arXiv:hep-ph/9510377}}.

\bibitem{Wise:1997sg}
M.~B. Wise, ``{Heavy quark physics: Course},'' in {\em {Les Houches Summer
  School in Theoretical Physics, Session 68: Probing the Standard Model of
  Particle Interactions}}, pp.~1051--1089.
\newblock 7, 1997.
\newblock \href{http://arxiv.org/abs/hep-ph/9805468}{{\ttfamily
  arXiv:hep-ph/9805468}}.

\bibitem{Manohar:2000dt}
A.~V. Manohar and M.~B. Wise, {\em {Heavy Quark Physics}}, vol.~10.
\newblock Cambridge University Press, 2000.

\bibitem{Manohar:1995xr}
A.~V. Manohar, ``{Effective field theories},'' in {\em {10th Lake Louise Winter
  Institute: Quarks and Colliders}}, pp.~274--315.
\newblock 6, 1995.
\newblock \href{http://arxiv.org/abs/hep-ph/9508245}{{\ttfamily
  arXiv:hep-ph/9508245}}.

\bibitem{Kaplan:1995uv}
D.~B. Kaplan, ``{Effective field theories},'' in {\em {17th National Nuclear
  Physics Summer School 2005}}.
\newblock 6, 1995.
\newblock \href{http://arxiv.org/abs/nucl-th/9506035}{{\ttfamily
  arXiv:nucl-th/9506035}}.

\bibitem{Rothstein:2003mp}
I.~Z. Rothstein, ``{TASI lectures on effective field theories},'' in {\em {TASI
  2002}}.
\newblock 8, 2003.
\newblock \href{http://arxiv.org/abs/hep-ph/0308266}{{\ttfamily
  arXiv:hep-ph/0308266}}.

\bibitem{Kaplan:2005es}
D.~B. Kaplan, ``{Five lectures on effective field theory},'' in {\em {7th
  Summer School in Nuclear Physics Symmetries}}.
\newblock 10, 2005.
\newblock \href{http://arxiv.org/abs/nucl-th/0510023}{{\ttfamily
  arXiv:nucl-th/0510023}}.

\bibitem{Petrov:2016azi}
A.~A. Petrov and A.~E. Blechman, \href{http://dx.doi.org/10.1142/8619}{{\em
  {Effective Field Theories}}}.
\newblock WSP, 2016.

\bibitem{Manohar:2018aog}
A.~V. Manohar, ``{Introduction to Effective Field Theories},'' in {\em {Les
  Houches summer school}: {EFT in Particle Physics and Cosmology}}.
\newblock 4, 2018.
\newblock \href{http://arxiv.org/abs/1804.05863}{{\ttfamily arXiv:1804.05863
  [hep-ph]}}.

\bibitem{Cohen:2019wxr}
T.~Cohen, ``{As Scales Become Separated: Lectures on Effective Field Theory},''
  in {\em TASI 2018}.
\newblock 2019.
\newblock \href{http://arxiv.org/abs/1903.03622}{{\ttfamily arXiv:1903.03622
  [hep-ph]}}.

\bibitem{Beneke:1997zp}
M.~Beneke and V.~A. Smirnov, ``{Asymptotic expansion of Feynman integrals near
  threshold},'' \href{http://dx.doi.org/10.1016/S0550-3213(98)00138-2}{{\em
  Nucl. Phys. B} {\bfseries 522} (1998) 321--344},
  \href{http://arxiv.org/abs/hep-ph/9711391}{{\ttfamily arXiv:hep-ph/9711391}}.

\bibitem{Smirnov:2002pj}
V.~A. Smirnov, ``{Applied asymptotic expansions in momenta and masses},'' {\em
  Springer Tracts Mod. Phys.} {\bfseries 177} (2002) 1--262.

\bibitem{Grishchuk:1990bj}
L.~Grishchuk and Y.~Sidorov, ``{Squeezed quantum states of relic gravitons and
  primordial density fluctuations},''
  \href{http://dx.doi.org/10.1103/PhysRevD.42.3413}{{\em Phys. Rev. D}
  {\bfseries 42} (1990) 3413--3421}.

\bibitem{Kehagias:2012pd}
A.~Kehagias and A.~Riotto, ``{Operator Product Expansion of Inflationary
  Correlators and Conformal Symmetry of de Sitter},''
  \href{http://dx.doi.org/10.1016/j.nuclphysb.2012.07.004}{{\em Nucl. Phys.}
  {\bfseries B864} (2012) 492--529},
\href{http://arxiv.org/abs/1205.1523}{{\ttfamily arXiv:1205.1523 [hep-th]}}.

\bibitem{Mata:2012bx}
I.~Mata, S.~Raju, and S.~Trivedi, ``{CMB from CFT},''
  \href{http://dx.doi.org/10.1007/JHEP07(2013)015}{{\em JHEP} {\bfseries 07}
  (2013) 015}, \href{http://arxiv.org/abs/1211.5482}{{\ttfamily arXiv:1211.5482
  [hep-th]}}.

\bibitem{Arkani-Hamed:2018kmz}
N.~Arkani-Hamed, D.~Baumann, H.~Lee, and G.~L. Pimentel, ``{The Cosmological
  Bootstrap: Inflationary Correlators from Symmetries and Singularities},''
\href{http://arxiv.org/abs/1811.00024}{{\ttfamily arXiv:1811.00024 [hep-th]}}.

\bibitem{Baumann:2019oyu}
D.~Baumann, C.~Duaso~Pueyo, A.~Joyce, H.~Lee, and G.~L. Pimentel, ``{The
  Cosmological Bootstrap: Weight-Shifting Operators and Scalar Seeds},''
  \href{http://arxiv.org/abs/1910.14051}{{\ttfamily arXiv:1910.14051
  [hep-th]}}.

\bibitem{Sleight:2019mgd}
C.~Sleight, ``{A Mellin Space Approach to Cosmological Correlators},''
  \href{http://dx.doi.org/10.1007/JHEP01(2020)090}{{\em JHEP} {\bfseries 01}
  (2020) 090}, \href{http://arxiv.org/abs/1906.12302}{{\ttfamily
  arXiv:1906.12302 [hep-th]}}.

\bibitem{Sleight:2019hfp}
C.~Sleight and M.~Taronna, ``{Bootstrapping Inflationary Correlators in Mellin
  Space},'' \href{http://dx.doi.org/10.1007/JHEP02(2020)098}{{\em JHEP}
  {\bfseries 02} (2020) 098}, \href{http://arxiv.org/abs/1907.01143}{{\ttfamily
  arXiv:1907.01143 [hep-th]}}.

\bibitem{Arkani-Hamed:2015bza}
N.~Arkani-Hamed and J.~Maldacena, ``{Cosmological Collider Physics},''
\href{http://arxiv.org/abs/1503.08043}{{\ttfamily arXiv:1503.08043 [hep-th]}}.

\bibitem{Arkani-Hamed:2018bjr}
N.~Arkani-Hamed and P.~Benincasa, ``{On the Emergence of Lorentz Invariance and
  Unitarity from the Scattering Facet of Cosmological Polytopes},''
\href{http://arxiv.org/abs/1811.01125}{{\ttfamily arXiv:1811.01125 [hep-th]}}.

\bibitem{Benincasa:2018ssx}
P.~Benincasa, ``{From the flat-space S-matrix to the Wavefunction of the
  Universe},''
\href{http://arxiv.org/abs/1811.02515}{{\ttfamily arXiv:1811.02515 [hep-th]}}.

\bibitem{Baumann:2020dch}
D.~Baumann, C.~Duaso~Pueyo, A.~Joyce, H.~Lee, and G.~L. Pimentel, ``{The
  Cosmological Bootstrap: Spinning Correlators from Symmetries and
  Factorization},'' \href{http://arxiv.org/abs/2005.04234}{{\ttfamily
  arXiv:2005.04234 [hep-th]}}.

\bibitem{Luke:1999kz}
M.~E. Luke, A.~V. Manohar, and I.~Z. Rothstein, ``{Renormalization group
  scaling in nonrelativistic QCD},''
  \href{http://dx.doi.org/10.1103/PhysRevD.61.074025}{{\em Phys. Rev. D}
  {\bfseries 61} (2000) 074025},
  \href{http://arxiv.org/abs/hep-ph/9910209}{{\ttfamily arXiv:hep-ph/9910209}}.

\bibitem{Green:2020ebl}
D.~Green and E.~Pajer, ``{On the Symmetries of Cosmological Perturbations},''
  \href{http://arxiv.org/abs/2004.09587}{{\ttfamily arXiv:2004.09587
  [hep-th]}}.

\bibitem{Luke:1992cs}
M.~E. Luke and A.~V. Manohar, ``{Reparametrization invariance constraints on
  heavy particle effective field theories},''
  \href{http://dx.doi.org/10.1016/0370-2693(92)91786-9}{{\em Phys. Lett. B}
  {\bfseries 286} (1992) 348--354},
  \href{http://arxiv.org/abs/hep-ph/9205228}{{\ttfamily arXiv:hep-ph/9205228}}.

\bibitem{Weinberg:2008hq}
S.~Weinberg, ``{Effective Field Theory for Inflation},''
  \href{http://dx.doi.org/10.1103/PhysRevD.77.123541}{{\em Phys. Rev. D}
  {\bfseries 77} (2008) 123541},
  \href{http://arxiv.org/abs/0804.4291}{{\ttfamily arXiv:0804.4291 [hep-th]}}.

\bibitem{Green:2013rd}
D.~Green, M.~Lewandowski, L.~Senatore, E.~Silverstein, and M.~Zaldarriaga,
  ``{Anomalous Dimensions and Non-Gaussianity},''
  \href{http://dx.doi.org/10.1007/JHEP10(2013)171}{{\em JHEP} {\bfseries 10}
  (2013) 171},
\href{http://arxiv.org/abs/1301.2630}{{\ttfamily arXiv:1301.2630 [hep-th]}}.

\bibitem{Raju:2012zr}
S.~Raju, ``{New Recursion Relations and a Flat Space Limit for AdS/CFT
  Correlators},'' \href{http://dx.doi.org/10.1103/PhysRevD.85.126009}{{\em
  Phys. Rev.} {\bfseries D85} (2012) 126009},
\href{http://arxiv.org/abs/1201.6449}{{\ttfamily arXiv:1201.6449 [hep-th]}}.

\bibitem{Green:2020whw}
D.~Green and R.~A. Porto, ``{Signals of a Quantum Universe},''
  \href{http://arxiv.org/abs/2001.09149}{{\ttfamily arXiv:2001.09149
  [hep-th]}}.

\bibitem{Bernardeau:2001qr}
F.~Bernardeau, S.~Colombi, E.~Gaztanaga, and R.~Scoccimarro, ``{Large scale
  structure of the universe and cosmological perturbation theory},''
  \href{http://dx.doi.org/10.1016/S0370-1573(02)00135-7}{{\em Phys. Rept.}
  {\bfseries 367} (2002) 1--248},
  \href{http://arxiv.org/abs/astro-ph/0112551}{{\ttfamily
  arXiv:astro-ph/0112551}}.

\bibitem{Boyanovsky:2004gq}
D.~Boyanovsky and H.~J. de~Vega, ``{Particle decay in inflationary
  cosmology},'' \href{http://dx.doi.org/10.1103/PhysRevD.70.063508}{{\em Phys.
  Rev.} {\bfseries D70} (2004) 063508},
\href{http://arxiv.org/abs/astro-ph/0406287}{{\ttfamily arXiv:astro-ph/0406287
  [astro-ph]}}.

\bibitem{McDonald:2006hf}
P.~McDonald, ``{Dark matter clustering: a simple renormalization group
  approach},'' \href{http://dx.doi.org/10.1103/PhysRevD.75.043514}{{\em Phys.
  Rev.} {\bfseries D75} (2007) 043514},
\href{http://arxiv.org/abs/astro-ph/0606028}{{\ttfamily arXiv:astro-ph/0606028
  [astro-ph]}}.

\bibitem{Burgess:2009bs}
C.~P. Burgess, L.~Leblond, R.~Holman, and S.~Shandera, ``{Super-Hubble de
  Sitter Fluctuations and the Dynamical RG},''
  \href{http://dx.doi.org/10.1088/1475-7516/2010/03/033}{{\em JCAP} {\bfseries
  1003} (2010) 033},
\href{http://arxiv.org/abs/0912.1608}{{\ttfamily arXiv:0912.1608 [hep-th]}}.

\bibitem{Arnowitt:1959ah}
R.~L. Arnowitt, S.~Deser, and C.~W. Misner, ``{Dynamical Structure and
  Definition of Energy in General Relativity},''
  \href{http://dx.doi.org/10.1103/PhysRev.116.1322}{{\em Phys. Rev.} {\bfseries
  116} (1959) 1322--1330}.

\bibitem{Hinterbichler:2012nm}
K.~Hinterbichler, L.~Hui, and J.~Khoury, ``{Conformal Symmetries of Adiabatic
  Modes in Cosmology},''
  \href{http://dx.doi.org/10.1088/1475-7516/2012/08/017}{{\em JCAP} {\bfseries
  1208} (2012) 017},
\href{http://arxiv.org/abs/1203.6351}{{\ttfamily arXiv:1203.6351 [hep-th]}}.

\bibitem{Hinterbichler:2013dpa}
K.~Hinterbichler, L.~Hui, and J.~Khoury, ``{An Infinite Set of Ward Identities
  for Adiabatic Modes in Cosmology},''
  \href{http://dx.doi.org/10.1088/1475-7516/2014/01/039}{{\em JCAP} {\bfseries
  01} (2014) 039}, \href{http://arxiv.org/abs/1304.5527}{{\ttfamily
  arXiv:1304.5527 [hep-th]}}.

\bibitem{Kinney:2005vj}
W.~H. Kinney, ``{Horizon crossing and inflation with large eta},''
  \href{http://dx.doi.org/10.1103/PhysRevD.72.023515}{{\em Phys. Rev. D}
  {\bfseries 72} (2005) 023515},
  \href{http://arxiv.org/abs/gr-qc/0503017}{{\ttfamily arXiv:gr-qc/0503017}}.

\bibitem{Namjoo:2012aa}
M.~H. Namjoo, H.~Firouzjahi, and M.~Sasaki, ``{Violation of non-Gaussianity
  consistency relation in a single field inflationary model},''
  \href{http://dx.doi.org/10.1209/0295-5075/101/39001}{{\em EPL} {\bfseries
  101} no.~3, (2013) 39001}, \href{http://arxiv.org/abs/1210.3692}{{\ttfamily
  arXiv:1210.3692 [astro-ph.CO]}}.

\bibitem{Martin:2012pe}
J.~Martin, H.~Motohashi, and T.~Suyama, ``{Ultra Slow-Roll Inflation and the
  non-Gaussianity Consistency Relation},''
  \href{http://dx.doi.org/10.1103/PhysRevD.87.023514}{{\em Phys. Rev. D}
  {\bfseries 87} no.~2, (2013) 023514},
  \href{http://arxiv.org/abs/1211.0083}{{\ttfamily arXiv:1211.0083
  [astro-ph.CO]}}.

\bibitem{Creminelli:2008es}
P.~Creminelli, S.~Dubovsky, A.~Nicolis, L.~Senatore, and M.~Zaldarriaga, ``{The
  Phase Transition to Slow-roll Eternal Inflation},''
  \href{http://dx.doi.org/10.1088/1126-6708/2008/09/036}{{\em JHEP} {\bfseries
  09} (2008) 036}, \href{http://arxiv.org/abs/0802.1067}{{\ttfamily
  arXiv:0802.1067 [hep-th]}}.

\bibitem{Dubovsky:2008rf}
S.~Dubovsky, L.~Senatore, and G.~Villadoro, ``{The Volume of the Universe after
  Inflation and de Sitter Entropy},''
  \href{http://dx.doi.org/10.1088/1126-6708/2009/04/118}{{\em JHEP} {\bfseries
  04} (2009) 118}, \href{http://arxiv.org/abs/0812.2246}{{\ttfamily
  arXiv:0812.2246 [hep-th]}}.

\bibitem{Lewandowski:2013aka}
M.~Lewandowski and A.~Perko, ``{Leading slow roll corrections to the volume of
  the universe and the entropy bound},''
  \href{http://dx.doi.org/10.1007/JHEP12(2014)060}{{\em JHEP} {\bfseries 12}
  (2014) 060}, \href{http://arxiv.org/abs/1309.6705}{{\ttfamily arXiv:1309.6705
  [hep-th]}}.

\bibitem{Barenboim:2016mmw}
G.~Barenboim, W.-I. Park, and W.~H. Kinney, ``{Eternal Hilltop Inflation},''
  \href{http://dx.doi.org/10.1088/1475-7516/2016/05/030}{{\em JCAP} {\bfseries
  05} (2016) 030}, \href{http://arxiv.org/abs/1601.08140}{{\ttfamily
  arXiv:1601.08140 [astro-ph.CO]}}.

\bibitem{Garcia:2019icv}
M.~A. Garcia, M.~A. Amin, S.~G. Carlsten, and D.~Green, ``{Stochastic Particle
  Production in a de Sitter Background},''
  \href{http://dx.doi.org/10.1088/1475-7516/2019/05/012}{{\em JCAP} {\bfseries
  05} (2019) 012}, \href{http://arxiv.org/abs/1902.09598}{{\ttfamily
  arXiv:1902.09598 [astro-ph.CO]}}.

\bibitem{Garcia:2020mwi}
M.~A. Garcia, M.~A. Amin, and D.~Green, ``{Curvature Perturbations From
  Stochastic Particle Production During Inflation},''
  \href{http://arxiv.org/abs/2001.09158}{{\ttfamily arXiv:2001.09158
  [astro-ph.CO]}}.

\end{thebibliography}\endgroup

\end{document}